\documentclass{siamltex}
\usepackage{amsmath,amssymb,mathabx}
\usepackage{graphicx,esint,ulem}

\usepackage[english]{babel}
\usepackage{amsmath}
\usepackage{amsfonts}
\usepackage{fullpage}
\usepackage{graphicx}
\usepackage{setspace}
\usepackage{float}
\usepackage{multirow}

\usepackage{fancyhdr}
\usepackage{time}
\newtheorem{remark}{Remark}[section]
\title{Discrete Solitary Waves in Systems with Nonlocal Interactions and the Peierls-Nabarro Barrier}
\author{M. Jenkinson and M.I. Weinstein}

\begin{document}

\maketitle

\begin{abstract}
We study a class of discrete focusing nonlinear Schr{\"o}dinger equations (DNLS) with general  nonlocal  interactions.
 We prove the existence of onsite and offsite discrete solitary waves, which bifurcate from the trivial solution at the endpoint frequency of the continuous spectrum of  linear dispersive waves. We also prove exponential smallness, in the frequency-distance to the bifurcation point, of the Peierls-Nabarro energy barrier (PNB), as measured by the difference in Hamiltonian or mass functionals evaluated on the onsite and offsite states.
 These results extend those of the authors for the case of nearest neighbor interactions to a large class of nonlocal short-range and long-range interactions. 
  The appearance of distinct onsite and offsite states is a consequence of the breaking of continuous spatial translation invariance. The PNB plays a role in the dynamics of energy transport in such nonlinear Hamiltonian lattice systems. 

Our  class of nonlocal interactions is defined in terms of coupling coefficients, $J_m$, where  $m\in\mathbb{Z}$ is the lattice site index, with $J_m\simeq m^{-1-2s}, s\in[1,\infty)$ and $J_m\sim e^{-\gamma|m|},\ s=\infty,\ \gamma>0,$ (Kac-Baker). For $s\ge1$, the bifurcation is seeded by solutions of the (effective / homogenized)
 cubic focusing nonlinear Schr{\"o}dinger equation (NLS). However, for $1/4<s<1$, the bifurcation is controlled by the fractional nonlinear Schr{\"o}dinger equation, FNLS, with $(-\Delta)^s$ replacing $-\Delta$. The proof is based on a Lyapunov-Schmidt reduction strategy applied to a momentum space formulation. The PN barrier
 bounds require appropriate uniform decay estimates for the discrete Fourier transform of DNLS discrete solitary waves. A key role is also played by non-degeneracy of the ground state of FNLS, recently proved by Frank, Lenzmann \& Silvestre.
\end{abstract}


\section{Introduction}

  In this paper we study the discrete, focusing and cubically nonlinear Schr{\"o}dinger equations with nonlocal interactions. We consider the both short-range and 
  long-range interactions.
 We prove the existence of onsite and offsite discrete solitary waves, which bifurcate from the trivial solution at the endpoint frequency of the continuous spectrum of  linear dispersive plane waves. This is the limit of long waves of small amplitude.  The profile of such states is, at leading order, expressible in terms of the ground state of the continuum, in general fractional, nonlinear Schr{\"o}dinger equation. Further, we prove exponential smallness, in the frequency-distance to the bifurcation point, of the Peierls-Nabarro energy barrier (PNB), as measured by the difference in Hamiltonian or mass functionals evaluated on the onsite and offsite states. 

  The appearance of  both onsite and offsite states is a discreteness effect, in particular a consequence of the breaking of continuous translation invariance, and is therefore not captured by continuum approximations (homogenization, effective media). The PNB has played a role in a physical understanding of the dynamics of energy transport in nonlinear Hamiltonian lattice systems since the pioneering work of Peyrard and Kruskal \cite{Peyrard_Kruskal_1984}.
 In particular, the PNB is interpreted as the energy a localized discrete wave  must expend (shed to dispersive radiation) in order to transit from one lattice site to the next.

  Such radiation damping is the mechanism through which a traveling discrete localized structure slows and eventually gets trapped and pinned to a lattice site; 
   see the discussion in the introduction to \cite{Jenkinson_Weinstein_2015} and, for example, see \cite{Kevrekidis_2009,Oxtoby_Barashenkov_2007,Kevrekidis_Weinstein_2000}. On the general subject of the role of radiation damping in extended Hamiltonian systems and applications see, for example, \cite{Weinstein_2015,Soffer_Weinstein_1999,Soffer_Weinstein_2004,Soffer_Weinstein_2005}.

 \subsection{DNLS and FNLS}  We consider the discrete and nonlocal nonlinear Schr{\"o}dinger equation (DNLS)
\begin{align}
& i \partial_t u_n(t) = - (\mathcal{L} u)_n(t)  - |u_n(t) |^2 u_n(t) \ ,\ n\in\mathbb{Z},  \ \ t \geq 0 \label{eqn:IVPdnlsnonlocal}
%\\
%& u_n(0) = f_n \in l^2(\mathbb{Z}). 
\end{align}
Here, $\mathcal{L} $ is a linear and nonlocal interaction operator:
 \begin{align}
(\mathcal{L} u)_n = \sum_{  m \in \mathbb{Z}   } J_{|m - n|} \ (u_m - u_n) . \label{eqn:Ldef}
\end{align}
The range of the nonlocal interaction is characterized by the rate of decay of the coupling sequence, $\{J_{m}\}$.
 Nonlocal interactions arising in applications typically have coupling sequences, $J_m=J_m^s$ with 
  polynomial decay $ J_m^s \simeq  m^{-1-2s},\  0<s<\infty$, or exponential (Kac-Baker) decay: $J_m^\infty \simeq e^{- \gamma m}$, $ \gamma > 0 $. 
We refer to the case $s > 1$ as the  {\it short-range interaction} case, and the case $0 < s < 1$ as {\it long-range interaction} case. The case $s = 1$ is the {\it critical or marginal range interaction}. 
The case $J_0=0, J_{\pm1} = 1$ and $J_m = 0, |m| \geq 1$, the operator $\mathcal{L}$ corresponds to the nearest-neighbor discrete Laplacian. 

DNLS is a Hamiltonian system, expressible in the form
\begin{align}
i\partial_t{ u} &= \frac{ \delta\mathcal{H}[u,\overline{u}] }{\delta \overline{u}},\ \ {\rm where}\label{dnls-hamsys}\\
\mathcal{H}[{u},\overline{u}] &=  \frac{1}{2 }  \sum_{n\in\mathbb{Z}} \sum_{  \substack{ m \in \mathbb{Z}  \\ m \neq n} } \  J_{|m - n|} | u_m - u_n|^2-\frac12 |u_n|^4\ .
\label{dnls-ham}\end{align}
We note the following basic result concerning $\mathcal{L}$ acting on 
$l^2(\mathbb{Z})$.
\begin{proposition} \label{prop:Lprops}
Assume $ J= \{J_m \}_{m \in \mathbb{Z}}$ is non-negative ($J_m\ge0$), symmetric ($J_m = J_{|m|}$), and $J \in l^1(\mathbb{Z})$. 
Then,
\begin{enumerate}
\item $\mathcal{L}$ is a bounded linear operator on $l^2(\mathbb{Z})$.
\item $\mathcal{L}$ is self-adjoint.
\item $-\mathcal{L}$ is non-negative.
\item The spectrum of $-\mathcal{L}$ is continuous and equal to $[0,M_\star]$, where $M_\star =4 \underset{q \in [- \pi, \pi]}{ \max}\sum_{m = 1}^{\infty} J_m \ \sin^2 \left( \frac{qm}{2}\right)$.
\end{enumerate}
\end{proposition}
\noindent {\it Proof:} Young's inequality implies boundedness. The other details of the proof are presented in Appendix \ref{appendix:operatorL}.

The initial value problem 
is globally well-posed, in the sense that for each ${f} = \{ f_n \}_{n \in \mathbb{Z}} \in l^2(\mathbb{Z})$ there exists a unique global solution $u(t) = \{ u_n(t) \}_{n \in \mathbb{Z}} \in C^1([0,\infty), l^2(\mathbb{Z}))$  \cite{Kirkpatrick_Lenzmann_2012}. 
Time translation invariance implies that  $\mathcal{H}[{u},\overline{u}]$ is time-invariant on solutions and the invariance $u\mapsto e^{i\theta}u$ implies the time-invariance of 
\begin{equation}
\mathcal{N}[{u},\overline{u}]=\sum_{n\in\mathbb{Z}} |u_n|^2 .
\label{N-l2}\end{equation}
%Well-posedness follows from a standard contraction mapping argument applied to the equivalent %integral equation formulation of the initial value problem.
%In particular, their proof relies on the inclusion $\|{f}\|_{l^\infty(\mathbb{Z})}\lesssim\|{ f}\|_{l^2(\mathbb{Z})}$.

   In \cite{Kirkpatrick_Lenzmann_2012} it was proved by
 Kirkpatrick, Lenzmann and Staffilani  that on any fixed time interval, $[0,T]$, the family of solutions which is parametrized by a discretization parameter tending  to zero, converges in some norm  to the solution of the initial value problem for the, in general, fractional nonlinear Schr{\"o}dinger equation (FNLS) with focusing cubic nonlinearity:
\begin{align}
& i \partial_t u(x,t) = (- \Delta_x)^{^{p}} \ u(x,t) - |u(x,t)|^2 u(x,t),\ \ x \in \mathbb{R}, \ t \in \mathbb{R}\ . \label{eqn:introNLS} 
\end{align}
where  $p=p(s)=\min{(1,s)}$ depends on the interaction range parameter, $s$. 
 (In terms of the Fourier transform, $\mathcal{F}$ and its inverse $\mathcal{F}^{-1}$, we define $(- \Delta)^{^{p}} v(x) = \mathcal{F}^{-1}\left[ |q|^p \mathcal{F}v(q)\right](x)$.) 
  Thus, if the interaction is sufficiently short-range  ($s\ge1$), the limiting equation is  the standard cubic
 nonlinear Schr{\"o}dinger equation, while if interactions are longer range ($0<s<1$) the limiting 
equation is the fractional nonlinear Schr{\"o}dinger equation (FNLS) with fractional Laplacian, $(-\Delta)^s$.
%%

%  Here, ``controlled by'' means the following: Let $\varepsilon>0$ and pick an arbitrary fixed time interval $[0,T]$, $T>0$. Then, there exists $h_0(T)$, such that  in an appropriate function space the solution $u^{(h)}(t)$ of the initial value problem \eqref{eqn:IVPdnlsnonlocal} is well-approximated, with error no larger than the order $\varepsilon$,  by a corresponding solution of the initial value problem for \eqref{eqn:introNLS}.

\begin{remark}[Discrete nonlinear systems as physical models]
Discrete nonlinear dispersive systems arise in nonlinear optics, {\it e.g.} \cite{Aceves_DeAngelis_1994a,Aceves_DeAngelis_1994b,Eisenberg_Silberberg_1998,Eisenberg_Silberberg_1999,Naethe_Vicencio_2011,Weinstein_Yeary_1996,Weinstein_1999}, the dynamics of biological molecules, {\it e.g.} \cite{Davydov_Kislukha_1973,Eilbeck_Lomdahl_1985}, and condensed matter physics. For example, they arise in the study of intrinsic localized modes in anharmonic crystal lattices, {\it e.g.} \cite{Barker_Sievers_1975,Takeno_Kisoda_1988}. See also \cite{Smerzi_Trombettoni_2002,Hennig_Dorignac_2010}. In these fields, discrete systems arise either as phenomenological models or as {\it tight-binding} approximations; see also \cite{Mackay_Schneider_2008,Pelinovsky_Schneider_2010}. There is also a natural interest in such systems as discrete numerical approximations of continuum equations.

Nonlocal discrete models such as nonlocal DNLS addressed in this paper are of particular interest in several of the fields above where atomic interactions occur at length scales considerably larger than those accounted for by models with nearest-neighbor coupling \cite{Gaididei_Mingaleev_1997,Mingaleev_Christiansen_1999,Davydov_1971,Kevrekidis_Gaididei_2001,Kac_Helfand_1963,Baker_1961}. 

A case of particular interest with respect to the coupling potential \eqref{eqn:Jdeftrue} is $s = 1$, which corresponds to dipole-dipole interactions in biopolymers \cite{Gaididei_Mingaleev_1997}; this is the threshold value of $s$ which marks the transition between the standard and fractional continuum Laplacian limits; see Theorem \ref{th:mainnonlocal}. The case of Coulomb interactions corresponding to $s = 0$ is also of interest in this context, though we do not address it here. 
The exponential (Kac-Baker) coupling sequence is of considerable interest in both models of statistical physics and again in the dynamics of lattice models of biopolymers \cite{Baker_1961,Kac_Helfand_1963,Gaididei_Mingaleev_1997}.

Continuum models with nonlocal effects such as FNLS also arise in several contexts including path-integral formulations of quantum mechanics \cite{Laskin_2002}, deep water internal and small-amplitude surface wave fluid dynamics \cite{Benjamin_1967,Ono_1975,Weinstein_1987}, semi-relativistic quantum mechanics (astrophysics) \cite{Lieb_Yau_1987,Frohlich_Lenzmann_2007}, and as the continuum limits of the aforementioned long-range discrete models \cite{Kirkpatrick_Lenzmann_2012,Tarasov_2006,Tarasov_Zaslavsky_2006,Gaididei_Mingaleev_1996}. 
\end{remark}
\medskip
  
Cubic FNLS has, for $1/4 < p \leq 1$ and for any frequency $\omega<0$, a standing wave solution:
\begin{align}
u_\omega(x,t) = e^{- i\omega t} \psi_{| \omega |}(x), \label{eqn:cbreather}
\end{align}
where $\psi_{| \omega |}(x)$ is the unique positive {\it nonlinear ground state} solution to the nonlinear eigenvalue problem
\begin{align}
(- \Delta_x)^p \psi_{_{| \omega |}}(x) - \psi_{_{| \omega |}}(x)^3\ =\ \omega\ \psi_{_{| \omega |}}(x), \quad u\in H^{p}(\mathbb{R}) \label{eqn:introtiNLS}
\end{align}
This solution is  radially symmetric and decreasing to zero at spatial infinity
\cite{Frank_Lenzmann_2013,Frank_Lenzmann_2015}; see also Proposition \ref{prop:Psinonlocal}.   Thus,  $\psi_{_{| \omega |}}(x) = \sqrt{ | \omega |} \ \psi_{_1} \left( \ |\omega|^{1/2p} \ x\right)=\sqrt{ | \omega |} \ \psi \left( \ |\omega|^{1/2p} \ x\right)$, where we adopt the abbreviated notation:  $\psi(x) \equiv \psi_1(x)$.

\begin{remark} \label{remark:srestrict}
Although the dynamics \eqref{eqn:IVPdnlsnonlocal} are defined for any choice of $\{ J_m \}_{m \in \mathbb{N}}$ in $l^1(\mathbb{N})$, we shall restrict our attention to the case $\{ J^s_m \}_{m \in \mathbb{N}}$ in \eqref{eqn:Jdeftrue} for $s > 1/4$. The constraint on the range of the nonlocal interaction is the range for which the limiting fractional NLS equation has nonlinear bound state  solutions. See Proposition \ref{prop:psinonlocal}, Theorem \ref{th:mainnonlocal}, and Remark \ref{remark:main2nonlocal}. 
\end{remark}

 A key recent advance in the analysis of fractional / nonlocal equations, which plays as central role in  the current work, are the results of Frank, Lenzmann and Silvestrie \cite{Frank_Lenzmann_2015} on the uniqueness and non-degeneracy of the ground state solitary waves of FNLS. Our bifurcation results are perturbative about the FNLS limit and are based on an application of an implicit function theorem 
 to a nonlinear mapping acting on a space of functions defined on momentum space. That the differential of this mapping, 
  evaluated at the FNLS ground state,  is an isomorphism between appropriate spaces is a consequence of this non-degeneracy. \medskip

\begin{remark} The continuum nonlinear Schroedinger equation (NLS), with local dispersion corresponding to $p=1$ is Galilean invariant. Hence, any solitary standing wave of NLS can be boosted to give a solitary traveling wave; see \cite{Jenkinson_Weinstein_2015}. 
In \cite{Krieger_Lenzmann_2013,Hong_Sire_2015}, non-deforming traveling solitary waves of FNLS $i \partial_t u(x,t) = (- \Delta)^p u(x,t) - |u(x,t)|^2 u(x,t), \quad 1/2 \leq p < 1$ were been shown to exist. This is in direct contrast with the discrete setting, where the PNB phenomenon occurs due to the breaking of continuous translational symmetries. In \cite{Krieger_Lenzmann_2013}, traveling waves of the form $u(x,t) = e^{i t} U_v(x - vt), \quad |v| < 1$ are shown to occur for $p = 1$. In \cite{Hong_Sire_2015}, waves of the form $u(x,t) = e^{- i t (|v|^{2 p} - \omega^{2p})} \ U_{\omega, v} \left( x - 2 t p |v|^{2p - 2} v \right), \quad \omega > 0, v \in \mathbb{R}$ for $1/2 < p < 1$, are constructed.
 \end{remark}

\subsection{Discrete nonlocal solitary standing waves near the continuum limit}

In analogy with the case of nearest-neighbor DNLS \cite{Jenkinson_Weinstein_2015}, we seek discrete solitary standing waves of DNLS with nonlocal interactions:
 \begin{align}
u_n(t) = e^{- i \omega t} \ G_n,\ \ n\in\mathbb{Z} \quad {\rm where}\ \omega < 0\ .
 \end{align}
Thus, $G=\{G_n\}_{n\in\mathbb{Z}}$ satisfies the nonlinear algebraic eigenvalue problem:
 \begin{align}
 - ( \mathcal{L}^s G )_n - |G_n|^2 G_n\ =\  \omega\ G_n,\ \ n\in\mathbb{Z};
  \quad \quad  G \in l^2(\mathbb{Z}), \label{eqn:tiDNLSnonlocal}
 \end{align}
 where $\mathcal{L}^s$ is defined by \eqref{eqn:Ldef} and $J_m^s\simeq m^{-1-2s}, $
  for $0<s<\infty$ and $J_m^\infty\simeq e^{-\gamma m},\ \gamma>0$ for $s = \infty$. The precise expressions for $J_m^s$ is displayed below in \eqref{eqn:Jdeftrue}.
\medskip

 The goal of this work is to obtain a precise understanding of the onsite symmetric and offsite symmetric discrete solitary standing waves of DNLS with short- and long-range nonlocal interactions. \medskip

 \begin{definition}[Onsite symmetric and offsite symmetric states in one spatial dimension]
\label{defn:onoff}\\
Let ${G} = \{ G_n \}_{n \in \mathbb{Z}}$.
\begin{enumerate}
\item A solution to equation \eqref{eqn:tiDNLSnonlocal} is referred to as onsite symmetric if for all $ n \in \mathbb{Z}$, it satisfies
$
G_{n} = G_{-n}.
$
 In this case, ${G}$ is symmetric about $n = 0$. \medskip

\item A solution to equation \eqref{eqn:tiDNLSnonlocal} is referred to as offsite symmetric or bond-centered symmetric if for all $n \in \mathbb{Z}$, it satisfies
$
 G_{n} = G_{-n + 1}.
$
In this case, ${G}$ is symmetric about the point halfway between $ n = 0$ and $n = 1$.
\end{enumerate}
\end{definition}

\begin{figure}[H]
  \begin{minipage}[c]{.5\textwidth}
    \includegraphics[width=\textwidth]{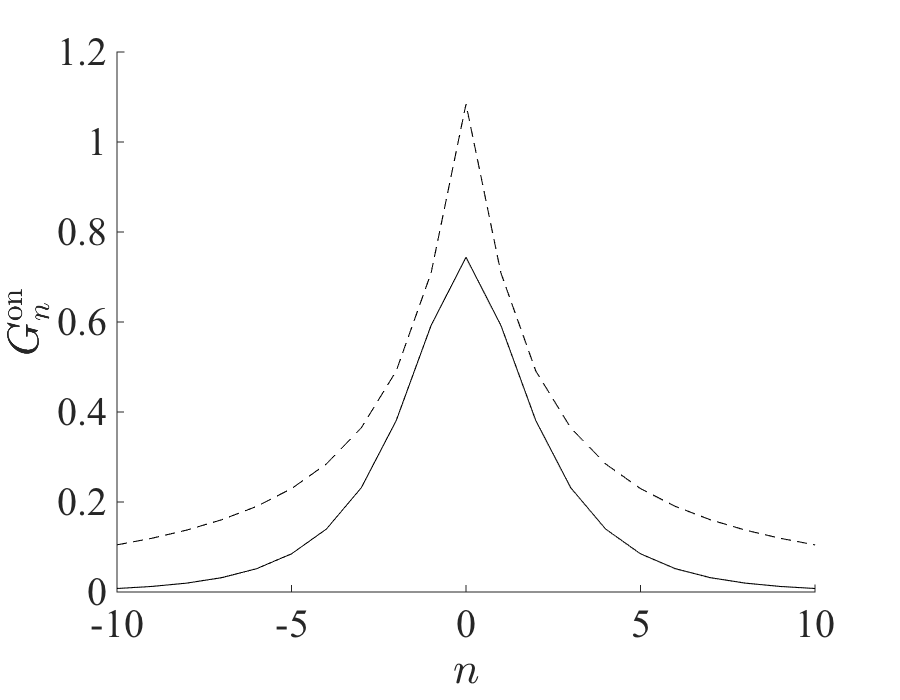}
  \end{minipage}\hfill
    \begin{minipage}[c]{.5\textwidth}
    \includegraphics[width=\textwidth]{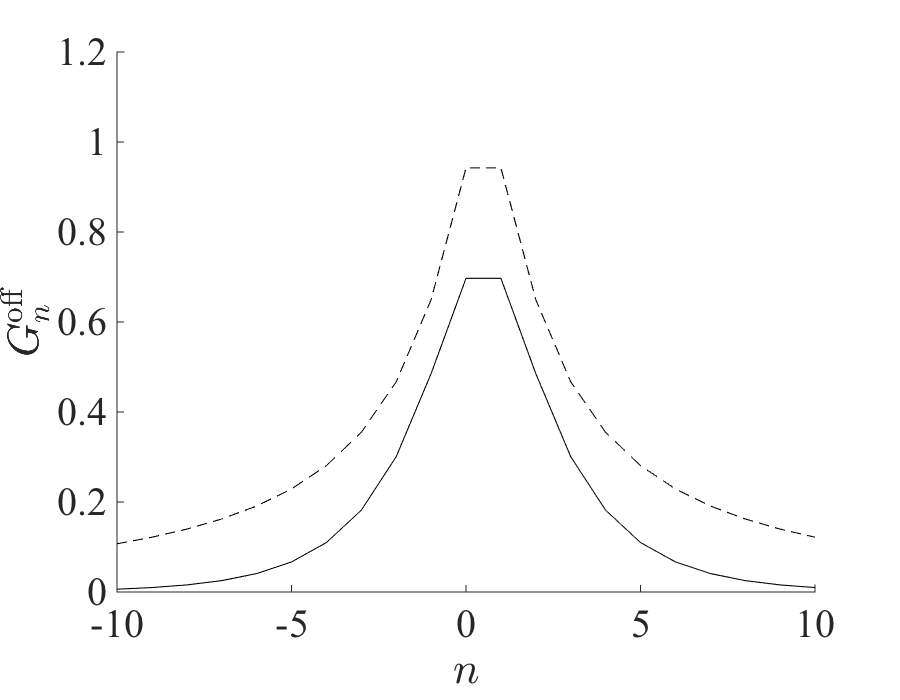}
  \end{minipage}\hfill
%    \begin{minipage}[c]{.33\textwidth}
%    \includegraphics[width=\textwidth]{bifurcations3.png}
%  \end{minipage}\hfill
%  \begin{minipage}[c]{0.22\textwidth}
 \caption{Left: on-site symmetric states of nonlocal discrete NLS for $\alpha = 0.5$, with short-range coupling ($s=1.5$, solid line) and long-range coupling ($s = 0.5$, dashed line). Right: off-site symmetric states of nonlocal discrete NLS for $\alpha = 0.5$, with short-range coupling ($s=1.5$, solid line) and long-range coupling ($s = 0.5$, dashed line). }  
%  \end{minipage}
\label{fig:onoff}
\end{figure}

We seek spatially localized onsite and offsite solutions of \eqref{eqn:tiDNLSnonlocal} in
 the {\it long wave limit} as $\omega \to 0$. Examples of such states are displayed in figure \ref{fig:onoff}. Here, there is large scale separation between the width of the discrete standing wave and the unit lattice spacing.
Solutions are expected to approach a (homogenized or averaged) continuum FNLS equation with fractional power $1/4 < p \leq 1$ to be determined.
 
\medskip

 We study the limit $\omega \to 0$ through the introduction of a single small parameter, $\alpha > 0$. 
 Introduce the continuous function where $\kappa_s(\alpha) = - \omega > 0$, where $\kappa_s(\alpha) \to 0$ as $\alpha \downarrow 0$ is appropriately chosen. Then,  $G =\{G_n\}_{n\in\mathbb{Z}}$ satisfies the nonlinear eigenvalue problem
\begin{align}
- \kappa_s(\alpha) \ G_n =  \ - (\mathcal{L}^s  G)_n - |G_n|^2 G_n ,\qquad  G \in l^2(\mathbb{Z})\ . 
%= & \ - \sum_{ \substack{m \in \mathbb{Z} \\ m \neq n} } J^s_{|m - n|}  (G_m - G_n) - %|G_n|^2 G_n\
\label{eqn:tiDNLSrescalednonlocal}
\end{align}

\noindent The precise form of $\kappa_s(\alpha)$ is displayed below in \eqref{eqn:kappa-def} and depends on whether the coupling interaction operator $\mathcal{L}^s$ is long-range ($1/4 < s < 1$), short-range ($1 < s \leq \infty$), or critical ($s = 1$). Thus, the study of nonlocal DNLS in the limit $\omega \to 0^-$ is equivalent to the study of $\alpha \to 0^+$. Onsite and offsite nonlinear bound states of \eqref{eqn:tiDNLSrescalednonlocal} arise as bifurcations of non-trivial localized states from the zero state at frequency $\kappa_s(0) =0$, the bottom of the continuous spectrum of $ - \mathcal{L}^s$ (Proposition \ref{prop:Lprops}).  

\begin{remark}
 In \cite{Kirkpatrick_Lenzmann_2012}, Kirkpatrick {\it et. al.} address the so-called {\it continuum limit, $h \to 0$} of the following nonlocal DNLS equation:
\begin{align}
& i \partial_t v_n(t) = - \frac{1}{ \kappa_s(h) } \ (\mathcal{L} v)_n(t)  - |v_n(t) |^2 v_n(t) \ ,\ n\in\mathbb{Z},  \ \ t \geq 0, \ \ h > 0. \label{eqn:KLSDNLS}
\end{align}
Here, $h > 0$ is the lattice spacing. This is a limit of interest in numerical computations. Solutions $v_m(t) = v^h_m(t) = v^h(mh, t)$ are expected to approach, as $h \to 0$, those of continuum FNLS, $u^{h = 0}(x,t)$, with fractional power $1/4 < p \leq 1$ to be determined. Equation \eqref{eqn:KLSDNLS} may be mapped to \eqref{eqn:IVPdnlsnonlocal} via the substitution $v_n(t) = \kappa_s( h )^{-1/2} \ u_n \left( \kappa_s( h )^{-1} \ t \right)$. 

Time-periodic bound states $v_n(t) = e^{- i \omega t} \ g_n, \ \ \omega < 0$ of \eqref{eqn:KLSDNLS} satisfy 
\begin{align}
\omega \ g_n =  - \frac{1}{ \kappa_s(h) } \ (\mathcal{L} g)_n - |g_n|^2 g_n, \qquad \{ g_n \}_{n \in \mathbb{Z}} \in l^2(\mathbb{Z}) . 
\end{align}
Substituting $g_n = \kappa_s(h)^{-1/2} G_n$ and taking $\kappa_s(\alpha) = \omega \ \kappa_s(h)$, we obtain \eqref{eqn:tiDNLSrescalednonlocal}. 
\end{remark}

\subsection{Main results}
 
\noindent Our main results concern the existence and energetic properties of localized solutions of \eqref{eqn:tiDNLSrescalednonlocal} for $\alpha$, and therefore $\kappa_s(\alpha)$, small: 

\medskip
 
\begin{enumerate}
\item {\bf Theorem \ref{th:mainnonlocal}; Bifurcation of onsite and offsite states of nonlocal DNLS:}\ Let the sequence of coupling coefficients, $\{J_m\}_{m \in \mathbb{Z}}$, of nonlocal DNLS be (1) the polynomially decaying sequence \eqref{eqn:Jdeftrue} with coupling decay rate $1/4 < s < \infty$, or (2) the exponentially decaying sequence \eqref{eqn:Jdeftrue} for which we set $s = \infty$. 
Then there exist families of onsite (vertex-centered) symmetric and offsite (bond-centered) symmetric solitary standing waves  of \eqref{eqn:tiDNLSrescalednonlocal} (in the sense of Definition \ref{defn:onoff}), which bifurcate for $\alpha \geq 0$ from the continuum limit ground state solitary wave of fractional NLS \eqref{eqn:introtiNLS} with fractional power $p = \min(1,s)$. \smallskip

\item {\bf Theorem \ref{th:PNnonlocal}; Exponential smallness of the Peierls-Nabarro barrier:}\  Let $\eta = \min(1,2s)$. Then, there exist positive constants $\alpha_0$ and $C>0$ such that for all $0 < \alpha < \alpha_0$, we have:
\begin{align}
& \Big|\ \mathcal{N} \left[G^{\alpha, \rm on} \right] - \mathcal{N} \left[ G^{\alpha, \rm off} \right] \ \Big| 
 \lesssim \frac{\kappa_s(\alpha)}{\alpha} \ e^{- C / \alpha^{\eta}} , \nonumber \\
&
\Big|\ \mathcal{H} \left[ G^{\alpha, \rm on} \right] - \mathcal{H} \left[ G^{\alpha, \rm off} \right] \ \Big| \lesssim   \frac{\kappa_s(\alpha)}{\alpha}   \ e^{- C / \alpha^{\eta}}  \ . 
\label{PN-bound}
\end{align}

\begin{remark} Theorem \ref{th:PNnonlocal} is subtle - to any polynomial order 
in the small parameter, $\alpha$, $G^{\alpha, \rm on} $ and $G^{\alpha, \rm off} $
are discrete translates of one another by a half integer, but differ at exponentially small order in $\alpha$. 
\end{remark}
%2) The $s=1/2$ corresponds to the case where the FNLS ground states is that of  the generalized Benjamin-Ono equation  $-\partial_x H u-u+u^3=0$, where $H$ denotes the Hilbert transform. In this case the ground state is given by 
% and its Fourier transform is , indicating that the PN barrier bound is sharp in this case.\\
% These can be improved on a bit. 
 
%Here, 
%\begin{align}
%\mathcal{N}[G] \equiv \mathcal{N} \left[G,\overline{G} \right] = \sum_{n \in \mathbb{Z}} |G_n|^2,  \label{eqn:power2nonlocal}
%\end{align}
%and
%\begin{align}
%\mathcal{H}[G] \equiv \mathcal{H} \left[G, \overline{G} \right] = \frac{1}{2} \sum_{n\in\mathbb{Z}} \sum_{  \substack{ m \in \mathbb{Z}  \\ m \neq n} } \  J^s_{|m - n|} | G_m - G_n|^2-\frac12 | G_n|^4 \label{eqn:hamil2nonlocal}
%\end{align}
%are the corresponding square $l^2(\mathbb{Z})$ norm (Mass/Power/Charge) and Hamiltonian of \eqref{eqn:tiDNLSrescalednonlocal} (for effective lattice spacing $h = 1$). 
\end{enumerate}

%
%\medskip
%
%
%Throughout this paper, we use $\psi_{\alpha^{2p}}$ for $1/4 < p \leq 1$ to refer to the unique {\it positive}, symmetric solution to the one-dimensional ($d = 1$) fractional nonlinear Schr\"{o}dinger equation with cubic nonlinearity (FNLS) and frequency $\alpha^{2p} > 0$:
%\begin{align}
%FNLS: \hspace{1cm} 0 = \alpha^{2p} \ \psi_{\alpha^{2p} }(x) + (- \Delta_x)^p \ \psi_{\alpha^{2p} }(x) -  \left( \psi_{\alpha^{2p} }(x) \right)^3, \hspace{1cm} x \in \mathbb{R}. \label{eqn:tiNLSpropnonlocal}
%\end{align}
%We also frequently use the convention, for the case where $\alpha = 1$, $\psi(x) \equiv \psi(x)$ to mean the solution to
%\begin{align}
%0 =   \psi (x) + (- \Delta_x )^p \ \psi  (x) -  \psi (x)  ^3, \hspace{1cm} x \in \mathbb{R}. \label{eqn:tiNLSprop1nonlocal}
%\end{align}
%We establish the properties of $\psi_{\alpha^{2p} }$ and its continuous Fourier transform in Section \ref{section:psinonlocal}
%

\begin{figure}[H]
  \begin{minipage}[c]{.5\textwidth}
    \includegraphics[width=\textwidth]{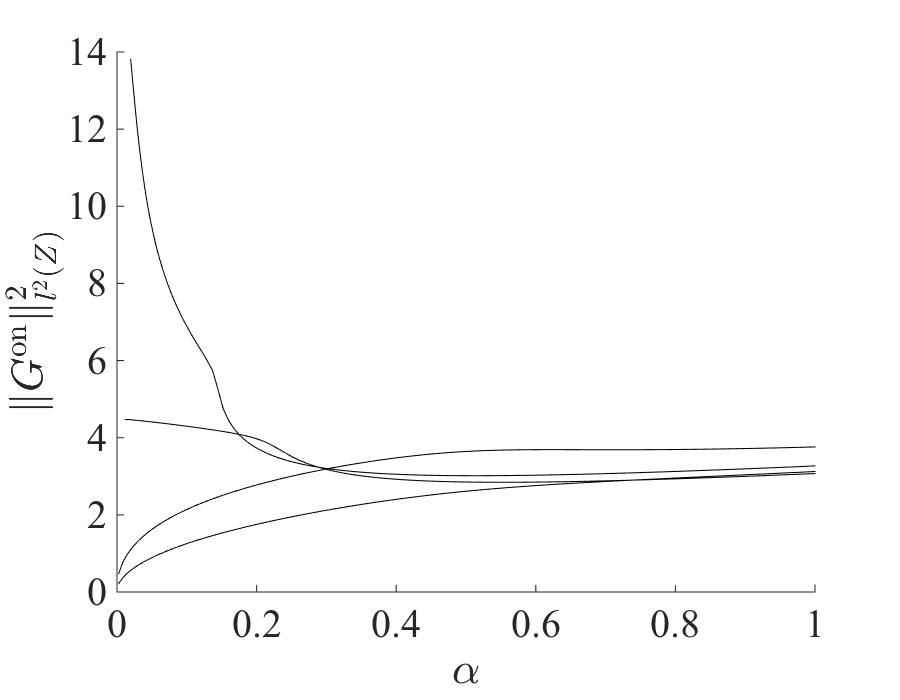}
  \end{minipage}\hfill
    \begin{minipage}[c]{.5\textwidth}
    \includegraphics[width=\textwidth]{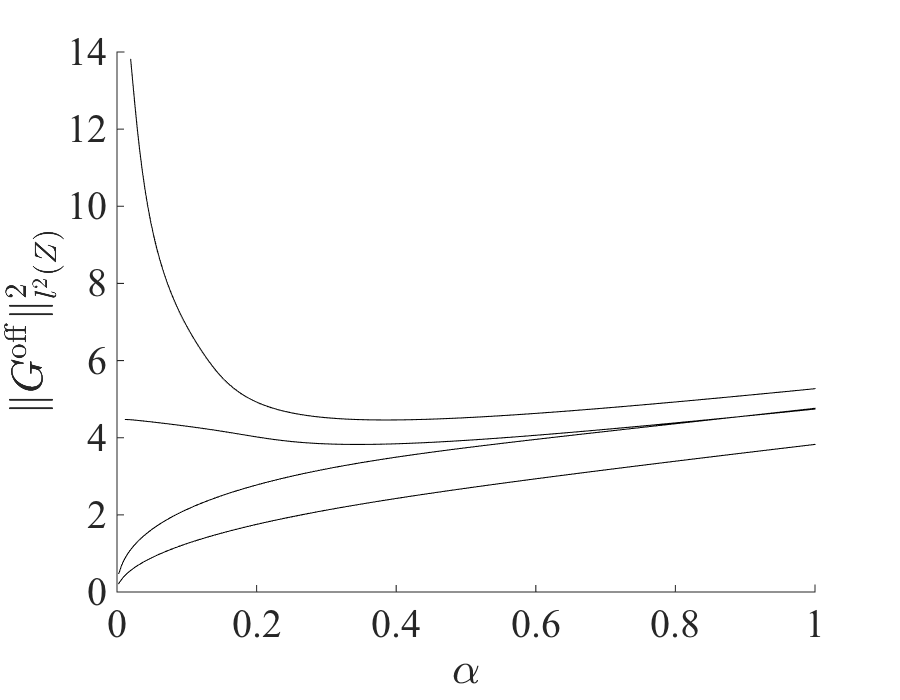}
  \end{minipage}\hfill
%    \begin{minipage}[c]{.33\textwidth}
%    \includegraphics[width=\textwidth]{bifurcations3.png}
%  \end{minipage}\hfill
%  \begin{minipage}[c]{0.22\textwidth}
 \caption{Left: bifurcation of onsite symmetric solutions of nonlocal discrete NLS for $s = 5, 1, 0.5, 0.375$ (respectively from bottom to top along the vertical axis). Right: bifurcation of offsite symmetric solutions of nonlocal discrete NLS for $s = 5, 1, 0.5, 0.375$ (respectively from bottom to top along the vertical axis).} \label{fig:bifurcations}
%  \end{minipage}
\end{figure}

\subsection{Strategy of proofs of Theorem \ref{th:mainnonlocal} and Theorem \ref{th:PNnonlocal}} \label{section:nonlocalstrategy}

The limit $\kappa_s(\alpha) \downarrow 0$ for $\alpha \to 0$ in \eqref{eqn:tiDNLS2nonlocal} is related to the continuum FNLS limit. In order to compare 
 the spatially discrete and spatially continuous problems,  it is  natural to work with the discrete and continuous Fourier transforms respectively;  both are functions of a continuous (momentum) variable.

 Let $\widehat{ g}(q)=\mathcal{F}_{_D}[ g](q)$ denote the discrete Fourier transform on $\mathbb{Z}$ of the sequence $g=\{g_n\}_{n\in\mathbb{Z}}$ and 
  let $\widetilde{f}(q)=\mathcal{F}_{_C}[f](q)$ denote the continuous Fourier transform on $\mathbb{R}$ of $f:\mathbb{R}\to\mathbb{C}$; see Section \ref{section:introDFT} for definitions and Appendix \ref{appendix:DFT}
    for a discussion of key properties.  The following proposition characterizes onsite symmetric and offsite symmetric states on the integer lattice, $\mathbb{Z}$, in terms of the discrete Fourier transform. 
     The proof follows from a direct calculation using the definition of the discrete Fourier transform and its inverse \cite{Jenkinson_Weinstein_2015}.  \medskip

\begin{proposition}
\label{prop:off-on}
\begin{itemize}
\item[(a)] Let $G = \{G_n \}_{n \in \mathbb{Z}}\in l^2(\mathbb{Z})$ be real and onsite symmetric in the sense of Definition \ref{defn:onoff}. Then, $\widehat{G}(q)$, the discrete Fourier transform of $G$, is real-valued and symmetric. Conversely, if $\widehat{G}(q)$ is real and symmetric, then ${\mathcal{F}^{-1}_{_D}[ \widehat{G} ]}$, its inverse discrete Fourier transform, is real and onsite symmetric.

\item[(b)] If $G = \{G_n \}_{n \in \mathbb{Z}} \in l^2(\mathbb{Z})$ is real and offsite symmetric in the sense of Definition \ref{defn:onoff}, then
\begin{equation}
\widehat{G}(q) = e^{- i q/2} \widehat{K}(q),
\label{dft-off}
\end{equation}
 where $\widehat{K}(q)$ is real and symmetric. Conversely, if $\widehat{G}(q) = e^{- i q/2} \widehat{K}(q)$, where $\widehat{K}(q)$ is real and symmetric, then ${\mathcal{F}^{-1}_{_D}[\ \widehat{G}\ ]}$ is real and offsite symmetric.
\end{itemize}
\end{proposition}
\medskip

Motivated by Proposition \ref{prop:off-on}, we first rewrite the equation for a DNLS standing wave profile (onsite or offsite) $G=G^{\sigma,\alpha,s}\in l^2(\mathbb{Z})$,  \eqref{eqn:tiDNLSrescalednonlocal}, in discrete Fourier space
 for $\widehat{G}(q) = e^{-i\sigma q} \widehat{K}(q)$,
  where $K=\widehat{K}^{\sigma,\alpha}$ satisfies $K(q+2\pi)=\widehat{K}(q)$. 
 Here, $\sigma=0$ corresponds to the onsite case and $\sigma=1/2$ to the offsite case. Note that $\widehat{K}(q)$ is determined by its restriction, $\widehat{\phi}(q)$, to
  the fundamental cell $q\in\mathcal{B}=[-\pi,\pi]$  (Brillouin zone); we often suppress the dependence on $\alpha$, $s$, and $\sigma$ for notational convenience.
  
  We obtain the following equation for $\widehat{\phi}(q)$, defined for $ q \in \mathbb{R}$:
  \begin{align}
& [\kappa_s(\alpha) + the(q)] \ \widehat{\phi}(q) - \frac{\chi_{_{\mathcal{B}}}(q)}{4\pi^2}  
\  \left(\ \widehat{\phi} *  \widehat{\phi} *  \widehat{\phi}\ \right) (q) 
\nonumber\\
&\qquad\qquad - \ 
\frac{\chi_{_{\mathcal{B}}}(q)}{4\pi^2}  
\sum_{m = \pm1} \ e^{2 m \pi i \sigma} \ \left(\ \widehat{\phi} *  \widehat{\phi} *  \widehat{\phi}\ \right) (q - 2 m \pi) = 0.
\label{eqn:phieqnnonlocal0}
\end{align}
where $\chi_{_{\mathcal{B}}}(q)$ is the characteristic function on $\mathcal{B}= [- \pi, \pi]$. 
 Here, $M^s(q)$ is the discrete Fourier symbol of the operator $- \mathcal{L}^s$:
$
 M^s(q) = 4 \ \sum_{m = 1}^{\infty} \ J^s_{m} \ \sin^2 (q m /2) \ ; 
$
see \eqref{LC-def} and Lemma \ref{lemma:nonlocaltransform}. In particular, we take
 \begin{align}
  J^s_m \equiv \left\{ \begin{array}{ll}  C^{-1}_s \ m^{-1- 2s} & \quad : \quad 1/4 < s < \infty  \\
  & \\
 C_{\infty}^{-1} \ e^{- \gamma \ m} & \quad : \quad s = \infty; \quad \gamma > 0,  \quad {\rm constant} . \end{array} \right.  \label{eqn:Jdeftrue}
\end{align}
The positive constant, $C_s $, is chosen so that the expansion of $M^s(q)$ , for small $q$, leads to an effective limiting ($\alpha\downarrow0$) continuum equation in which $(-\Delta)^{^p}$ has a coefficient equal to one (unit effective mass). 
  
Toward a formal determination of a limiting equation, we introduce rescalings of the momentum: $Q=q/\alpha$ and $\widehat{\Phi}(Q)= \rho(\alpha) \ \widehat{\phi}(q/\alpha)$, where $\rho(\alpha)$ is to be determined. Substitution into   \eqref{eqn:phieqnnonlocal0} and dividing by $\kappa_s(\alpha) \ \rho(\alpha)$, we obtain the following equation for $\widehat{\Phi}(Q)$:
\begin{align}
\mathcal{D}^{\sigma,\alpha}[\widehat{\Phi}] & \equiv [1 + M_{\alpha}^s(Q)] \ \widehat{\Phi}(Q) - \frac{\chi_{_{\mathcal{B}_{_{\alpha}}}}(Q)}{4\pi^2}  \left( \frac{\alpha^2 \ \rho(\alpha)^2}{\kappa_s(\alpha)} \right)
\  \left(\ \widehat{\Phi} *  \widehat{\Phi} *  \widehat{\Phi}\ \right) (Q) 
\nonumber\\
&\qquad\qquad +  \left( \frac{\alpha^2 \ \rho(\alpha)^2}{\kappa_s(\alpha)} \right) R_1^{\sigma} [ \widehat{\Phi} ](Q).  \label{eqn:Phieqn3-intro}
\end{align}
 Here, $\chi_{_{\mathcal{B}_{_{\alpha}}}}(Q)$ is characteristic function of $\mathcal{B}_\alpha$,  $ M^s_{\alpha}(Q) \equiv \frac{1}{\kappa_s(\alpha)} M^s(Q \alpha)$ 
% \begin{align} 
% =  \frac{4}{\kappa_s(\alpha)  }  \sum_{m = 1}^{\infty} J^s_m \   \sin^2 (Q m \alpha / 2 )  ,  \label{eqn:M2def-intrononlocal}
% \end{align}
 and
 \begin{align}
R_1^{\sigma} [ \widehat{\Phi} ](Q) &= -\frac{\chi_{_{\mathcal{B}_{_{\alpha}}}}( Q)}{4 \pi^2}\ \sum_{m = \pm 1} \ e^{2 m \pi i \sigma} \ \left(\ \widehat{\Phi} * \widehat{\Phi} * \widehat{\Phi}\ \right)  ( Q - 2 m \pi / \alpha).  \label{eqn:R1rewrite-intrononlocal}
\end{align}
Note that the dependence of the nonlinear operator, $\widehat{\Phi}\mapsto\mathcal{D}^{\sigma,\alpha}[\widehat{\Phi}]$ on $\sigma$ designates the case of onsite or offsite states.
 Note that for  $\alpha |Q| \ll 1$:
\begin{align}
M^s (Q \alpha) \sim \left\{ \begin{array}{ll} \alpha^{2s} \ |Q|^{2s} & : 1/4 < s < 1 \\
(- \log(\alpha)) \alpha^2 \ |Q|^2 & : s = 1 \\
\alpha^2 \ |Q|^2 & : 1 < s \leq \infty \end{array} \right. . 
\end{align}
Therefore, choosing $\kappa_s(\alpha)$ to depend on the nonlocality parameter, $s$:
\begin{align}
\kappa_s(\alpha) =   \left\{ \begin{array}{lll} \alpha^{2s}   & : 1/4 < s < 1 & \qquad {\rm (long-range)} \\
(- \log(\alpha)) \alpha^2  & : s = 1 & \qquad {\rm (critical/marginal)} \\
\alpha^2  & : 1 \le s \leq \infty  & \qquad {\rm (short-range)} \end{array} \right. ,
\label{eqn:kappa-def}
\end{align}
we obtain the formal scaled convergence as $\alpha \downarrow 0$, 
\begin{align}\label{formal-scaled-conv}
M^s_{\alpha}(Q)   &\rightarrow    |Q|^{2\min(1,s)},\ \textrm{or equivalently}\ 
%as 
%\begin{align}
%M_{\alpha}^s(Q) = \frac{1}{\kappa_s(\alpha)} M^s( Q \alpha) \quad \rightarrow  \quad \left\{ \begin{array}{ll} |Q|^{2s}  & : 1/4 < s < 1 \\
%|Q|^2 & : 1 \le s \leq \infty \end{array} \right.  \ ,
%\end{align}
\mathcal{F}^{-1}M^s_{\alpha}(Q)\mathcal{F} \to (-\Delta)^{\min(1,s)}
\end{align}
Returning to equation \eqref{eqn:Phieqn3-intro}, we obtain a formal balance of terms in the equation $\alpha \downarrow 0$ by choosing $\rho=\rho_s(\alpha)$ so that
$
\frac{\alpha^2 \ \rho_s(\alpha)^2}{\kappa_s(\alpha)} = 1 $ or equivalently $ \rho_s(\alpha) = \sqrt{ \frac{\kappa_s(\alpha)}{\alpha^2} }.
$
The formal limit of \eqref{eqn:Phieqn3-intro} is then 
\begin{equation}
(1+|Q|^{2p} )\ \widetilde{\psi}(Q)\ -\ 
 \frac{1}{4 \pi^2}\ \left(\widetilde{\psi}*\widetilde{\psi}*\widetilde{\psi}\right)(Q)=0,\ \ \textrm{where}\ 
  p=p(s) \equiv \min(1,s)
\label{FT-NLS}\end{equation}
  the equation for $\widetilde{\psi}(Q)$, the Fourier transform on $\mathbb{R}$ of the continuum FNLS solitary wave, $\psi(x)$. Here,  we have used: (a)  $M^s_\alpha(Q)\to|Q|^{2p}$, for bounded $Q$,  as $\alpha\to0$, (b) $\mathcal{B}_\alpha\to\mathbb{R}$  as $\alpha\to0$, and
(c)   $\chi_{_{\mathcal{B}_{{\alpha}}}}(Q) R_1^{\sigma}[\widehat\Phi](Q)\to0$ as $\alpha\to0$ for $\widehat{\Phi}(Q)$ localized; see Lemma \ref{lemma:expconvononlocal}.
We therefore expect, for small $\alpha$, that $\widehat{\Phi}^{\sigma,\alpha}\approx \widetilde{\psi}(Q)$.

Therefore, the onsite and offsite discrete standing solitary waves, $G^{\alpha,{\rm on}} = G^{\alpha, \sigma = 0}$ and $G^{\alpha,{\rm off}} = G^{\alpha, \sigma = 1/2} $ can be constructed via rescaling $Q=q/\alpha$: 
$
\widehat{G}(q) = e^{- i q \sigma} \widehat{\phi}(q) = e^{- i q \sigma} \ \sqrt{ \frac{\kappa_s(\alpha)}{\alpha^2} } \ \widehat{\Phi} \left( \frac{q}{\alpha} \right) \sim e^{- i q \sigma} \ \sqrt{ \frac{\kappa_s(\alpha)}{\alpha^2} } \   \widetilde{\psi}\left( \frac{q}{\alpha} \right), $ $q \in \mathcal{B},
$
where $\sigma=0, 1/2$, respectively and inverting the discrete Fourier transform.

The proof of bounds on the Peierles-Nabbaro (PN) barrier  (Theorem \ref{th:PNnonlocal})
follows the outline \cite{Jenkinson_Weinstein_2015} for the case of nearest-neighbor interaction. The PN barrier bounds depend on  exponential decay bounds, uniformly in $\alpha$ small, for $\widehat{\Phi^{\alpha,\sigma}}(Q), \ \sigma=0,1/2$. The extension of these bounds to the nonlocal case is presented in Appendix \ref{appendix:expogeneral}. 
%We obtain $\widehat{\Phi^{\alpha,\sigma}}(Q) \sim e^{- C |Q|}$ for $1/2 \leq s \leq \infty$ and %$\widehat{\Phi^{\alpha,\sigma}}(Q) \sim e^{- C |Q|^{2s}}$ for $1/4 < s < 1/2$. These decay rates %govern the size of the estimates in Theorem \ref{th:PNnonlocal}. 

%The above convergence argument, as $\alpha\downarrow0$, for fixed scaled quasimomentum, $Q\in [-\pi/\alpha,\pi/\alpha]$. 

Reduction to $p-$ FNLS employs a  Lyapunov-Schmidt reduction strategy. We first solve for the high frequency
 components of $\widehat{E_J^{\alpha, \sigma}}\left( Q\right)$ for $\lambda_s(\alpha)^{-1} \le |Q|\le \pi/\alpha$, ($\lambda_s(\alpha) > \alpha$ and $\lambda_s(\alpha) \downarrow 0$ as $\alpha \to 0$ )  in terms 
of those for $0\le |Q|\le \lambda_s(\alpha)^{-1}$ (low frequency components of $\widehat{E_J^{\alpha, \sigma}}$ ). This yields a closed system for the  low-frequency components, which we study perturbatively about the continuum FNLS limit using the implicit function theorem.

\medskip

  \subsection{Extensions}

\subsubsection{General coupling sequences} We note that our results for non-local coupling sequences
 $\{J^s_{|m}|\}_{m\in\mathbb{Z}}$ can be extended to general non-local coupling sequences
  $\{\mathcal{J}^s_{|m|}\}_{m\in\mathbb{Z}}$, with asymptotic decay rate $\sim |m|^{-1-2s}$. If $1/4 < s \leq 1$, one may only extend under appropriate assumptions on the rate of decay of $\mathcal{J}^s_{|m|}/J^s_{|m|}-1$ as $|m|\to\infty$. 

The results may also be extended to coupling sequences $\{\mathcal{J}^{\infty}_{|m|}\}_{m\in\mathbb{Z}}$ with asymptotic decay faster than any algebraic power in $m$. Note that this case includes any coupling sequence with compact support, including that of the nearest-neighbor (centered-difference) Laplacian.

\subsubsection{Higher dimensional cubic lattices}
In \cite{Jenkinson_Weinstein_2015}, our results apply to cubic DNLS with nearest-neighbor coupling in dimensions $d = 1,2,3$. This range of spatial  dimensions corresponds to those for which the continuum NLS equation has a non-trivial ground state, as seen using classical virial identities. For the current context of nonlocal lattice equations,  we expect our results on the one-dimensional cubic nonlocal DNLS to  extend to dimensions, $d$, satisfying: $2 \leq d < 4p$;
 see Remark \ref{remark:pohoone}.

\subsection{Outline of the paper} 
In Section \ref{section:preliminaries}, we provide a summary of basic facts about the discrete and continuous Fourier transforms and also provide a list of notations and conventions used throughout the paper.

In Section \ref{section:psinonlocal}, we summarize the properties of the continuum FNLS solitary standing wave.

In Section \ref{section:mainresults}, we state our result on the bifurcation of onsite and offsite solitary waves, Theorem \ref{th:mainnonlocal}. The exponentially small bound on the PN-barrier, the $l^2$ energy difference between onsite and offsite solutions, is stated in Theorem \ref{th:PNnonlocal}.

Sections \ref{section:1dproofnonlocal} through \ref{section:rescalinglow0nonlocal} contain the proof of Theorem 3.1. 

Sections \ref{section:nonlocalleadingorderproof} and \ref{section:rescalinglow0nonlocal} construct the corrector	 in the rigorous asymptotic expansion of the solution to DNLS.

In Section \ref{section:nonlocalleadingorderproof}, we derive equations for the corrector. We also construct the high frequency component of the corrector as a functional of the low frequency components and the small parameter $\alpha$.

In Section \ref{section:rescalinglow0nonlocal}, we construct the low frequency component of the corrector and map our asymptotic expansion back to the solution to DNLS, completing the proof of Theorem \ref{th:mainnonlocal}.

In Section \ref{section:higherorder}, we outline the steps required to prove Theorem \ref{th:SJalpha-error}, which generalizes the leading order expansion given in Theorem \ref{th:mainnonlocal} to a higher order expansion in the small parameter $\alpha$. 

There are several appendices. Appendix \ref{appendix:subadd} contains an elementary and very useful subadditivity lemma. In Appendix \ref{appendix:DFT}, we discuss properties of the discrete Fourier transform. In Appendix \ref{appendix:operatorL}, we discuss spectral properties of the operator $\mathcal{L}$. In Appendix \ref{appendix:IFT}, we provide a formulation of the implicit function theorem, which we can apply directly to our setting. In Appendix \ref{appendix:asymptotics}, we discuss the asymptotic properties of several special functions, which are related to the dispersion relation for discrete and nonlocal NLS. Appendix \ref{appendix:expogeneral} addresses the important exponential decay properties of solitary waves in Fourier ( momentum) space.

 \subsection{Acknowledgements}
 This work was supported in part by National Science Foundation grants  DMS-1412560 and DGE-1069420 (the Columbia Optics and Quantum Electronics IGERT), and a
grant from the Simons Foundation (\#376319, MIW). The authors thank O. Costin, R. Frank,  and P. Kevrekides for very stimulating discussions.
 
\subsection{Preliminaries, Notation and Conventions}
\label{section:preliminaries}

\subsubsection{Discrete Fourier Transform} \label{section:introDFT}

For a sequence
  ${f} = \{ f_n \}_{n \in \mathbb{Z}} \in l^1(\mathbb{Z}) \cap l^2(\mathbb{Z})$
    , define the discrete Fourier transform (DFT) 
\begin{align}
\widehat{f}(q) = \mathcal{F}_{_D} [{f} ](q) \equiv \sum_{n \in \mathbb{Z}}  f_n e^{- i q \cdot n} \ ,\ q \in \mathbb{R} . \label{eqn:DFT}
 \end{align}
 Since $\widehat{f}(q + 2 \pi  ) = \widehat{f}(q)$,  we shall view $\widehat{f}(q)$ as being defined on the torus $\mathbb{T} \simeq \mathcal{B}/ \pi \mathbb{Z}$, where $\mathcal{B}$ is fundamental period cell  ({\it Brillouin zone}),
$
\mathcal{B} \equiv [-\pi, \pi] .$
 $\widehat{f}$ is completely determined by its values on $\mathcal{B}$.
 We shall also make use of the scaled {\it Brillouin zone},
$
 \mathcal{B}_\alpha\ \equiv \Big[-\frac{\pi}{\alpha}, \frac{\pi}{\alpha}\Big]$,  with $ \mathbb{T} \simeq \mathcal{B}_{\alpha} \Big/ \frac{\pi}{\alpha} \mathbb{Z}.
$
The inverse discrete Fourier transform is given by
 \begin{align}
 f_n = \left( \mathcal{F}^{-1}_{_D} [\widehat{f}] \right)_n = \frac{1}{2 \pi} \int_{\mathcal{B}} \widehat{f}(q) e^{i q \cdot n} dq\ . \label{eqn:invDFT}
 \end{align}
A summary of key properties of the discrete Fourier transform is included in Appendix \ref{appendix:DFT}. 
For $a$ and $b$ in $L^1_{_{\rm loc}}(\mathbb{R})$, we define the \emph{convolution on $\mathcal{B}_\alpha$} by
\begin{align}
    \left(a *_{_{\alpha}} b\right)(q) = \int_{\mathcal{B}_{\alpha}} a(q-\xi) b(\xi) d\xi, \label{eqn:periodicconvolution}
\end{align}

\medskip
\begin{proposition}
\label{prop:pseudoperiodicity}
Let $a(q)$,$b(q)$, and $c(q)$ denote $2 \pi \sigma / \alpha -$ pseudo-periodic functions. Then,  $
a *_{\alpha} b = b *_{\alpha} a$ and   $(a *_{\alpha} b) *_{\alpha} c = a *_{\alpha} ( b *_{\alpha} c).
$
\end{proposition}
\medskip

\subsubsection{Continuous Fourier Transform on $\mathbb R$}

\noindent For $f \in L^2(\mathbb{R})$, the continuous Fourier transform and its inverse are given by
\begin{align}
\widetilde{f}(q) = \mathcal{F}_{_C} [ f ](q) \equiv \int_{\mathbb{R}} f(x) e^{- i q \cdot x} dx, \qquad {\rm and} \qquad
 f(x) = \mathcal{F}_{_C}^{-1} [\widetilde{f}](x) = \frac{1}{2 \pi} \int_{\mathbb{R}} \widetilde{f}(q) e^{i q \cdot x} dq\ .
 \end{align}
 \medskip

\noindent The standard convolution of functions on  $\mathbb{R}$ is defined by
\begin{align}
(F* G) (q) = \int_{\mathbb{R}} F(q-\xi) G (\xi) d \xi\ =\ \int_{\mathbb{R}} F(\xi) G (q - \xi) d \xi\ =\ (G*F)(q)
\end{align}

Since our differential equations are cubically nonlinear, we must often work with the expression for a triple convolution:
\begin{equation}
\left( F*G*H\right)(q)\ =\ \int \int F(q-\xi-\zeta) G(\xi) H(\zeta)\ d\xi d\zeta 
\label{triple-conv}
\end{equation}

\subsubsection{Notations, Conventions and a useful Lemma}
\begin{enumerate}
\item A function $F(q), \ q \in \mathbb{R}$ is  $2 \pi-$ {\it periodic } if for all $q$, $F(q) = F(q + 2 \pi )$. For $\sigma\in\mathbb R$, a function $F(q), \ q \in \mathbb{R}$ is  $2 \pi \sigma-$ {\it pseudo-periodic } if  $F(q) = e^{2 \pi i \sigma} F(q + 2 \pi)$ for all $q\in\mathbb R$. 
\\

 \item  The inner product on $L^2(\mathcal{B})$ is given by 
$
 \left\langle f, g \right\rangle = \left\langle f, g \right\rangle_{L^2(\mathcal{B})} = \int_{\mathcal{B}} \overline{f(q)} \ g(q) \ dq,
$
 where $\overline{f}$ is the complex conjugate of $f$. \\

\item $ \| {f} \|^2_{l^2(\mathbb{Z})}  = \sum_{n \in \mathbb{Z}} | f_n |^2 . $ \\

\item $L^2(A)$, the space of functions satisfying $\| f \|_{L^2(A)} = \left( \int_A | f(x) |^2 dx \right)^{1/2} < \infty $. \\

\item $H^1(A)$, the space of functions satisfying $\| f \|_{L^2(A)} = \left( \| f \|^2_{L^2(A)} + \| \nabla f \|^2_{L^2(A)}  \right)^{1/2} < \infty $.

\item $L^{2,a}(A)$, the space of functions satisfying $\| f \|_{L^{2,a}(A)} = \left(\int_A (1 + |q|^2)^a |f(q)|^2 dq \right)^{1/2} < \infty$, where $L^{2,0}(A) = L^2(A)$. \\

\item $H^a_{_{\rm even}}$ and $L^{2,a}_{_{\rm even}}$, are subspaces of $H^a$ and $L^{2,a}$ consisting of functions, $f$, which satisfy $f(x) = f(-x)$. We also refer to these as symmetric functions. \\

\item $H^a(\mathbb{R})$, the space of functions such that $\widetilde{f} = \mathcal{F}_{_C}[f] \in L^{2,a}(\mathbb{R})$, with $\left\| f \right\|_{H^a(\mathbb{R})} \equiv \| \widetilde{f} \|_{L^{2,a}(\mathbb{R})}$. \\

\item For $f:\mathbb{Z}\to\mathbb{C}$, $f=\{f_n\}_{n\in\mathbb{Z}}$, define the forward  difference operator 
$
\left( \delta f\right)_n = \left( \delta_1 f \right)_n = f_{n + 1} - f_n.
$

%\item For $d = 1$ and $f= \{ f_n \}_{n \in \mathbb{Z}}$, the one-dimensional discrete Laplacian is given by:
% \begin{align*}
%\left( \delta^2 f \right)_n = f_{n + 1} + f_{n - 1} - 2 f_n\ .
% \end{align*}
% For general dimension, $d$:
% \begin{align*}
%\left( \delta^2 f\right)_n =  \sum_{|j-n|=1} f_j - 2d\ f_n\ .
% \end{align*} 
% The summation is over nearest neighbor lattice points.\\

\item $\chi_{_A}(x) = \left\{
     \begin{array}{lr}
       1 & : x \in A \\
       0 & : x \notin A
     \end{array}
   \right.$, the indicator function for a set $A$, and
    $ \overline{\chi}_{_A} = 1 - \chi_{_A} $. \\

\item $f(\alpha) = \mathcal{O}(\alpha^{\infty})$ if for all $n \geq 1$, $ f(\alpha) = \mathcal{O}(\alpha^n)$. \\

\item For $a, b \in \mathbb{R}$, we write $ a \lesssim b$ if there exists a constant $ C > 0$, independent of $a$ and $b$, such that $a \leq C \ b$. \\
\item We shall frequently derive norm bounds of the type $\|f\|_X\lesssim \|g\|_Y$. Here, the implied constant is independent of $f$ and $g$ but may depend on the spaces $X$ and $Y$.\\
\item Generic constants are denoted by $C_1, C_2$ etc.\\
\item We shall often write, for example,  $\| |Q|^m f(Q;\alpha)\|_{L^2(\mathbb{R}_Q)}$ rather than 
$\| |\cdot|^m f(\cdot;\alpha)\|_{L^2(\mathbb{R})}$
\end{enumerate}
\medskip

We shall make frequent use of the following \medskip

\begin{lemma}\label{lemma:Ha-algebra}
\begin{enumerate}
\item $f\in H^a(\mathbb{R})$ if and only if  $\widetilde{f}\in L^{2,a}(\mathbb{R})$.
\item Fix $a > d/2$. Then, if  $f_1, f_2\in H^a(\mathbb{R})$ then, the product $f_1 f_2\in H^a(\mathbb{R})$.  Moreover, we have
\begin{align}
\left\| f_1 f_2 \right\|_{H^a(\mathbb{R})} \lesssim \left\| f_1 \right\|_{H^a(\mathbb{R})} \ \left\| f_2 \right\|_{H^a(\mathbb{R})} .
\label{eqn:algebraA}
\end{align}
\item  Fix $a > d/2$.  If $\widetilde{f_1},\ \widetilde{f_2}\in L^{2,a}(\mathbb{R})$ then, their convolution $\widetilde{f_1} * \widetilde{f_2}\in L^{2,a}(\mathbb{R})$
 and we have:
\begin{align}
\left\| \widetilde{f_1} * \widetilde{f_2} \right\|_{L^{2,a}(\mathbb{R})} \lesssim \left\| \widetilde{f_1} \right\|_{L^{2,a}(\mathbb{R})} \ \left\| \widetilde{f_2} \right\|_{L^{2,a}(\mathbb{R})}. \label{eqn:algebraB}
\end{align}
\end{enumerate}
The implied constants in \eqref{eqn:algebraA} and \eqref{eqn:algebraB} are independent of $f_1$ and $f_2$, but depend on $a$. 
\end{lemma}
\medskip

Finally, we find it convenient to record certain  $\alpha$- and $s$-dependent functions, $\mathfrak{e}_s, \mathfrak{E}_s, \mathfrak{d}_s \to 0$ as $\alpha \to 0$. These arise throughout the our analysis:
\medskip

\begin{align}\label{frak-e}
\mathfrak{e}_s(\alpha) \equiv \left\{ \begin{array}{ll}  \alpha^{3/2 - s} & : 1/4 < s < 1 \\
 \left( \frac{\alpha}{- \log(\alpha)} \right)^{1/2} & : s = 1 \\
 \alpha^{ 2s - 3/2} & : 1  < s < 2 \\
 (- \log(\alpha)) \alpha^{5/2} & : s = 2 \\
 \alpha^{5/2} & : 2 < s \leq \infty.
 \end{array} \right. , \quad \ \
 \mathfrak{E}_s(\alpha) = \left\{ \begin{array}{ll} \alpha^{2r (1 - s)} + \alpha^{2s (1 - r)} & : s < 1 \\
\frac{\log(- \log(\alpha))}{(- \log(\alpha))} + \frac{1}{(\log(\alpha))^2} & : s = 1 \\
\alpha^{2r (s - 1)} + \alpha^{2 (1 - r)}   & : 1 < s < 2 \\
(- \log(\alpha)) \alpha^{2r} + \alpha^{2 (1 - r)}   & : s = 2 \\
\alpha^{2r}  + \alpha^{2 (1 - r)}  & : 2 < s \leq \infty 
\end{array} \right. ,
 \end{align}

\begin{align}
\mathfrak{d}_s(\alpha) = \left\{ \begin{array}{ll} \alpha^{2r (1 - s)} & : s < 1 \\
\frac{\log(- \log(\alpha))}{(- \log(\alpha))} & : s = 1 \\
\alpha^{2r (s - 1)} & : 1 < s < 2 \\
(- \log(\alpha)) \alpha^{2r} & : s = 2 \\
\alpha^{2r} & : 2 < s \leq \infty 
\end{array} \right. .  \label{eqn:gamma3} 
\end{align}  
Here, $0 < r < 1$.

\section{Properties of the continuum solitary wave of FNLS, $\psi_{ | \omega|  }$, on $\mathbb{R}$} \label{section:psinonlocal}
The following results summarize properties of the FNLS ground state solitary wave (``soliton'') and its Fourier transform on $\mathbb{R}$. See, for example,  references \cite{Frank_Lenzmann_2013,Frank_Lenzmann_2015}. 

 A central role is played by the characterization of the continuum FNLS standing wave, and  the  linearized operator about a solution 
  of \eqref{eqn:introtiNLS} $\omega = -1$:
\begin{equation}
f\mapsto L_+ f\equiv
\left(\ 1 + (-  \Delta)^p\ -  3 (\psi (x))^2\ \right) f(x) .
\label{L+defnonlocal}
\end{equation}

\medskip

\begin{definition} 
Let $1/4 < p \leq 1$. Let $\psi_1 \in H^p(\mathbb{R})$ denote a solution of \eqref{eqn:introtiNLS} governing FNLS solitary standing waves with frequency $\omega = -1$.  If $\psi_1$ is real-valued, $\psi_1\ge0$, $\psi_1\nequiv0$, and   $L_+$ (with $\psi=\psi_1$) has exactly one negative eigenvalue (counting multiplicity), {\it i.e.} $\psi_1$ has Morse index equal to one, then
we say that $\psi_1$ is a ground state solution \eqref{eqn:introtiNLS}.

\end{definition}

\medskip 

\begin{remark}
Ground states of \eqref{eqn:introtiNLS} may be constructed via unconstrained minimization of an appropriate functional defined on $H^p(\mathbb{R})$, which characterizes the optimal constant in a fractional Gagliardo-Nirenberg-Sobolev inequality \cite{Frank_Lenzmann_2013,Frank_Lenzmann_2015}. 
\end{remark}
\medskip

\begin{theorem}[FNLS Ground State]
\label{prop:psinonlocal}

Let $1/4 < p \leq 1$ and consider the equation \eqref{eqn:introtiNLS} governing FNLS solitary standing waves. For $\omega = -1$, there exists a unique ground state solution solution $\psi_1(x)$ to \eqref{eqn:introtiNLS} which is real-valued, symmetric about $x = 0$ and decaying to zero at infinity. Therefore, the ground state of \eqref{eqn:introtiNLS} with arbitrary frequency $\omega<0$ is given by $\psi_{|\omega|}(x)= \sqrt{ \omega}  \ \psi_1( |\omega|^{1/2p} x)$. 
Moreover, the following properties hold:

\begin{enumerate}
\item $\psi_1(x) \in H^{2p + 1}(\mathbb{R})$ and satisfies the decay estimate:
 %   \begin{align}
$    | \psi_{1 }(x) | \lesssim (1 + |x|^{1 + 2p})^{-1}$. 
%    \end{align}
% \item $ \psi_{| \omega |   }(x) =  \sqrt{ |\omega | } \ \psi_1 ( | \omega| ^{1/2p} x) $.

\item For all $k\ge0$, $\psi_1(x) \in H^{k}(\mathbb{R})$. 
 
\item 
Any positive solution of \eqref{eqn:introtiNLS} is of the form
%\begin{equation}
$\psi_{| \omega | }(x - x_0) $
%\end{equation}
for some  $x_0 \in \mathbb{R}$.
\end{enumerate}
\end{theorem}
\bigskip

\begin{remark} \label{remark:pohoone}
A well-known  argument based on ``Pohozaev'' / virial identities shows that the equation $(-\Delta)^p u+u-u^m=0$ does not admit non-trivial $H^{2p + 1}(\mathbb{R}^d) \cap L^{m + 1}(\mathbb{R}^d)$ solutions when $m \geq (d+2 p )/(d-2 p) $ (where if $d - 2p \leq 0$, all $m > 0$ are admissible). For the cubic case $m=3$ in dimension $d = 1$, this implies $p > 1/4$; see \cite{Strauss_1977,Sulem_Sulem_1999,Ros-Oton_Serra_2014}.
\end{remark}

\medskip

The following result concerning $\widetilde{\psi_{| \omega | }} = \mathcal{F}_{_C} [ \psi_{| \omega | }]$, the Fourier transform of $\psi_{| \omega | }$, is a consequence of  Proposition \ref{prop:psinonlocal} and Lemma \ref{lemma:expogeneral} of Appendix \ref{appendix:expogeneral}. 

\medskip
\begin{proposition}[Fourier transform of FNLS Ground State] 
\label{prop:Psinonlocal}
{\ }

\noindent Fix $1/4 < p \leq 1$. The Fourier transform, $\widetilde{\psi_{| \omega | }}(Q) = \mathcal{F}_{_C} [ \psi_{| \omega | }](Q)$, satisfies the equation
\begin{align}
    \left(\ |\omega|\ +\ |Q|^{2p}\ \right) \ \widetilde{\psi_{| \omega | }}(Q)\ -\ \frac{1}{4 \pi^2}\ \widetilde{\psi_{| \omega | }} * \widetilde{\psi_{| \omega | }} * \widetilde{\psi_{| \omega | }}(Q)\ =\ 0,\qquad  \widetilde{\psi_{| \omega | }}\in L^{2,a}(\mathbb{R}).
\label{Psi-eqnnonlocal}    \end{align}
Moreover,  $\widetilde{\psi_{| \omega | }}$ satisfies the following properties:
\begin{enumerate}
\item  Scaling: 
\begin{align}
 \widetilde{\psi_{| \omega | }}(Q) = | \omega|^{(p - 1)/2p} \ \widetilde{\psi_1} \left(\frac{Q}{ | \omega |^{1/2p} }\right) =  | \omega|^{(p - 1)/2p} \ \widetilde{\psi} \left(\frac{Q}{ | \omega |^{1/2p} }\right)  . \label{eqn:Psi-omeganonlocal}
 \end{align}
\item Exponential decay: Let $a > 1/2$ and set $\eta=\eta(s)=\min(1,2s)$. 
Then, there exists a positive constant $\mu = \mu [   \|   \widetilde{\psi}  \|_{L^{2,a}(\mathbb{R} )} ] $ such that
\begin{equation}
\left\| e^{ \mu  |Q|^\eta } \ \widetilde{\psi}(Q) \right\|_{L^{2,a}(\mathbb{R}_Q)} \lesssim \left\|   \widetilde{\psi}(Q) \right\|_{L^{2,a}(\mathbb{R}_Q)}. 
\label{Psi-boundnonlocal}
\end{equation}

 \end{enumerate}
\end{proposition}

\noindent We require, in particular,  a characterization of the $L^2(\mathbb{R})-$ kernel of $L_+$.  This was obtained in \cite{Kwong_1989} for the case $p = 1$ and more generally in the important recent work \cite{Frank_Lenzmann_2013,Frank_Lenzmann_2015}. In Fourier space, we define the associated operator
\begin{equation}
\widetilde{f}(Q)\mapsto \widetilde{L_+} \widetilde{f}(Q)\equiv  \left(\ 1 + |Q|^{2p}- \frac{3}{4 \pi^2}\ \ \widetilde{\psi} * \widetilde{\psi} * \ \right)\widetilde{f}(Q) 
\label{L+Fouriernonlocal}
\end{equation}

\begin{proposition} 
\label{prop:Lplusnonlocal}
\begin{enumerate}
\item The continuous spectrum of $L_+$ is given by the half-line $[1,\infty)$.
\item Zero is an isolated eigenvalue of $L_+$  with corresponding eigenspace, 
 $ {\rm kernel}(L_+)={\rm span}\Big\{\partial_{x} \psi (x) \Big\}.$
\item Zero is an isolated eigenvalue of $\widetilde{L_+}$  with corresponding eigenspace,   ${\rm kernel}(\widetilde{L_+})={\rm span}\Big\{Q\ \widetilde{\psi}(Q) \Big\}$.
%the kernel of $\widetilde{L_+} $  is spanned by the functions $Q\ \widetilde{\psi}(Q)$.
\item $L_+: H^{a}_{_{\rm even}}(\mathbb{R})\to H^{a-2p}_{_{\rm even}}(\mathbb{R})$ is an isomorphism. 
\item $\widetilde{L_+}: L^{2,a}_{_{\rm even}}(\mathbb{R})\to L^{2,a-2p}_{_{\rm even}}(\mathbb{R})$ is an isomorphism.
\end{enumerate}
\end{proposition}

\section{Main results} \label{section:mainresults}
For $1/4 < s < \infty$, set $J_m^s=C_s^{-1}m^{-1-2s}$ giving
\begin{align}
(\mathcal{L}_s G)_n = \frac{1}{C_s} \ \sum_{ \substack{ m \in \mathbb{Z}  \\ m \neq n} }  \frac{ (G_m - G_n) }{|m - n|^{1 + 2s}}, \qquad {\rm where} \qquad   C_s = \left\{ \begin{array}{ll}  - 2 \Gamma(- 2 s) \cos( \pi s) & : 1/4 < s < 1 \\  1 & : s = 1 \\ \zeta(2s - 1) & : s > 1 \end{array} \right. .
\label{LC-def}
\end{align}
For $s = \infty$, set $J_m^s=C_\infty^{-1} e^{-\gamma m},\ \gamma > 0$ giving
\begin{align}
(\mathcal{L}_s G)_n = \frac{1}{C_{\infty}} \ \sum_{ \substack{ m \in \mathbb{Z}  \\ m \neq n} } e^{- \gamma \  |m - n|}  (G_m - G_n) , \qquad {\rm where} \qquad C_{\infty} = \frac{ {\rm exp}(\gamma) (  {\rm exp}(\gamma) + 1) }{ ({\rm exp}(\gamma) - 1)^3} . 
\end{align}

\begin{theorem}({\it Nonlocal DNLS solitary waves on $\mathbb{Z}$})
\label{th:mainnonlocal}
Consider the nonlinear eigenvalue problem governing real-valued  standing waves of the discrete nonlocal DNLS \eqref{eqn:tiDNLSrescalednonlocal}:
\begin{align}
& - \kappa_s(\alpha) \  G^{\alpha}_n = - (\mathcal{L}_s G^{\alpha})_n - (G^{\alpha}_n)^3,
\qquad G^{\alpha} \in l^2(\mathbb{Z}). \label{eqn:tiDNLS2nonlocal}
\end{align}
where $1/4<s\le\infty$ and $\kappa_s(\alpha)$ is the $s-$ dependent scaling displayed in \eqref{eqn:kappa-def}. 
%\begin{align}
%\kappa_s(\alpha) = \left\{ \begin{array}{lll} \alpha^{2s} & : 1/4 < s < 1 & \qquad (long-range)  \\ (- \log(\alpha) ) \alpha^2 & : s = 1 & \qquad (critical/marginal) \\ \alpha^2 & : 1 < s \leq \infty & \qquad (short-range) \end{array} \right. 
%\end{align} 
Let $p = p(s)= \min(s,1)$ and denote by  $\psi(x) \equiv \psi_1 (x) $ denote the ground state of the $p(s)$-FNLS equation, \eqref{eqn:introtiNLS}; see Proposition \ref{prop:psinonlocal}. 

 Then, there exists  $\alpha_0 = \alpha_0[s] > 0$ such that for all $0 < \alpha < \alpha_0$, nonlocal DNLS has  two real-valued solitary wave solutions to \eqref{eqn:tiDNLS2nonlocal}. These are onsite (lattice point- centered) and offsite (bond-centered) symmetric states. To leading order in $\alpha$, these are given by $\sigma=0, 1/2$ translates of $\psi( \alpha x)$ sampled on $\mathbb{Z}$.   The leading order expansions with correctors are given by:
  
 \noindent {\bf Onsite symmetric \ (vertex-centered):}
 \[ G^{\alpha, {\rm on} }_n =  \kappa_s(\alpha)^{1/2} \ \psi \left( \alpha \ n  \right) + \mathcal{E}^{\alpha, {\rm on } }_n, \quad \quad \quad  n \in \mathbb{Z} \]

  \noindent {\bf Offsite symmetric \ (bond-centered):}
 \[  G^{\alpha, {\rm off} }_n =  \kappa_s(\alpha)^{1/2} \ \psi \left( \alpha \ [n - 1/2]  \right)  + \mathcal{E}^{\alpha, {\rm off} }_n, \quad \quad \quad  n \in \mathbb{Z}\ . \] 
Here, 
$\left\|  \mathcal{E}^{\alpha, {\rm on}} \right\|_{l^2(\mathbb{Z})}
 + \left\|  \mathcal{E}^{\alpha, {\rm off}} \right\|_{l^2(\mathbb{Z})}  \lesssim \mathfrak{e}_s(\alpha)\to0$
 as $\alpha\to0$, where the $s-$ dependent rate $\mathfrak{e}_s(\alpha)$ is
  is given in \eqref{frak-e}. Note: the relative $l^2(\mathbb{Z})$ norm of the corrector to the leading term, $\mathfrak{e}_s(\alpha)\div\left(\kappa_s(\alpha)/\alpha\right)^{1\over2}\to0$ as $\alpha\to0$.  
%
%\begin{align}
%\begin{array}{lll}
%& \textrm{\rm Onsite symmetric \ (vertex-centered):} \quad  G^{\alpha, {\rm on} }_n =  \kappa_s(\alpha)^{1/2} \ \psi \left( \alpha \ n  \right) + \mathcal{E}^{\alpha, {\rm on } }_n, \quad \quad \quad  n \in \mathbb{Z} \\
%&\\
%& \textrm{\rm Offsite symmetric \ (bond-centered):}  \\
%& \quad G^{\alpha, {\rm off} }_n =  \kappa_s(\alpha)^{1/2} \ \psi \left( \alpha \ [n - 1/2]  \right)  + \mathcal{E}^{\alpha, {\rm off} }_n, \quad \quad \quad  n \in \mathbb{Z} \\
%%& \textrm{with the following bounds on functions in }l^2(\mathbb{Z}) : \\
% &   \quad \left\|   \psi \left( \alpha \ n  \right) \right\|_{l^2(\mathbb{Z}_n)} \sim \alpha^{-1/2}, \quad \quad  \quad  \left\|  \mathcal{E}^{\alpha, {\rm on}} \right\|_{l^2(\mathbb{Z})} \lesssim \mathfrak{e}(\alpha), \\
% & \\
%
%%& \textrm{ with the following bounds: } \\
%   &  \quad \quad \left\|   \psi \left( \alpha \ [n - 1/2] \right) \right\|_{l^2(\mathbb{Z}_n)} \sim \alpha^{-1/2},  \quad \quad \quad \left\| \mathcal{E}^{ \alpha, {\rm off}}  \right\|_{l^2(\mathbb{Z})} \lesssim \mathfrak{e}(\alpha).
%\end{array} \label{eqn:solutionsnonlocal}
%\end{align}
 \end{theorem}
 \medskip

 \begin{remark} \label{remark:main2nonlocal}
In Theorem \ref{th:mainnonlocal}, we assume $s > 1/4$.
For cubic nonlinearity in one space dimension do not expect a bifurcation of discrete solitary waves from at zero frequency for $s \leq 1/4$. Indeed, the $\kappa_s(\alpha) \to 0$ rescaled-limit of such bifurcating states is the solitary standing wave solution of continuum fractional NLS, satisfying  $(- \Delta_x)^p u+u-u^3=0$. A well-known  argument based on ``Pohozaev'' / virial identities \cite{Strauss_1977,Sulem_Sulem_1999,Ros-Oton_Serra_2014} implies that the equation $(-\Delta_x)^p u +u-u^m =0$ has  $H^{2p + 1}(\mathbb{R}^d)$ solutions  only if $m<(d+2p)/(d-2p)$. For the case of cubic nonlinearity, $m = 3$, with $d = 1$, this implies $p = \min(1,s) > 1/4$.
\end{remark}
 
 \bigskip

\begin{theorem} 
\label{th:PNnonlocal} [Exponential smallness of Peierls-Nabarro barrier]
Let $\eta = \min(2s, 1)$. There exist constants $\alpha_0 > 0, C >0$ and $D>0$ such that for all $0 < \alpha < \alpha_0$,
\begin{align}
& \bigg| \mathcal{N} [ G^{\alpha,  {\rm off}} ] - \mathcal{N} [ G^{\alpha,  {\rm on} } ] \bigg| =  \left|  \| G^{\alpha, {\rm off} } \|^2_{l^2(\mathbb{Z})} - \|G^{\alpha, {\rm on} } \|^2_{l^2(\mathbb{Z})} \right|\ \le\ D \left( \frac{\kappa_s(\alpha)}{\alpha} \right)  \ e^{- C / \alpha^{\eta}} ,\quad \textrm{and} \nonumber \\
  &  \bigg| \mathcal{H} [ G^{\alpha,  {\rm off} } ] - \mathcal{H} [ G^{\alpha,  {\rm on} } ] \bigg| \ \le \ D \left( \frac{\kappa_s(\alpha)}{\alpha} \right)  \ e^{- C / \alpha^{\eta}}. \label{eqn:pntheoremnonlocal}
\end{align}
\end{theorem}

\medskip

\section{Beginning of the proof of Theorem \ref{th:mainnonlocal}  ; formulation of nonlocal DNLS in Fourier space}
\label{section:1dproofnonlocal}

The proof follows the general strategy of our study of the  local (nearest-neighbor) DNLS  \cite{Jenkinson_Weinstein_2015}  and refer to it when convenient. 
%Recall that
%\begin{align}
%(\mathcal{L}^s G)_n =     \sum_{ \substack{ m \in \mathbb{Z}  \\ m \neq n} } J^s_{|m  - n|} \   (G_m - G_n)  ,
%\end{align}
%where we set
% \begin{align}
%  J^s_m = \left\{ \begin{array}{ll} \frac{1}{C_s \ m^{1 + 2s}} & : 1/4 < s < \infty  \\
%  & \\
% \frac{1}{C_{\infty}} \ e^{- \gamma \ m} & : s = \infty \end{array} \right. . 
%\end{align}
%for $\gamma > 0$ arbitrary and
%\begin{align}
%C_s = \left\{ \begin{array}{ll}  - 2 \Gamma(- 2 s) \cos( \pi s) & : 1/4 < s < 1 \\  1 & : s = 1 \\ \zeta(2s - 1) & : 1 < s \leq \infty \\  \frac{ {\rm exp}(\gamma) (  {\rm exp}(\gamma) + 1) }{ ({\rm exp}(\gamma) - 1)^3} & : s = \infty \end{array} \right. . \label{eqn:Csdef}
%\end{align}
%We also set
%\begin{align}
%\kappa_s(\alpha) = \left\{ \begin{array}{ll} \alpha^{2s} & : 1/4 < s < 1 \\ (- \log(\alpha) ) \alpha^2 & : s = 1 \\ \alpha^2 & : 1 < s \leq \infty \end{array} \right. \label{eqn:kappadef}
%\end{align}
%\medskip
Applying the discrete Fourier transform
 \eqref{eqn:DFT} to equation \eqref{eqn:tiDNLS2nonlocal}, governing $G=G^{\alpha, s}$, we obtain an equivalent equation for the discrete Fourier transform, $\widehat{G}(q)=\widehat{G^{\alpha}}(q)$:
\begin{align}
     \widehat{DNLS}[\widehat{G}](q)\ \equiv\ &\big[ \kappa_s(\alpha) +  M^s(q) \big] \widehat{G}(q) - \left(\frac{1}{2 \pi} \right)^2 \widehat{G} *_{_1} \widehat{G} *_{_1} \widehat{G}(q) = 0, \nonumber\\
 &    \widehat{G}(q + 2\pi) = \widehat{G}(q).\ \label{eqn:tiDNLS2realfouriernonlocal}
\end{align}
Here, $M^s(q)$ denotes the (discrete) Fourier symbol of the operator $\mathcal{L}^s$ (see Lemma \ref{lemma:nonlocaltransform}),
\begin{align}
M^s(q) \ \widehat{G}(q)\ =\ - \widehat{ (\mathcal{L}^s G)}(q) =  4 \sum_{m = 1}^{\infty} J^s_m \sin^2(q m /2) \  \widehat{G}(q), \  \label{eqn:Mdefnonlocal}
\end{align}
and  $f*_{_1}g$ denotes the convolution on $\mathcal{B} = \mathcal{B}_1$; see \eqref{eqn:periodicconvolution} and Appendix \ref{appendix:DFT}. 

 By Proposition \ref{prop:off-on}  
\begin{enumerate}
\item $G$ is onsite symmetric iff $\widehat{G}(q)= \widehat{K}(q)$, where $ \widehat{K}(q)$ is real and symmetric, and
\item $G$ is offsite symmetric iff $\widehat{G}(q)=e^{-iq/2} \widehat{K}(q)$, where $\widehat{K}(q)$ is real and symmetric.
\end{enumerate}
Hence we seek $\widehat{G}(q)$ in the form
\begin{align}
\widehat{G^{\sigma}}(q)\ &=\ e^{-i\sigma q}\ \widehat{K^{\sigma}}(q),\ \ \sigma=0,1/2\label{Gaqnonlocal}
\\ \widehat{K^{\sigma}}(q)&= \widehat{K^{\sigma}}(-q),\ \ \overline{\widehat{K^{\sigma}}(q)}=\widehat{K^{\sigma}}(q)
 \label{Hrealsymnonlocal}
\end{align}

\noindent The following result (Proposition 4.3 of \cite{Jenkinson_Weinstein_2015}) states that it suffices 
to solve \eqref{eqn:tiDNLS2realfouriernonlocal}, projected onto $\mathcal{B}=[-\pi,\pi]$.
 
 \begin{proposition} \label{prop:phieqnnonlocal}
 Let $\widehat{\phi^{\sigma}}(q)=\chi_{_{\mathcal{B}}}(q)\ \widehat{\phi^{\sigma}}(q)$ satisfy:
\begin{align}
& [\alpha^2 + M^s(q)] \ \widehat{\phi^{\sigma}}(q) - \frac{\chi_{_{\mathcal{B}}}(q)}{4\pi^2}  
\  \left(\ \widehat{\phi^{\sigma}} *  \widehat{\phi^{\sigma}} *  \widehat{\phi^{\sigma}}\ \right) (q) 
\nonumber\\
&\qquad\qquad - \ 
\frac{\chi_{_{\mathcal{B}}}(q)}{4\pi^2}  
\sum_{m = \pm1} \ e^{2 m \pi i \sigma} \ \left(\ \widehat{\phi^{\sigma}} *  \widehat{\phi^{\sigma}} *  \widehat{\phi^{\sigma}}\ \right) (q - 2 m \pi) = 0,
\label{eqn:phieqnnonlocal}
\end{align}
%where
%$M^s (q) \widehat{G}(q)\ =\ 4 \sum_{m = 1}^{\infty} J_m^s \ \sin^2(q m /2) \ \widehat{G}(q) = - ( \widehat{\mathcal{L}^s G} ) (q) $; see \eqref{eqn:Mdefnonlocal}. 
Then \eqref{eqn:tiDNLS2realfouriernonlocal} is solved by $\widehat{G}(q) = e^{- i q \sigma} \ \widehat{K^\sigma}(q)$,  where 
\begin{align}
\widehat{K^\sigma}(q) = \sum_{m \in \mathbb{Z}} \chi_{_{\mathcal{B}}}(q - 2 m \pi) \ \widehat{\phi^\sigma}(q - 2m \pi) e^{2 m \pi i \sigma},
\end{align}
\end{proposition}

\subsection{Rescaled equation for $\widehat{\phi^{\sigma}}$} \label{section:rescaledphinonlocal}
As discussed in Section \ref{section:nonlocalstrategy}, we expect that for $\alpha \ll 1$:  
\begin{align}
\widehat{\phi^{\sigma}}(q) \sim \left( \frac{\kappa_s(\alpha)^{1/2}}{\alpha^p} \right) \  \widetilde{\psi_{\alpha^{2p}}}(q) =  \left( \frac{\kappa_s(\alpha)^{1/2}}{\alpha} \right)  \ \widetilde{\psi} \left( \frac{q}{\alpha} \right), \quad \quad 
\end{align}
see \eqref{eqn:introtiNLS} and Proposition \ref{prop:psinonlocal}.
We therefore study \eqref{eqn:phieqnnonlocal} using a rescaling which makes explicit the relation between
DNLS and the continuum (FNLS) limit for $\alpha$ small.\\
{\bf Rescalings:}
\begin{align}
& \textrm{\rm Rescaled momentum:} \quad \quad & Q \equiv q/\alpha,\ \ 
Q\in \mathcal{B}_{\alpha} = [-\pi/\alpha,\pi/\alpha],  \nonumber \\
& \textrm{\rm Rescaled projection:} \quad \quad  &  \chi_{_{\mathcal{B}_{_{\alpha}}}}(Q) \equiv \chi_{_{\mathcal{B}}}(Q \alpha) = \chi_{_{ \left[ - \frac{\pi}{\alpha}, \frac{\pi}{\alpha} \right] }}(Q), 
 \nonumber \\
& \textrm{\rm Rescaled wave:} \quad \quad  &  \widehat{\Phi^{\sigma}}(Q) \equiv  \left( \frac{\alpha}{\kappa_s(\alpha)^{1/2}} \right) \widehat{\phi^{\sigma}}(Q \alpha) =   \left( \frac{\alpha}{\kappa_s(\alpha)^{1/2}} \right)  \widehat{\phi^{\sigma}}( q), \nonumber \\
& \textrm{\rm Rescaled discrete Fourier symbol:} \quad \quad  & M_{\alpha}^s(Q) \equiv \frac{1}{\kappa_s(\alpha)} M^s(Q \alpha) =  \frac{4}{\kappa_s(\alpha)}  \sum_{m = 1}^{\infty} J_m^s \ \sin^2 ( Q m \alpha/2) \label{eqn:M2defnonlocal}
\end{align}
The following proposition is a reformulation of Proposition \ref{prop:phieqnnonlocal}  in terms of the rescaled quasi-momentum, $Q$:

\begin{proposition}
\label{prop:Phieqnnonlocal}
Equation \eqref{eqn:phieqnnonlocal} for $\widehat{\phi^{\sigma}}(q)$ on $q \in\mathcal{B}$ is equivalent to the following equation for  $\widehat{\Phi}(Q)= \chi_{_{\mathcal{B}_{\alpha}}}(Q) \widehat{\Phi^{\sigma}}(Q)$, supported on
$\mathcal{B}_\alpha=[-\pi/\alpha,\pi/\alpha]$:
\begin{align}
&\mathcal{D}^{\sigma,\alpha}[\widehat{\Phi}](Q)\ \equiv\ [1 + M^s_\alpha(Q)] \ \widehat{\Phi}(Q) - \frac{\chi_{_{\mathcal{B}_\alpha}}(Q)}{4\pi^2}  
\  \left(\ \widehat{\Phi} *  \widehat{\Phi} *  \widehat{\Phi}\ \right) (Q) 
\ +\ R_1^{\sigma} [ \widehat{\Phi} ] (Q) = 0\ ,
\label{eqn:Phieqnnonlocal}
\end{align}
Here, $R_1^{\sigma} [ \widehat{\Phi} ]$ contains the $\pm1$-sideband contributions:
\begin{align}
R_1^{\sigma} [ \widehat{\Phi} ] (Q) &\equiv  -  \frac{\chi_{_{\mathcal{B}_{_{\alpha}}}}( Q)}{4 \pi^2} \ \sum_{m = \pm 1} \ e^{2 m \pi i \sigma} \  \left(\ \widehat{\Phi} *  \widehat{\Phi} *  \widehat{\Phi}\ \right)  ( Q - 2 m \pi / \alpha). \label{eqn:R1defnonlocal}
\end{align}
\end{proposition}

\noindent {\bf Proof of Proposition \ref{prop:Phieqnnonlocal}:} We use the following lemma to rescale the convolutions in \eqref{eqn:phieqnnonlocal}. 
\medskip
\begin{lemma}
\label{lemma:convresc}
Suppose that $\widehat{a}(q) = \widehat{A}(Q)$, $\widehat{b}(q) = \widehat{B}(Q)$, and $\widehat{c}(q) = \widehat{C}(Q)$, where $Q=q/\alpha$.  Then
\begin{align}
\left( \widehat{a} * \widehat{b} * \widehat{c} \right)  (q)  = \alpha^2 \ \left( \widehat{A} * \widehat{B} * \widehat{C} \right) (Q) .
\end{align}
\end{lemma}
Applying the rescalings \eqref{eqn:M2defnonlocal} and Lemma \ref{lemma:convresc} to \eqref{eqn:phieqnnonlocal} and  then dividing by $ \kappa_s(\alpha)^{3/2}/\alpha$, we obtain \eqref{eqn:Phieqnnonlocal}. $\Box$   \\

The next proposition states the sense in which $M_{\alpha}^s(Q) \to |Q|^{2p}$ ($p=\min(1,s)$) as $\alpha \to 0$. This facilates the solution of   $\mathcal{D}^{\sigma, \alpha} \left[ \widehat{\Phi} \right](Q) = 0$ for $\widehat{\Phi}^{\sigma,\alpha}(Q)$, perturbatively about the solution of the  limiting $p(s)-$ FNLS limit equation \eqref{eqn:Psi-omeganonlocal}.
%
%  $ \left( 1 + |Q|^{2p} \right) \ \widetilde{\psi}(Q) - (2 \pi)^{-2} \ \left( \widetilde{\psi} * \widetilde{\psi} * \widetilde{\psi} \right)(Q) = 0$. \medskip

\begin{proposition} \label{prop:symbolexpo}
Let $0 < \eta \leq 1$ and fix $p = \min(1,s)$. Suppose  $e^{ C |Q|^{\eta}} \widehat{g}(Q) \in L^{2,a}(\mathbb{R}_Q)$ for some $C>0$.
% Let $\mathfrak{e}_s(\alpha)$ be as defined in \eqref{eqn:gamma111}. 
 Then, as $\alpha\to0$,
\begin{align}
\left\| \chi_{_{\mathcal{B}_{_{\alpha}}}}(Q) \ \left[ M_{\alpha}^s(Q)  - |Q|^{2p} \right] \widehat{g}(Q) \ \right\|_{L^{2,a}(\mathbb{R}_Q)} \lesssim \left( \frac{\alpha }{ \kappa_s(\alpha)} \right)^{1/2} \ \mathfrak{e}_s(\alpha). \label{eqn:e1est}
\end{align}
\end{proposition}

 \noindent {\bf Proof of Proposition \ref{prop:symbolexpo}:} The heart of the matter is Proposition \ref{prop:symbolconvnonlocal1} which states that for  $|Q| \leq \pi / \alpha$ ($Q\in\mathcal{B}_\alpha$), there exists a function $f_s(Q;\alpha)$ such that $\max_{Q\in\mathcal{B}_\alpha}| f_s(Q; \alpha) | \lesssim 1$ and
 \begin{align}
 M^s_{\alpha}(Q) = \frac{1}{C_s \ \kappa_s(\alpha)} M^s(Q \alpha) = |Q|^{2p} + \left\{ \begin{array}{ll} \alpha^{2 - 2s} \ f_s( Q; \alpha) \ |Q|^2 & : \quad s < 1\\ 
& \\
 \hline \\
 \frac{1}{- \log(\alpha)} \ f_s(Q ; \alpha) \ \left( \frac{3}{2} - \log(|Q|) \right) \ |Q|^2 & : \quad s = 1 \\
& \\
 \hline \\
 \alpha^{2s - 2} \ f_s(Q; \alpha) \ |Q|^{2s} & : \quad 1 < s < 2 \\
 & \\
 \hline \\
 (- \log(\alpha)) \ \alpha^2 \ f_s(Q; \alpha) \ |Q|^4 & : \quad s = 2 \\
& \\
 \hline \\
 \alpha^2 \ f_s(Q; \alpha) \ |Q|^4 & : \quad 2 < s \leq \infty \end{array} \right. . \label{eqn:symbolexpoest1}
 \end{align}
Using \eqref{eqn:symbolexpoest1}, $\chi_{_{\mathcal{B}_{_{\alpha}}}}(Q) \ \left[ M_{\alpha}^s(Q)  - |Q|^{2p}\right]\widehat{g}(Q)$ can be bounded by $(\alpha / \kappa_s(\alpha) )^{1/2} \ \mathfrak{e}_s(\alpha)$ (see \eqref{frak-e}) times a constant in terms of  
$
\left\| \ |Q|^{2j} \ \widehat{g}(Q) \right\|_{L^{2,a}(\mathbb{R}_Q)}, \quad \left\| \ (\log(|Q|)) \  |Q|^{2j} \ \widehat{g}(Q) \right\|_{L^{2,a}(\mathbb{R}_Q)}
$ ($j=1,2$), 
which is finite by the hypothesis on $ \widehat{g}$. 
 This completes the proof of Proposition \ref{prop:symbolexpo}. $\Box$ \\

\section{Bifurcation analysis and the proof of Theorem \ref{th:leadingordernonlocal}} \label{section:nonlocalleadingorderproof}

 Our main results on bifurcation are derived from the following:
\begin{theorem} \label{th:leadingordernonlocal}
Fix  $a > 1/2$. Consider the nonlinear eigenvalue problem \eqref{eqn:tiDNLS2nonlocal} for onsite symmetric ($\sigma=0$) and offsite symmetric ($\sigma=1/2$) bound states of nonlocal DNLS with interaction parameter $s\in(1/2,\infty]$. Let $p = \min(1,s)$.  Let $\psi$ denote the ground state of the p-FNLS equation, \eqref{Psi-eqnnonlocal}, $\psi + (- \Delta)^p \psi - \psi^3 = 0$, and let $\widetilde{\psi}$ be its continuous Fourier transform. Then there exist a constant $\alpha_0 = \alpha_0[a, \sigma, s] > 0$, and a unique, real-valued function $\widehat{E
^{\alpha, \sigma}} \in L_{_{\rm even}}^{2,a}(\mathbb{R})$, with $ \widehat{E^{\alpha, \sigma}} = \chi_{_{\mathcal{B}_{_{\alpha}}}}  \widehat{E^{\alpha, \sigma}}$,  defined for all $0 < \alpha < \alpha_0$,  such that 
\begin{align}
\widehat{\Phi^{\alpha,\sigma}}( Q) = \chi_{_{\mathcal{B}_{_{\alpha}}}}(Q) \  \widetilde{\psi} (Q)  + \widehat{E^{\alpha, \sigma}}(Q), \label{eqn:leadingordernonlocal}
\end{align}
solves \eqref{eqn:Phieqnnonlocal}, with  corrector bound:
$
\left\| \widehat{E^{\alpha, \sigma}} \right\|_{L^{2,a}(\mathbb{R})} \lesssim ( \alpha  / \kappa_s(\alpha) )^{1/2} \ \mathfrak{e}_s(\alpha) \to 0$ $ \alpha\to0$. $\mathfrak{e}_s(\alpha)$ is displayed in \eqref{frak-e}.
%Here,
% \begin{align}
%\mathfrak{e}_s(\alpha) = \left\{ \begin{array}{ll} \alpha^{2 (1 - s)} & : s < 1 \\
%\frac{1}{(- \log(\alpha)} & : s = 1 \\
%\alpha^{2 ( s - 1)} & : 1 < s < 2 \\
%(- \log(\alpha)) \ \alpha^2  & : s = 2 \\
%\alpha^2 & : s > 2
%\end{array} \right. .  \label{eqn:gamma111} 
%\end{align} 
\end{theorem}

\subsection{Equation for the remainder, $\widehat{E^{\alpha, \sigma}}$}
 For $\widehat{\Phi}=\widehat{\Phi^{\alpha,\sigma}}$ we take the ansatz
\begin{align}
\widehat{\Phi}( Q) = \chi_{_{\mathcal{B}_{_{\alpha}}}}(Q) \  \widetilde{\psi} (Q)  + \widehat{E^{\alpha, \sigma}}(Q) \equiv S(Q)  + \widehat{E}(Q). \label{eqn:ansatzsumnonlocal}
\end{align} 
Here, $S(Q) = \chi_{_{\mathcal{B}_{_{\alpha}}}}(Q) \  \widetilde{\psi} (Q) $. 
To prove Theorem \ref{th:leadingordernonlocal},  we derive an equation for 
 $\widehat{E}(Q)$ and then construct and bound its solution.
Recall, by Proposition \ref{prop:Phieqnnonlocal}, that $\widehat{\Phi}$ must satisfy $\mathcal{D}^{\sigma,\alpha}[\widehat{\Phi}](Q)=0$;  
see \eqref{eqn:Phieqnnonlocal}.
Substitution of the ansatz \eqref{eqn:ansatzsumnonlocal} into this equation yields the required equation for $\widehat{E}(Q)$. \\

\begin{proposition} \label{prop:Eeqnnonlocal}
$\mathcal{D}^{\sigma,\alpha}[\widehat{\Phi}]=0$ is equivalent to the following equation for $\widehat{E}(Q)=\chi_{_{\mathcal{B}_{_{\alpha}}}}(Q) \ \widehat{E}(Q)$:
\begin{align}
\left[ 1 + M_{\alpha}^s( Q) \right] \ \widehat{E}( Q ) - 3 \ \chi_{_{\mathcal{B}_{_{\alpha}}}}( Q)  \   \frac{1}{(2 \pi)^2}\
\widetilde{\psi} * \widetilde{\psi} * \widehat{E} ( Q) = \mathcal{R}_1^{\sigma} \left[ \alpha, \widehat{E} \right] (Q), \quad \widehat{E} \in L^{2,a}(\mathbb{R}).\label{eqn:Eeqnnonlocal}
\end{align}
where 
\begin{align}
\mathcal{R}_1^{\sigma} \left[ \alpha, \widehat{E} \right]  \equiv \mathcal{D}^{\sigma,\alpha}[S]  + R_{_{\rm L}}^{\sigma} \left[ \alpha, \widehat{E} \right] + R_{_{\rm NL}}^{\sigma} \left[ \alpha, \widehat{E} \right] . \label{eqn:HJdefnonlocal}
\end{align}
$R_{_{\rm L}}^{\sigma} $ is linear in $\widehat{E}$ and of order $\mathcal{O}(1)$ in $\alpha$ on $ Q \in \mathcal{B}_{_{\alpha}}$. 
$R_{_{\rm NL}}^{\sigma} $ contains terms which are nonlinear in $\widehat{E}$:
\begin{align}
& R_{_{\rm L}}^{\sigma} \left[ \alpha, \widehat{E} \right] (Q) \equiv   \chi_{_{\mathcal{B}_{_{\alpha}}}}( Q)  \ \frac{3}{4 \pi^2} \ \bigg[ \ \sum_{m = -1}^1 \ e^{2 m \pi i \sigma} \   \left(S * S * \widehat{E}\right) (Q - 2 m \pi / \alpha)  - \ \left(\widetilde{\psi_1} * \widetilde{\psi_1}  * \widehat{E}\right)(Q) \bigg], \nonumber \\
& R_{_{\rm NL}}^{\sigma} \left[ \alpha, \widehat{E} \right] (Q) \equiv \chi_{_{\mathcal{B}_{_{\alpha}}}}( Q)  \  \left( \frac{1}{2 \pi} \right)^2 \ \sum_{m = -1}^1 \ e^{2 m \pi i \sigma} \ \bigg[ 3  \ \left(S * \widehat{E} * \widehat{E}\right) (Q - 2 m \pi / \alpha)  \nonumber \\
& \hspace{7cm} + \left(\widehat{E} * \widehat{E} * \widehat{E}\right) (Q - 2 m \pi / \alpha) \bigg]. \label{eqn:RLNLdefnonlocal}
\end{align}
\end{proposition}

\begin{remark}
Note that the operator on the left-hand side has formal limit $\widetilde{L_+}$, where $L_+$ is the linearized continuum FNLS operator displayed in Proposition \ref{prop:Lplusnonlocal}.
 
\end{remark}
\medskip

\subsection{Coupled system for high and low frequency components of $\widehat{E}=\widehat{E^{\alpha, \sigma}}$} \label{section:decompnonlocal}
We now embark on the construction   $\widehat{E}=\widehat{E^{\alpha, \sigma}} \in L^{2,a}(\mathbb{R})$ for $\alpha > 0$ sufficiently small. Our strategy is to reformulate the equation for $\widehat{E}$ as an equivalent coupled system for its high and low frequency components. 

Let $r$ be such that $0 < r < 1$. Define the spectral cutoff parameter:
\begin{align}
\lambda(\alpha) = \lambda_s(\alpha) \equiv \left\{ \begin{array}{ll} \alpha^{1 - r} & : s \neq 1 \\
 \frac{1}{(- \log(\alpha))} & : s = 1 \end{array} \right. .
\label{lambda-alpha}\end{align}
Note that $\lambda_s(\alpha) \to 0$ as $\alpha \to 0$. 
 
Next, define the sharp spectral cutoffs onto low and high frequency regimes:
\begin{align}
& \chi_{_{\rm lo}}(Q) =  \chi \left( |Q| \leq \frac{1}{\lambda_s(\alpha)} \right) \ \ {\rm and}\ \  \chi_{_{\rm hi}}(Q) =  \chi \left( |Q| > \frac{1}{\lambda_s(\alpha)} \right), \label{lo-hi-cutnonlocal}
\end{align}
where $1 = \chi_{_{\rm lo}}(Q) + \chi_{_{\rm hi}}(Q)$.
Note that $\chi_{_{\rm lo}}\ \chi_{_{\mathcal{B}_{_{\alpha}}}}= \chi_{_{\rm lo}} $, while $\chi_{_{\rm hi}} \ \chi_{_{\mathcal{B}_{_{\alpha}}}} =  \chi_{_{\mathcal{B}_{_{\alpha}}}} - \chi_{_{\rm lo}}$. 
For general $\widehat{A}$, defined on $\mathbb{R} $, introduce its localizations near and away from $Q =0$:
\begin{align}
&\widehat{ A}{_{_{\rm lo}}}( Q )\ =\ \left(\chi_{_{\rm lo}} \widehat{A}\right)( Q ) \equiv \chi_{_{\rm lo}}( Q ) \widehat{A}(Q), \\
&\widehat{ A}{_{_{\rm hi}}}( Q ) = \left(\chi_{_{\rm hi}} \widehat{A}\right)( Q ) \equiv \chi_{_{\rm hi}}( Q ) \widehat{A}(Q),
\end{align}
In particular, we use   $\chi_{_{\rm lo}}$ and $\chi_{_{\rm hi}}$ to localize $\widehat{E}=\widehat{E^{\alpha, \sigma}}$, where  $| Q | \leq \lambda_s(\alpha)$ and  where $ |Q| > \lambda_s(\alpha)$:
\begin{align}
& \widehat{E_{_{\rm lo}}}( Q ) = \chi_{_{\rm lo}}( Q ) \widehat{E}( Q )\ \ {\rm and}\ \  \widehat{E_{_{\rm hi}}}( Q ) = \chi_{_{\rm hi}}(Q ) \widehat{E}( Q ),
\end{align}
and therefore 
$ \widehat{E}( Q ) = \widehat{E_{_{\rm lo}}}( Q ) + \widehat{E_{_{\rm hi}}}( Q ) . $

\noindent {\bf NOTE:} {\it Since the analysis for the cases $\sigma = 0$ (onsite) and $\sigma = 1/2$ (offsite) in sections \ref{section:decompnonlocal} - \ref{section:solutionlownonlocal} are very similar, in order to keep the notation less cumbersome we omit the superscripts $\alpha$ and $\sigma$ when the context is clear, and shall instead write:}
\begin{align}
\widehat{E}( Q )=\widehat{E}^{\alpha, \sigma}( Q ), \quad \quad
\widehat{E_{_{\rm lo}}}( Q )=\widehat{E^{\alpha, \sigma}_{_{\rm lo}}}( Q ), \quad \quad
\widehat{E_{_{\rm hi}}}( Q )=\widehat{E^{\alpha, \sigma}_{_{\rm hi}}}( Q ). \quad \quad
\label{eqn:drop-supernonlocal} \end{align}

The following Proposition is obtained by applying the spectral projections $\chi_{_{\rm lo}}$ and $\chi_{_{\rm hi}}$ to \eqref{eqn:Eeqnnonlocal}.

\bigskip

\begin{proposition}
\label{prop:highlownonlocal}
If  $\widehat{E}$ is a solution of \eqref{eqn:Eeqnnonlocal} then its  low and high frequency components, $\widehat{E_{_{\rm lo}}}$ and $\widehat{E_{_{\rm hi}}}$,  satisfy the coupled system:
\begin{flalign}
& \textbf{Low Frequency Equation:} \nonumber \\
&  \hspace{2cm} \left[ 1 + M_{\alpha}^s(Q) \right] \  \widehat{E_{_{\rm lo}}} ( Q ) - \chi_{_{\rm lo}}( Q )   \frac{3}{4 \pi^2}   \left( \widetilde{\psi} * \widetilde{\psi} * \widehat{E_{_{\rm lo}}}  (Q) + \widetilde{\psi} * \widetilde{\psi} * \widehat{E_{_{\rm hi}}}  (Q) \right) & \nonumber \\ 
& \hspace{7cm} = \chi_{_{\rm lo}}( Q ) \ \mathcal{R}_1^{\sigma} \left[ \alpha, \widehat{E_{_{\rm lo}}} + \widehat{E_{_{\rm hi}}} \right] ( Q), \label{eqn:lownonlocal} \\
& \textbf{High Frequency Equation:} \nonumber \\
&  \hspace{2cm} \left[ 1 + M_{\alpha}^s(Q) \right] \  \widehat{E_{_{\rm hi}}} ( Q ) - \chi_{_{\rm hi}}( Q )  \ \chi_{_{\mathcal{B}_{_{\alpha}}}}(Q) \  \frac{3}{4 \pi^2}   \left( \widetilde{\psi} * \widetilde{\psi} * \widehat{E_{_{\rm lo}}}  (Q) + \widetilde{\psi} * \widetilde{\psi} * \widehat{E_{_{\rm hi}}}  (Q) \right) & \nonumber \\
& \hspace{7cm} = \chi_{_{\rm hi}}( Q ) \ \mathcal{R}_1^{\sigma} \left[ \alpha, \widehat{E_{_{\rm lo}}} + \widehat{E_{_{\rm hi}}} \right] ( Q).   \label{eqn:highnonlocal}
\end{flalign}
Here, $\mathcal{R}_1^{\sigma}$ is defined in \eqref{eqn:HJdefnonlocal}. Conversely, if 
 $(\widehat{E_{_{\rm lo}}}, \widehat{E_{_{\rm hi}}})$ solves \eqref{eqn:lownonlocal}-\eqref{eqn:highnonlocal}, then 
 $\widehat{E}\equiv \widehat{E_{_{\rm lo}}} + \widehat{E_{_{\rm hi}}}$ solves 
 \eqref{eqn:Eeqnnonlocal}.
\end{proposition} 
\medskip

We solve \eqref{eqn:lownonlocal} and \eqref{eqn:highnonlocal} via the Lyapunov-Schmidt reduction strategy used in \cite{Jenkinson_Weinstein_2015} ; see  Section \ref{section:nonlocalstrategy}.

\subsection{Reduction to a closed equation for the low frequency projection,  $\widehat{E_{_{\rm lo}}}$}
\label{subsect:lyapschmnonlocal}

We first solve for $\widehat{E_{_{\rm hi}}} =
\widehat{E_{_{\rm hi}}} \left[ \alpha , \widehat{E_{_{\rm lo}}} \right]$ as a functional of $\widehat{E_{_{\rm lo}}}$, by viewing the equation of $\widehat{E_{_{\rm hi}}}$ as depending on $\alpha \in \mathbb{R}, |\alpha| \ll 1 $ and $\Gamma( Q ) = \widehat{E_{_{\rm lo}}}( Q ) \in L^{2,a}(\mathbb{R})$. 
\begin{align}
\left[ 1 + M_{\alpha}^s(Q) \right] \ \widehat{E_{_{\rm hi}}}( Q)  - \chi_{_{\rm hi}}( Q) \ \chi_{_{\mathcal{B}_{_{\alpha}}}}(Q)  \  \frac{3}{4 \pi^2}   \bigg( \widetilde{\psi} * \widetilde{\psi} *  \Gamma (Q) + \widetilde{\psi} * \widetilde{\psi} *  \widehat{E_{_{\rm hi}}}(Q)  \bigg) \nonumber \\
- \chi_{_{\rm hi}}(Q) \ \mathcal{R}_1^{\sigma} [ \alpha, \Gamma + \widehat{E_{_{\rm hi}}} ] ( Q ) = 0\ . \label{eqn:EhatGamnonlocal}
\end{align}
In the following proposition, we construct the mapping $\widehat{E_{_{\rm lo}}} \mapsto \widehat{E_{_{\rm hi}}} \left[ \alpha, \widehat{E_{_{\rm lo}}} \right]$. 
 
\begin{proposition}
\label{prop:ifthighnonlocal} Set $0 < r < 1$, $p = \min(s,1)$, and $\eta = \min(2s, 1)$.  
\begin{enumerate}
\item There exist constants $\alpha_0, \beta_0 >  0$,  such that for all $\alpha \in (0, \alpha_0)$, equation \eqref{eqn:EhatGamnonlocal} defines a mapping
$
(\alpha, \Gamma) \mapsto \widehat{E_{_{\rm hi}}}[\alpha, \Gamma],
$
 $\widehat{E_{_{\rm hi}}}: [0,1] \times B_{_{   \beta_0 }}(0)  \rightarrow L^{2,a}(\mathbb{R}) $
where $B_{_{  \beta_0 }}(0) \subset L^{2,a}(\mathbb{R})$ such that  $\widehat{E_{_{\rm hi}}}[\alpha, \Gamma]$
 is the unique solution to  \eqref{eqn:EhatGamnonlocal} (see also \eqref{eqn:highnonlocal}). Moreover, if $\Gamma \in L^{2,a}_{_{\rm even}}$, then $\widehat{E_{_{\rm hi}}}[\alpha, \Gamma] \in L^{2,a}_{_{\rm even}}$.\
 \item The mapping is $C^1$ with respect to $\Gamma$, and there exists $C>0$, such that  for all $(\alpha, \Gamma) \in [0, \alpha_0) \times B_{_{\beta_0 }}(0)$
\begin{align}
& \left\| \widehat{E_{_{\rm hi}}}[\alpha, \Gamma] \right\|_{L^{2,a}(\mathbb{R})} \lesssim \  \left\{ \begin{array}{ll} \alpha^{2p (1 - r)}   \ \left\| \Gamma \right\|_{L^{2,a}(\mathbb{R})} +  e^{- C / \lambda_s(\alpha)^{\eta}  } & : s \neq 1 \\
\frac{1}{(- \log(\alpha)}    \ \left\| \Gamma \right\|_{L^{2,a}(\mathbb{R})} +  e^{- C / \lambda_s(\alpha)^{\eta}  } & : s = 1 \end{array} \right. 
 \label{eqn:Ehighestlyapnonlocal} \\
&\nonumber\\
& \left\| \ D_{\Gamma} \widehat{E_{_{\rm hi}}}[\alpha, \Gamma] \ \right\|_{L^{2,a}(\mathbb{R}) \longrightarrow L^{2,a}(\mathbb{R})} \lesssim  \left\{ \begin{array}{ll} \alpha^{2p (1 - r)}  & : s \neq 1 \\
\frac{1}{(- \log(\alpha))} & : s = 1 \end{array} \right. .   \label{eqn:DEhighestnonlocal}
\end{align}
Here, $\lambda_s(\alpha) \to 0$ as $\alpha \to 0$ and is given by \eqref{lambda-alpha}. The implicit constants depend only on $\alpha_0$ and $\beta_0$. 

\item $\widehat{E_{_{\rm hi}}}[\alpha, \Gamma]$ is supported on $ Q \in  \left[ - \frac{\pi}{\alpha}, - \frac{1}{\lambda_s(\alpha)} \right) \cap \left( \frac{1}{\lambda_s(\alpha)}, \frac{\pi}{\alpha} \right]$;  $\widehat{E_{_{\rm hi}}}[\alpha, \Gamma]( Q ) $  $= \chi_{_{\rm hi}} \ \chi_{_{\mathcal{B}_{_{\alpha}}}} \ \widehat{E_{_{\rm hi}}}[\alpha, \Gamma]$. \\

\end{enumerate}
\end{proposition}

\noindent {\bf Proof of Proposition \ref{prop:ifthighnonlocal}:} The proof follows that given in \cite{Jenkinson_Weinstein_2015} for nearest-neighbor DNLS. We summarize the proof  and  elaborate where the proof differs. Since $0< 1 \leq 1 + M_{\alpha}^s(Q)  $ for all  $\alpha$ positive and small, we may rewrite \eqref{eqn:EhatGamnonlocal}  as
\begin{align}
\widehat{E_{_{\rm hi}}}(Q ) -   \chi_{_{\rm hi}}( Q )  \ [ 1 + M^s_{\alpha}(Q) ]^{-1} \bigg( \chi_{_{\mathcal{B}_{_{\alpha}}}}(Q) \  \frac{3}{4 \pi^2}  \ \left[ \widetilde{\psi} * \widetilde{\psi} *  \Gamma (Q) + \widetilde{\psi} * \widetilde{\psi} * \widehat{E_{_{\rm hi}}}(Q) \right]  \nonumber \\
- \mathcal{R}_1^{\sigma} [ \alpha, \Gamma + \widehat{E_{_{\rm hi}}}]( Q ) \bigg) = 0. \label{eqn:highrewritenonlocal}
\end{align}
 The following results (Propositions \ref{prop:Mnonlocal}, \ref{prop:RFbdnonlocal} and Lemma \ref{lemma:expconvononlocal}) are the essential ingredients required to extend the arguments of   \cite{Jenkinson_Weinstein_2015} to the DNLS with non-local interactions. 

\begin{proposition}
\label{prop:Mnonlocal}
For any function $\widehat{f} \in L^{2,a}(\mathbb{R})$, we have
\begin{align}
\left\| \  \chi_{_{\rm hi}} \ \chi_{_{\mathcal{B}_{_{\alpha}}}} \  \left[ 1 + M_{\alpha}^s(Q) \right]^{-1} \  \widehat{f} \ \right\|_{L^{2,a}(\mathbb{R}_Q)} \lesssim  \left\{ \begin{array}{ll} \alpha^{2p (1 - r)} \  \left\| \ \widehat{f} \ \right\|_{L^{2,a}(\mathbb{R})} & : s \neq 1 \\
\frac{1}{(- \log(\alpha)} \left\| \ \widehat{f} \ \right\|_{L^{2,a}(\mathbb{R})} & : s = 1 \end{array} \right. . 
\end{align}
\end{proposition}
\medskip
\noindent {\bf Proof of Proposition \ref{prop:Mnonlocal}:} We require the following lemma.

\begin{lemma} \label{lemma:Mbdnonlocal}
Let $p = \min(1,s)$. There exists a constant, $C=C_s$, such that for $s \neq 1$ and $Q \in \mathcal{B}_{_{\alpha}}$, 
$
M^s_{\alpha}(Q)  
%\frac{4}{ \kappa_s(\alpha)} \sum_{m = 1}^{\infty} J^s_m \sin^2 \left( Q  m \alpha / 2 \right)
  \geq C \ | Q |^{2p}$. 
If  $s = 1$, then $ M^s_{\alpha}(Q)\geq \frac{C}{(- \log(\alpha) )} \ | Q |^{2}$. 
\end{lemma}\\

We defer the proof of Lemma \ref{lemma:Mbdnonlocal} until the end of this section. Recall that $\chi_{_{\rm hi}}(Q)$ projects onto the set $|Q| \geq \frac{1}{\lambda_s(\alpha)}$, where $\lambda_s(\alpha)\to0$, is given by \eqref{lambda-alpha}.
%\begin{align}
%\lambda(\alpha) =  \left\{ \begin{array}{ll} \alpha^{1 - r} & : s \neq 1 \\
% \frac{1}{(- \log(\alpha))} & : s = 1 \end{array} \right. .
%\end{align}
Now let $s \neq 1$. From Lemma \ref{lemma:Mbdnonlocal}, we have 
$
\chi_{_{\rm hi}}(Q) \ \chi_{_{\mathcal{B}_{_{\alpha}}}} (Q)  \ \left[ 1 + M_{\alpha}^s(Q) \right]   \geq C  \  \chi_{_{\rm hi}}(Q) \  \chi_{_{\mathcal{B}_{_{\alpha}}}}(Q) \ |Q|^{2p}  \geq C\lambda_s(\alpha)^{-2p} = C \ \alpha^{2p (r - 1)}. 
$
Next, let $s = 1$. Then Lemma \ref{lemma:Mbdnonlocal} gives
$
\chi_{_{\rm hi}}(Q) \ \chi_{_{\mathcal{B}_{_{\alpha}}}} (Q) \ \left[ 1 + M_{\alpha}^s(Q) \right]   \geq  \chi_{_{\rm hi}}(Q) \ \chi_{_{\mathcal{B}_{_{\alpha}}}} (Q) \  \frac{C}{(- \log(\alpha) )} \ | Q |^{2}  \geq \frac{C}{(- \log(\alpha))} \frac{1}{\lambda_s(\alpha)^2} = C \ (- \log(\alpha)). 
$
%In turn,
%\begin{align}
%\chi_{_{\rm lo}}(Q) \ \chi_{_{\mathcal{B}_{_{\alpha}}}} (Q)  \ \left[ 1 + M_{\alpha}^s(Q) \right]^{-1} %\lesssim \left\{ \begin{array}{ll} \alpha^{2p (1 - r)} & : s \neq 1 \\
%\frac{1}{(- \log(\alpha))} & : s = 1 \end{array} \right. . 
%\end{align}
Proposition \ref{prop:Mnonlocal} now follows. $\Box$ \\

We also require the following lemma, stated generally due to its broader  use, which establishes  exponential smallness of ``shifted" ($m = \pm 1$ by our convention) convolutions of exponentially decaying functions. \\

\begin{lemma}
\label{lemma:expconvononlocal}
Fix $0 < \eta \leq 1$ and $a > 1/2$. Let $\widehat{f} \in L^{2,a}(\mathbb{R})$ be such that $e^{\ C |Q|^\eta} \widehat{f}(Q) \in L^{2,a}(\mathbb{R}_Q)$ for some $C > 0$. Then, (a) for   $ m = \pm 1$, 
\begin{align}
\bigg\| \ \chi_{_{\mathcal{B}_{\alpha}}}(Q) \  \left(\widehat{f} * \widehat{f} * \widehat{f}\right)(Q - 2 m \pi/\alpha) \ \bigg\|_{L^{2,a}(\mathbb{R}_Q)} \lesssim \  e^{- C \pi^{\eta} / \alpha^{\eta}} \ \left\|  
\ e^{ C |Q|^{\eta} } \ \widehat{f}(Q) \ \right\|^3_{L^{2,a}(\mathbb{R}_Q)} \label{eqn:expconvo2nonlocal}
\end{align}
and (b)
\begin{align}
\bigg\| \ \chi_{_{\rm hi}}(Q) \  \left(\widehat{f} * \widehat{f} * \widehat{f}\right)(Q) \ \bigg\|_{L^{2,a}(\mathbb{R}_Q)} \lesssim \  e^{- C /(\lambda_s(\alpha))^\eta)} \ \left\|  
\ e^{ C |Q|^{\eta} } \ \widehat{f}(Q) \ \right\|^3_{L^{2,a}(\mathbb{R}_Q)} \label{eqn:expconvo2nonlocal}
\end{align}

\end{lemma}

\noindent {\bf Proof of Lemma \ref{lemma:expconvononlocal}:} We first prove (a).  For any $\xi, \zeta \in \mathbb{R}$ we have  $|Q - 2 m \pi / \alpha |^{\eta} \leq |Q - \xi - \zeta - 2 m \pi / \alpha|^{\eta} + |\xi |^{\eta} + | \zeta |^{\eta}$, since  $0<\eta\le1$ (Lemma \ref{lemma:tri}). Therefore, 
$
1  \leq e^{- C |Q - 2 m \pi / \alpha |^{\eta}} \ e^{C |Q - \xi - \zeta - 2 m \pi / \alpha |^{\eta}}  \ e^{C | \xi  |^{\eta}} \ e^{C | \zeta  |^{\eta}} \ . 
$
Moreover, if  $|Q| \leq \pi / \alpha$,  it follows that $|Q - 2 m \pi/\alpha| \geq \pi / \alpha$
and therefore 
$
\chi_{_{\mathcal{B}_{\alpha}}}(Q) e^{- C |Q - 2 m \pi / \alpha|^{\eta}} \leq e^{- C \pi^{\eta} / \alpha^{\eta}}. $
Now applying \eqref{triple-conv} and distributing the exponentials we have
\begin{align*}
&\chi_{_{\mathcal{B}_{\alpha}}}(Q)\left(\widehat{f} * \widehat{f} * \widehat{f}\right)(Q - 2 m \pi/\alpha)
= \chi_{_{\mathcal{B}_{\alpha}}}(Q) \int\int \widehat{f}(Q- 2 m \pi/\alpha-\xi-\zeta) \widehat{f}(\xi) \widehat{f}(\zeta)\ d\xi d\zeta \\
&\quad =  
\chi_{_{\mathcal{B}_{\alpha}}}(Q)\int\int e^{- C |Q - 2 m \pi / \alpha |^{\eta}} \cdot 
e^{C |Q  - 2 m \pi / \alpha - \xi - \zeta|^{\eta}}  \widehat{f}(Q- 2 m \pi/\alpha-\xi-\zeta) 
\ e^{C | \xi  |^{\eta}} \widehat{f}(\xi)\ \ e^{C | \zeta  |^{\eta}} \widehat{f}(\zeta)\ d\xi d\zeta 
\end{align*}
Therefore, 
\begin{align*}
&\Big|\ \chi_{_{\mathcal{B}_{\alpha}}}(Q)\left(\widehat{f} * \widehat{f} * \widehat{f}\right)(Q - 2 m \pi/\alpha)\ \Big|\\
&\quad\le e^{- C \pi^{\eta} / \alpha^{\eta}}\ 
\int  e^{C | \zeta  |^{\eta}} |\widehat{f}(\zeta)|\  d\zeta\ \int   e^{C |Q - 2 m \pi / \alpha - \xi - \zeta  |^{\eta}}  |\widehat{f}(Q - 2 m \pi/\alpha -\xi-\zeta)|
\ e^{C | \xi  |^{\eta}} |\widehat{f}(\xi)|\ d\xi\  \\
&\quad\  e^{- C \pi^{\eta} / \alpha^{\eta}}\ \left(\widehat{f}_\sharp*\widehat{f}_\sharp*\widehat{f}_\sharp)\right)(Q),\qquad \textrm{where}\ \  
\widehat{f}_\sharp(Q)\equiv e^{C | Q  |^{\eta}} |\widehat{f}(Q)|.
\end{align*}
To complete the proof of (a) we integrate over all $Q\in\mathbb{R}$ and apply Lemma \ref{lemma:Ha-algebra} twice. The proof of (b) is very similar. Here, $m=0$, but $\chi_{\rm hi}$ localizes the integrand on the set $|Q|\ge1/\lambda_s(\alpha)$. On this set we then use
 $
1  \leq e^{- C |Q|^\eta} \ e^{C |Q - \xi - \zeta |^{\eta}}  \ e^{C | \xi  |^{\eta}} \ e^{C | \zeta  |^{\eta}} 
\le e^{-C/(\lambda_s(\alpha))^\eta} e^{C |Q - \xi - \zeta |^{\eta}}  \ e^{C | \xi  |^{\eta}} \ e^{C | \zeta  |^{\eta}}
$ to complete the proof.
 $\Box$ 

Finally, we bound the forcing terms in $\chi_{_{\rm hi}}\mathcal{R}_1^{\sigma} [ \alpha, \Gamma + \widehat{E_{_{\rm hi}}}] $ which drive $\widehat{E}_{\rm hi}$.

\begin{proposition}\label{prop:RFbdnonlocal}
Assume $s>1/4$. Let $p = \min(1,s)$, $\eta = \min(1, 2s)$ and let the spectral cutoff parameter, $\lambda_s(\alpha)$, be given as in \eqref{lambda-alpha}. Let the operator $\mathcal{D}^{\sigma,\alpha}$ be defined in \eqref{eqn:Phieqnnonlocal} and set $S(Q)=\chi_{_{\mathcal{B}_\alpha}}(Q)\widetilde{\psi}(Q)$. Then, there exists $C>0$, such that 
$
\left\| \chi_{_{\rm hi}}(Q) \ \left[ 1 + M_{\alpha}^s( Q ) \right]^{-1} \ \mathcal{D}^{\sigma,\alpha}[S](Q)   \ \right\|_{L^{2,a}(\mathbb{R}_Q)} \lesssim  \ e^{- C / \lambda(\alpha)^{\eta} }$. 
\end{proposition}

\noindent {\bf Proof of Proposition \ref{prop:RFbdnonlocal}:} Recall from Proposition \ref{prop:Psinonlocal} that for some $\mu>0$,  $\|e^{\mu|Q|^\eta}\widetilde{\psi}(Q)\|_{L^{2,a}(\mathbb{R}_Q)}\lesssim1$, where $\eta = \min(2s, 1)$. 
By the definition of
 $ \mathcal{D}^{\sigma,\alpha}[S]$ displayed in  \eqref{eqn:Phieqnnonlocal} we have
\begin{align}
&  \left[ 1 + M_{\alpha}^s( Q ) \right]^{-1} \ \mathcal{D}^{\sigma,\alpha}[S] \nonumber \\
 & = S(Q) -  \frac{\chi_{_{\mathcal{B}_{_{\alpha}}}}( Q)}{4 \pi^2} \  \left[ 1 + M_{\alpha}^s( Q ) \right]^{-1}  \  \sum_{m = -1,0,1} \ e^{2 m \pi i \sigma} \  \left(  S * S * S \right) ( Q - 2 m \pi / \alpha) \ .
\end{align}
To conclude the proof, we observe that the $m = \pm 1$ terms  and the $m=0$ term are bounded, respectively,  by parts (a) and (b) of  Lemma \ref{lemma:expconvononlocal}.
%For the $S(Q)$ term, we note that $\chi_{_{\rm hi}}(Q)$ projects onto the set $|Q| \geq \frac{1}{\lambda(\alpha)}$, such that we have for any $C > 0$, 
%\begin{align}
%\chi_{_{\rm hi}}(Q) \ e^{- C |Q|^{\eta} } \leq e^{- C / \lambda(\alpha)^{\eta} }, \label{eqn:hiexpobd}  
%\end{align}
%such that $\| \chi_{_{\rm hi}} S \|_{L^{2,a}(\mathbb{R})} \lesssim e^{- C / \lambda(\alpha)^{\eta}} \ \| e^{ C |Q|^{\eta}} \ S(Q) \|_{L^{2,a}(\mathbb{R}_Q)}$. 
%Following the proof of Lemma \ref{lemma:expconvononlocal} and using Proposition \ref{prop:Mnonlocal}, we similarly have
%\begin{align}
%\left\| \chi_{_{\rm hi}}(Q) \ \left[ 1 + M_{\alpha}^s( Q ) \right]^{-1} ( S * S * S ) (Q)  \ \right\|_{L^{2,a}(\mathbb{R}_Q)} \lesssim \left\| \chi_{_{\rm hi}}(Q) \  ( S * S * S ) (Q)  \ \right\|_{L^{2,a}(\mathbb{R}_Q)} \nonumber \\
%\lesssim e^{- C / \lambda(\alpha)^{\eta}} \ \left\| e^{C |Q|^{\eta}} S(Q) \right\|_{L^{2,a}(\mathbb{R}_Q)}^3. 
%\end{align} $\Box$
%
%\bigskip

\bigskip
We conclude this section with the proof of Lemma \ref{lemma:Mbdnonlocal}.
  Recall the expression for $M^s_{\alpha}(Q)$ in \eqref{eqn:M2defnonlocal} and $\kappa_s(\alpha)$ in \eqref{eqn:kappa-def}. For $s > 1$, $\kappa_s(\alpha) = \alpha^2$ and therefore
$
4 \sin^2 (q/2) \geq \frac{4}{\pi^2}  |q|^2,\ q \in [- \pi, \pi]
% \label{eqn:sinbd}
$
gives
$ M^s_{\alpha}(Q)   \geq \frac{4}{ \alpha^2} J_1 \sin^2( Q \alpha / 2) \geq \frac{4 J_1}{ \pi^2} \ |Q|^2. 
$
Next assume $s = 1$. Then $\kappa_s(\alpha) = (- \log(\alpha)) \ \alpha^2$, $C_s = 1$, and similarly, 
$
M^s_{\alpha}(Q) 
% = \frac{4}{C_s \ \alpha^2 (- \log(\alpha) )} \sum_{m = 1}^{\infty} \frac{ \sin^2 \left( Q  m \alpha / 2 \right) }{m^{1 + 2s}} \nonumber \\
 \geq \frac{4}{\alpha^2 (- \log(\alpha))} \sin^2( Q \alpha / 2) \geq \frac{4}{\pi^2  (- \log(\alpha))} \ |Q|^2. 
$
Finally, suppose that $s < 1$ (thus $p(s)=\min(1,s)=s$). We work with $M^s(q)$ in the original variable $q \in \mathcal{B} = [-\pi, \pi]$, 
%\begin{align}
%M^s(q) =  \frac{1}{C_s} \ \sum_{m = 1}^{\infty} \frac{4 \sin^2 \left( q  m  / 2 \right) }{m^{1 + 2s}}, 
%\end{align}
%which we work with out of convenience. 
First consider $|q| \geq 1$, where $|q|^2 \geq |q|^{2s}$. Then
\begin{align}
M^s(q) =  \frac{1}{C_s} \ \sum_{m = 1}^{\infty} \frac{4 \sin^2 \left( q  m  / 2 \right) }{m^{1 + 2s}} \geq \frac{4}{C_s} \sin^2 (q /2 ) \geq \frac{4}{C_s \ \pi^2} \ |q|^2 \geq \frac{4}{C_s \ \pi^2} |q|^{2s}, \qquad \ 1 \leq |q| \leq \pi. \label{eqn:sleq11}
\end{align}
\bigskip

For $0 \leq |q| < 1$, Proposition \ref{prop:symbolconvnonlocal1} (and Remark \ref{remark:euler} for $s = 1/2$) give
\begin{align}
M^s(q) & =   |q|^{2s} + \frac{2}{C_s} \ \sum_{j = 1}^{\infty} \frac{ \zeta(1 + 2s - 2j) }{ (2j)! } \ (-1)^{j + 1} \ |q|^{2j} \nonumber \\
& =  |q|^{2s}  \left( 1 + \frac{2}{C_s} \ \sum_{j = 1}^{\infty} \frac{ \zeta(1 + 2s - 2j) }{ (2j)! } \ (-1)^{j + 1} \ |q|^{2j -2s} \right) .  \label{eqn:Mseriesfactor}
\end{align}
If we let
$
|q| \leq q_* \equiv \min \left\{ 1,  \left( \frac{4}{C_s} \ \sum_{j = 1}^{\infty} \frac{| \zeta(1 + 2s - 2j) | }{ (2j)! }  \right)^{- \frac{ 1}{2 - 2s}} \right\},
$
where $q_* > 0$, 
\begin{align}
\bigg| \frac{2}{C_s} \ \sum_{j = 1}^{\infty} \frac{ \zeta(1 + 2s - 2j) }{ (2j)! } \ (-1)^{j + 1} \ |q|^{2j -2s} \bigg| & \leq \frac{2}{C_s} \ \sum_{j = 1}^{\infty} \frac{ | \zeta(1 + 2s - 2j) | }{ (2j)! }  \ |q|^{2j -2s} \nonumber \\
& \leq 
\frac{2}{C_s} \ \sum_{j = 1}^{\infty} \frac{ | \zeta(1 + 2s - 2j) | }{ (2j)! }  \ |q|^{2 -2s} \leq \frac{1}{2}, 
\end{align}
such that by \eqref{eqn:Mseriesfactor}, 
\begin{align}
M^s(q) =  |q|^{2s}  \left( 1 + \frac{2}{C_s} \ \sum_{j = 1}^{\infty} \frac{ \zeta(1 + 2s - 2j) }{ (2j)! } \ (-1)^{j + 1} \ |q|^{2j -2s} \right)  \geq \frac{ |q|^{2s} }{2}, \qquad |q| \leq q_*.  \label{eqn:sleq12}
\end{align}
It remains to address $q_* \leq |q| < 1$ (if $q_* = 1$, we are done for $s < 1$). On this interval, we need to find $C_* > 0 $ such that 
\begin{align}
M^s(q) = \frac{1}{C_s} \ \sum_{m = 1}^{\infty} \frac{4 \sin^2 \left( q  m  / 2 \right) }{m^{1 + 2s}} \geq \frac{4}{C_s}  \sin^2 (q /2 ) \geq \frac{4}{C_s \ \pi^2} \ |q|^2 \geq C_* \ |q|^{2s}, \qquad q_* \leq |q| < 1. \label{eqn:sleq13}
\end{align}
where we have again used $\sin^2(q /2) \geq \frac{1}{\pi^2} |q|^2$. Take
$
C_* \equiv \underset{ q_* \leq |q| < 1 }{ \min} \ \left( \frac{4 \  |q|^{2 - 2s} }{C_s \ \pi^2} \right) = \frac{4 \  |q_* |^{2 - 2s} }{C_s \ \pi^2}  > 0,
$ and  let 
$
C \equiv \min \left\{ \frac{4}{C_s \ \pi^2}, \frac{1}{2}, C_* \right\}.  
$
Therefore, from \eqref{eqn:sleq11}, \eqref{eqn:sleq12}, and \eqref{eqn:sleq13}, and with the rescaling $Q = q/\alpha$, there exists some $C>0$ such that we have for $s < 1$, 
\begin{align}
& M^s(q) =  \frac{1}{C_s} \ \sum_{m = 1}^{\infty} \frac{4 \sin^2 \left( q  m  / 2 \right) }{m^{1 + 2s}} \geq C \ |q|^{2s}, \qquad \ q \in \mathcal{B} = [-\pi , \pi]  \nonumber \\
\Longrightarrow \qquad & M_{\alpha}^s(Q) = \frac{1}{C_s \ \alpha^{2s} } \ \sum_{m = 1}^{\infty} \frac{4 \sin^2 \left( Q  m  \alpha / 2 \right) }{m^{1 + 2s}} \geq C \ |Q|^{2s}, \qquad \ Q \in \mathcal{B}_{_{\alpha}} = \left[ - \frac{\pi}{\alpha}, \frac{\pi}{\alpha} \right].
 \end{align}
This completes the proof of Lemma \ref{lemma:Mbdnonlocal}. $\Box$

\section{Solution of the low frequency equation for $\widehat{E_{_{\rm lo}}}$} \label{section:rescalinglow0nonlocal}

\subsection{Equation for $\widehat{E_{_{\rm lo}}}$ as a perturbation of the continuum FNLS limit }
\label{section:rescalinglownonlocal}
Insertion of the map $\widehat{E_{_{\rm hi}}}[\alpha, \widehat{E_{_{\rm lo}}} ]$ (Proposition \ref{prop:ifthighnonlocal})  into equation \eqref{eqn:lownonlocal} yields a  closed equation for $\widehat{E_{_{\rm lo}}}$:
\begin{align}
&\left[ 1 + M_{\alpha}^s(Q) \right] \  \widehat{E_{_{\rm lo}}} ( Q ) - \chi_{_{\rm lo}}( Q )   \frac{3}{4 \pi^2}   \widetilde{\psi} * \widetilde{\psi} * \widehat{E_{_{\rm lo}}}  (Q)  & \nonumber \\ 
&  \quad  = \chi_{_{\rm lo}}( Q ) \ \mathcal{R}_1^{\sigma} \left[ \alpha, \widehat{E_{_{\rm lo}}} + \widehat{E_{_{\rm hi}}} [ \alpha, \widehat{E_{_{\rm lo}}} ] \right] ( Q) +   \chi_{_{\rm lo}}( Q )   \frac{3}{4 \pi^2}   \widetilde{\psi} * \widetilde{\psi} * \left( \widehat{E_{_{\rm hi}}}  [ \alpha, \widehat{E_{_{\rm lo}}} ] \right) (Q) .  \label{eqn:closednonlocal}
\end{align}
Here, $\mathcal{R}_1^{\sigma}[\alpha,\widehat{E}]$ is displayed in \eqref{eqn:HJdefnonlocal}.

Now note that the operator on the left-hand-side  of \eqref{eqn:closednonlocal} has a formal $\alpha\downarrow0$ limit equal to the linearized continuum FNLS operator, $\widetilde{L_+}$. Hence,   we now reexpress \eqref{eqn:closednonlocal} as a small $\alpha$ perturbation of this limit:

\begin{proposition}
\label{prop:rescalenonlocal}
Let $p = p(s)= \min(s,1)$. There exists $0\le\alpha_1\le\alpha_0$ and $0 < r < 1$ such that:
\begin{enumerate}
\item Equation  \eqref{eqn:closednonlocal}  for $\widehat{E_{_{\rm lo}}}$ may be rewritten as
\begin{align}
\widetilde{L_+} \widehat{E_{_{\rm lo}}} (Q) &= \mathcal{R}_2^{\sigma}[\alpha, \widehat{E_{_{\rm lo}}}](Q), \ \ {\rm where}\label{eqn:lowrescalednonlocal}\\
&\nonumber\\
\mathcal{R}_2^{\sigma} \left[ \alpha, \widehat{E_{_{\rm lo}}} \right](Q) & \equiv \chi_{_{\rm lo}}( Q ) \ \mathcal{R}_1^{\sigma} \left[ \alpha, \widehat{E_{_{\rm lo}}} + \widehat{E_{_{\rm hi}}} [ \alpha, \widehat{E_{_{\rm lo}}} ] \right] ( Q)  + R_{_{\rm pert}} \left[ \alpha, \widehat{E_{_{\rm lo}}} \right] (Q) , \ \ {\rm and}\label{eqn:Htildedefnonlocal}\\
R_{_{\rm pert}} \left[ \alpha, \widehat{E_{_{\rm lo}}} \right](Q) & \equiv 
%  \left[ \  | Q |^{2p} - M^s_{\alpha}(Q)  \ \right] \ \widehat{E_{_{\rm lo}}}(Q) -  \chi_{_{\rm hi}}(Q) \  %\frac{3}{4 \pi^2}  \   \widetilde{\psi} * \widetilde{\psi} * \widehat{E_{_{\rm lo}}}  (Q) \nonumber \\
  \chi_{_{\rm lo}} (Q) \ \left[ \  | Q |^{2p} - M^s_{\alpha}(Q)  \ \right] \ \widehat{E_{_{\rm lo}}}(Q) -  \chi_{_{\rm hi}}(Q) \  \frac{3}{4 \pi^2}  \   \widetilde{\psi} * \widetilde{\psi} * \widehat{E_{_{\rm lo}}}  (Q). \label{eqn:Rpertdefnonlocal}
\end{align}
 
\item 
$\mathcal{R}_2^{\sigma} \left[\alpha, \widehat{E_{_{\rm lo}}} \right] (Q) $ is continuous at $(0,0) \in [0,\alpha_1) \times L^{2,a}(\mathbb{R})$. Furthermore, the mapping $\widehat{E_{_{\rm lo}}} \mapsto \mathcal{R}_2^{\sigma} \left[ \alpha, \widehat{E_{_{\rm lo}}} \right]$ is Fr\'{e}chet differentiable, where  $D_{\widehat{E_{_{\rm lo}}}} \mathcal{R}_2^{\sigma}[ \alpha, \widehat{E_{_{\rm lo}}} ]$ is displayed in \eqref{eqn:DHtildenonlocal} below. Finally, we have the bounds
 \begin{align}
&  \left\| \mathcal{R}_2^{\sigma}[\alpha, \widehat{E_{_{\rm lo}}}] \right\|_{L^{2,a-2p }(\mathbb{R})} \lesssim \ \left( \frac{\alpha}{\kappa_s(\alpha)} \right)^{1/2} \ \mathfrak{e}_s(\alpha) +\ \left\| \widehat{E_{_{\rm lo}}} \right\|_{L^{2,a}(\mathbb{R})}\ +\
  \left\| \widehat{E_{_{\rm lo}}} \right\|^2_{L^{2,a}(\mathbb{R})}\ +\   \left\| \widehat{E_{_{\rm lo}}} \right\|^3_{L^{2,a}(\mathbb{R})} , \label{eqn:Hestimatenonlocal} \\
 & \left\| D_{\widehat{E_{_{\rm lo}}}} \mathcal{R}_2^{\sigma} \left[  \alpha, \widehat{E_{_{\rm lo}}} \right] \right\|_{_{L^{2,a}(\mathbb{R}) \to L^{2,a-2p}(\mathbb{R})}} \lesssim  \ \mathfrak{E}_s(\alpha) \  + \ \left\| \widehat{E_{_{\rm lo}}} \right\|_{L^{2,a}(\mathbb{R})}\ +\  \left\| \widehat{E_{_{\rm lo}}} \right\|^2_{L^{2,a}(\mathbb{R})}, \label{eqn:Hdiffestimatenonlocal}
 \end{align}
 where $\mathfrak{e}_s(\alpha)$ and $\mathfrak{E}_s(\alpha)$ are defined in \eqref{frak-e}.
% \footnote{?? banish all $\frak{e}$'s  to the notation section???}
%%  \begin{align}
%%\mathfrak{e}_s(\alpha) = \left\{ \begin{array}{ll} \alpha^{2 (1 - s)} & : s < 1 \\
%%\frac{1}{(- \log(\alpha)} & : s = 1 \\
%%\alpha^{2 ( s - 1)} & : 1 < s < 2 \\
%%(- \log(\alpha)) \ \alpha^2  & : s = 2 \\
%%\alpha^2 & : s > 2
%%\end{array} \right. .  \label{eqn:gamma11} 
%%\end{align}  
%%and
% \begin{align}
%\mathfrak{E}_s(\alpha) = \left\{ \begin{array}{ll} \alpha^{2r (1 - s)} + \alpha^{2s (1 - r)} & : s < 1 \\
%\frac{\log(- \log(\alpha))}{(- \log(\alpha))} & : s = 1 \\
%\alpha^{2r (s - 1)} + \alpha^{2 (1 - r)}   & : 1 < s < 2 \\
%(- \log(\alpha)) \alpha^{2r} + \alpha^{2 (1 - r)}   & : s = 2 \\
%\alpha^{2r}  + \alpha^{2 (1 - r)}  & : s > 2
%\end{array} \right. .  \label{frak-e} 
%\end{align}  
\end{enumerate}
 
  \end{proposition}
 
 \medskip
 
 Proposition \ref{prop:rescalenonlocal} is proven in Sections \ref{section:derivationphilononlocal} through \ref{section:Hanalysisnonlocal}. In Section \ref{prop:rescalenonlocal}, we derive equation  \eqref{eqn:lowrescalednonlocal}. In Section \ref{subsection:toolslownonlocal}, we provide tools used in Section \ref{section:Hanalysisnonlocal} to prove estimates \eqref{eqn:Hestimatenonlocal} and \eqref{eqn:Hdiffestimatenonlocal}. 
 
 \medskip

We then proceed to solve \eqref{eqn:lowrescalednonlocal} in Proposition \ref{prop:solutiontolownonlocal} of Section \ref{section:solutionlownonlocal} using the strategy detailed in Section \ref{subsection:strategy}. 
 
 \subsubsection{Derivation of equation \eqref{eqn:lowrescalednonlocal} for $\widehat{E_{_{\rm lo}}}$} \label{section:derivationphilononlocal}
 
We follow the derivation of the low-frequency equation found in \cite{Jenkinson_Weinstein_2015}, and rewrite equation \eqref{eqn:closednonlocal} in the form \eqref{eqn:lowrescalednonlocal}.
First, we use that $
 \ \chi_{_{\rm hi}} = 1 - \chi_{_{\rm lo}}  \ $
to get
\begin{align}
\chi_{_{\rm lo}}( Q ) \ \widetilde{\psi} * \widetilde{\psi} * \widehat{E_{_{\rm lo}}}  (Q)  =   \widetilde{\psi} * \widetilde{\psi} * \widehat{E_{_{\rm lo}}}  (Q) - \chi_{_{\rm hi}}(Q) \   \widetilde{\psi} * \widetilde{\psi} * \widehat{E_{_{\rm lo}}}  (Q). \label{eqn:loderive1nonlocal}
\end{align}
Since $\chi_{_{\rm hi}}  \widehat{E_{_{\rm lo}}} = 0$,  we may write:
\begin{align}
M^s_{\alpha}(Q) \ \widehat{E_{_{\rm lo}}} (Q)  
%=  |Q|^2 \  \widehat{E_{_{\rm lo}}} (Q) + \left[ M_{\alpha}(Q) - | Q |^2 \right] \ \widehat{E_{_{\rm %lo}}}(Q) \nonumber \\ 
& = |Q|^{2p} \  \widehat{E_{_{\rm lo}}} (Q) + \chi_{_{\rm lo}} \left[ M^s_{\alpha}(Q) - | Q |^{2p} \right] \ \widehat{E_{_{\rm lo}}}(Q) .  \label{eqn:loderive2nonlocal}
\end{align}
 We now express the left-hand side of \eqref{eqn:closednonlocal} via \eqref{eqn:loderive1nonlocal} and \eqref{eqn:loderive2nonlocal} in the form:
$
%\left[ 1 + M^s_{\alpha}(Q) \right] \  \widehat{E_{_{\rm lo}}} ( Q ) - \chi_{_{\rm lo}}( Q )   \frac{3}{4 %\pi^2}   \widetilde{\psi} * \widetilde{\psi} * \widehat{E_{_{\rm lo}}}  (Q)\ 
% =\ 
 \widetilde{L_+} \widehat{E_{_{\rm lo}}}(Q) - R_{_{\rm pert}} \left[ \alpha, \widehat{E_{_{\rm lo}}} \right] (Q),  \label{eqn:LHSrescalednonlocal}
$
where the linear operator $\widehat{E_{_{\rm lo}}}\mapsto R_{_{\rm pert}} \left[ \alpha, \widehat{E_{_{\rm lo}}} \right]$ is given in \eqref{eqn:Rpertdefnonlocal}.
%Here,  we have defined the (linear in $\widehat{E_{_{\rm lo}}}$) operator
%\begin{align}
%R_{_{\rm pert}} \left[ \alpha, \widehat{E_{_{\rm lo}}} \right](Q) & \equiv   \left[ \  | Q |^{2p} - M^s_{\alpha}(Q)  \ \right] \ \widehat{E_{_{\rm lo}}}(Q) -  \chi_{_{\rm hi}}(Q) \  \frac{3}{4 \pi^2}  \   \widetilde{\psi} * \widetilde{\psi} * \widehat{E_{_{\rm lo}}}  (Q) \nonumber \\
%& =   \chi_{_{\rm lo}} (Q) \ \left[ \  | Q |^{2p} - M^s_{\alpha}(Q)  \ \right] \ \widehat{E_{_{\rm lo}}}(Q) -  \chi_{_{\rm hi}}(Q) \  \frac{3}{4 \pi^2}  \   \widetilde{\psi} * \widetilde{\psi} * \widehat{E_{_{\rm lo}}}  (Q). \label{eqn:Rpertdefnonlocal}
%\end{align}
Equation \eqref{eqn:closednonlocal} may be written in the  desired form \eqref{eqn:lowrescalednonlocal}, where $\mathcal{R}_2^{\sigma} \left[ \alpha, \widehat{E_{_{\rm lo}}} \right]$ is given by \eqref{eqn:Htildedefnonlocal}. Finally, the expression for $D_{\widehat{E_{_{\rm lo}}}} \mathcal{R}_2^{\sigma}[ \alpha, \widehat{E_{_{\rm lo}}} ] $ is displayed in \eqref{eqn:DHtildenonlocal} and is bounded, together with  $\mathcal{R}_2^{\sigma}[ \alpha, \widehat{E_{_{\rm lo}}} ]$ in Section \ref{section:Hanalysisnonlocal}.

\subsubsection{Strategy for solving  the low frequency equation} \label{subsection:strategy}
 We study the solvability of the low frequency equation \eqref{eqn:lowrescalednonlocal}: $\mathcal{M}[\alpha, \widehat{E_{_{\rm lo}}}]\equiv \widetilde{L_+} \widehat{E_{_{\rm lo}}} -  \mathcal{R}_2^{\sigma}[\alpha, \widehat{E_{_{\rm lo}}}]=0$ by applying a following variant of the implicit function theorem
  (a consequence of Theorem \ref{th:IFT} of  Appendix \ref{appendix:IFT}) \cite{Jenkinson_Weinstein_2015}.

\begin{lemma} \label{lemma:generallinearopnonlocal}
 Consider the equation 
 \begin{align}
\mathcal{M}[\alpha, f]\equiv \mathfrak{L} f - \mathcal{R}[\alpha, f]=0,\ \ f\in L^{2,a}_{_{\rm even}}(\mathbb{R}). 
\label{tosolve}
\end{align}
\begin{enumerate}
\item  $\mathfrak{L}: L^{2,a}_{_{\rm even}}(\mathbb{R}) \longmapsto L^{2,a-2p}_{_{\rm even}}(\mathbb{R})$ be an isomorphism for $0 < p \leq 1$
\item 
 $\mathcal{R}: [0, \alpha_1)_{\alpha} \times L_{_{\rm even}}^{2,a}(\mathbb{R}) \longrightarrow L^{2,a-2p}_{_{\rm even}}(\mathbb{R})$  is continuous at $(0,0)$,  Fr\'{e}chet differentiable with respect to $f \in L_{_{\rm even}}^{2,a}(\mathbb{R}) $
 \item  $\mathcal{R}[0,0] = 0$ and  satisfies the bounds:
\begin{align}
& \left\|  \mathcal{R}[\alpha, f]  \right\|_{L^{2,a-2p}(\mathbb{R})} \lesssim   \mathfrak{K}(\alpha) + \| f \|_{L^{2,a}(\mathbb{R})} + \| f \|^2_{L^{2,a}(\mathbb{R})} + \| f \|^3_{L^{2,a}(\mathbb{R})}  , \label{eqn:HIFTestnonlocal} \\
& \left\| D_f \mathcal{R}[\alpha, f]  \right\|_{L^{2,a}(\mathbb{R}) \to L^{2,a-2p}(\mathbb{R})} \lesssim   \mathfrak{K}(\alpha)  + \|f \|_{L^{2,a}(\mathbb{R})} + \| f \|^2_{L^{2,a}(\mathbb{R})}  \label{eqn:HIFTestdiffnonlocal},
\end{align}
for some continuous function $\mathfrak{K}(\alpha) \geq 0$, satisfying $\mathfrak{K}(0) = 0$, $\mathfrak{K}(\alpha) \to 0$ as $\alpha \to 0$. 
\end{enumerate}

 Then there exists a constant $\alpha_2 \le \alpha_1$ such that for all $0<\alpha < \alpha_2$, the equation \eqref{tosolve}, $\mathcal{M}[\alpha, f]=0$, has a unique solution $f = f[\alpha] \in L^{2,a}_{_{\rm even}}(\mathbb{R})$ satisfying 
\begin{align}
\| f[\alpha] \|_{L^{2,a}(\mathbb{R})} \lesssim \mathfrak{K}(\alpha). \label{eqn:sigmaestnonlocal}
\end{align}
\end{lemma}

In order to construct 
$\widehat{E_{_{\rm lo}}}$ by application of Lemma \ref{lemma:generallinearopnonlocal}, we must  verify the hypotheses for 
 $\mathcal{M}[\alpha, f]\equiv \widetilde{L_+} f -  \mathcal{R}_2^{\sigma}[\alpha, f]$
  with $\mathfrak{L}= \widetilde{L_+} $ and $\mathcal{R}=\mathcal{R}_2^{\sigma}[\alpha, f]$. 
Note,  by  Proposition \ref{prop:Lplusnonlocal}, that the mapping  $\widetilde{L_+}: L^{2,a}_{_{\rm even}}(\mathbb{R})\to L^{2,a-2p}_{_{\rm even}}(\mathbb{R})$ is an isomorphism. Hence it remains to verify that $(\alpha,f)\mapsto\mathcal{R}_2^{\sigma}[\alpha, f]$ satisfies the necessary hypotheses. 
 This is done in the remainder of Section \ref{section:rescalinglow0nonlocal}, culminating in Proposition \ref{prop:solutiontolownonlocal}.

\subsubsection{Tools for estimation of  the low frequency equation}
\label{subsection:toolslownonlocal} 
 
To establish the properties of $\mathcal{R}_{2}^{\sigma}$
  given in Proposition \ref{prop:rescalenonlocal}, we require a number of general tools. Recall that $\chi_{_{\rm hi}}= 1- \chi_{_{\rm lo}}$,
where $ \chi_{_{\rm lo}}\equiv \chi_{_{ \left[ - \lambda(\alpha)^{-1},\lambda(\alpha)^{-1} \right]}}$, where $\lambda_s(\alpha)$ is defined in \eqref{lambda-alpha}.
%\begin{align}
%\lambda(\alpha) = \left\{ \begin{array}{ll} \alpha^{1 - r} & : s \neq 1 \\
% \frac{1}{(- \log(\alpha))} & : s = 1 \end{array} \right. , 
%\end{align}
% and $0 < r < 1$. 
%Also recall 
%$
%\left\| \widetilde{f_1} * \widetilde{f_2} \right\|_{L^{2,a}(\mathbb{R})} \lesssim \left\| \widetilde{f_1} \right\|_{L^{2,a}(\mathbb{R})} \ \left\| \widetilde{f_2} \right\|_{L^{2,a}(\mathbb{R})}. $
  Note first that since  $(1 + |Q|^2)^{-2p} \leq |Q|^{-2p}$, we have
\begin{lemma} \label{lemma:minus2nonlocal}
Let $0 < p \leq 1$. For any function $\widehat{g} \in L^{2,a-2p}(\mathbb{R})$ such that $ | Q |^{ - 2p}\ \widehat{g} \in L^{2,a}(\mathbb{R}_Q)$  , we have
\begin{align}
\left\| \widehat{g}   \right\|_{L^{2,a-2p }} \ \lesssim \ \left\| | Q |^{ - 2p} \widehat{g}(Q) \ \right\|_{L^{2,a}(\mathbb{R}_Q)}.
\end{align}
\end{lemma}

The bound $\chi_{_{\rm hi}}(Q) |Q|^{-2p} \leq \lambda(\alpha)^{2p}$ and Lemma \ref{lemma:minus2nonlocal} imply: 
\begin{lemma}
\label{lemma:ltwoasmallnonlocal}
Let $0 < p \leq 1$. For any function $\widehat{g} \in L^{2,a}(\mathbb{R})$, we have
\begin{align}
\left\| \ \chi_{_{\rm hi}}(Q) \   \widehat{g}(Q) \  \right\|_{L^{2,a-2p}(\mathbb{R}_Q)} \lesssim \ \lambda(\alpha)^{2p} \ \| \widehat{g} \|_{L^{2,a}(\mathbb{R})}.
\end{align}
\end{lemma}

\begin{lemma} \label{lemma:symbolconvlononlocal}
Let $0  < r < 1$ and $ p = \min(1,s)$. For any function $\widehat{g} \in L^{2,a}(\mathbb{R})$, we have
\begin{align}
\bigg\| \ \chi_{_{\rm lo}}(Q) \  \left[ \ | Q |^{2p} - M_{\alpha}^s( Q ) \ \right] \ \widehat{g}(Q) \ \bigg\|_{L^{2,a-2p }(\mathbb{R}_Q)} \ \lesssim \ \mathfrak{d}_s(\alpha) \ \left\| \ \widehat{g} \ \right\|_{L^{2,a}(\mathbb{R})}, \label{eqn:e3est}
\end{align}
where $\mathfrak{d}_s(\alpha) \to 0$ as $\alpha \to 0$ is displayed in \eqref{eqn:gamma3}. 
\end{lemma}

\noindent {\bf Proof of Lemma \ref{lemma:symbolconvlononlocal}:} Recall $p=p(s)=\min(s,1)$. Proposition \ref{prop:symbolconvnonlocal1} implies there exists  $f_s(Q;\alpha)$ such that $\sup_{Q\in\mathcal{B}_\alpha}| f_s(Q; \alpha) | \lesssim 1$ and  $M^s_{\alpha}(Q) - |Q|^{2p(s)}$
 is equal to $f_s(Q;\alpha)$ times a function of $\alpha$ and $Q$, displayed in 
 \eqref{eqn:symbolexpoest1}.
% \begin{align}
% M^s_{\alpha}(Q) - |Q|^{2p(s)} = \left\{ \begin{array}{ll} \alpha^{2 - 2s} \ f_s( Q; \alpha) \ |Q|^2 & : \quad s < 1 \\ 
% \hline 
% \frac{1}{- \log(\alpha)} \ f_s(Q ; \alpha) \ \left( \frac{3}{2} - \log(|Q|) \right) \ |Q|^2 & : \quad s = 1 \\
% \hline 
% \alpha^{2s - 2} \ f_s(Q; \alpha) \ |Q|^{2s} & : \quad 1 < s < 2 \\
% \hline 
% (- \log(\alpha)) \ \alpha^2 \ f_s(Q; \alpha) \ |Q|^4 & : \quad s = 2 \\
% \hline 
% \alpha^2 \ f_s(Q; \alpha) \ |Q|^4 & : \quad 2 < s \leq \infty \end{array} \right. . \label{eqn:symbolexpoest3a}
% \end{align}
Recall that $p=p(s)=\min(s,1)$ and $\lambda_s(\alpha) = \left\{ \begin{array}{ll} \alpha^{1 - r} & : s \neq 1 \\
 \frac{1}{(- \log(\alpha))} & : s = 1 \end{array} \right. $. Furthermore, recall that $\chi_{_{\rm lo}}(Q) = \chi \left( |Q| \leq \lambda(\alpha)^{-1} \right)$, and that $\lambda(\alpha)^{-1} \leq \pi/\alpha$ for $\alpha$ sufficiently small, implying  $\sup_Q\chi_{_{\rm lo}}(Q) \ | f_s(Q; \alpha) | \lesssim 1$.

 First, suppose that $s \neq 1$. We apply Lemma \ref{lemma:minus2nonlocal} to get
\begin{align}
 \bigg\| \ \chi_{_{\rm lo}}(Q) \  \left[ \ |Q|^{2p} - M_{\alpha}^s(Q) \ \right] \ \widehat{g}(Q) \ \bigg\|_{L^{2,a-{2p} }(\mathbb{R}_Q)}\ \lesssim \ \bigg\| \ \chi_{_{\rm lo}}(Q) \ |Q|^{-2p} \   \left[ \ |Q|^{2p} - M^s_{\alpha}(Q) \ \right] \ \widehat{g}(Q) \ \bigg\|_{L^{2,a  }(\mathbb{R}_Q)} . \label{eqn:minussymbolnonlocal}
\end{align}
Hence, by  \eqref{eqn:symbolexpoest1} to prove the desired bound it suffices to  bound the supremum of 
\begin{align}
\chi_{\rm lo}(Q)\ |Q|^{-2p} \   \left[ \ |Q|^{2p} - M^s_{\alpha}(Q) \ \right] \  = \chi_{\rm lo}(Q)\ \left\{ \begin{array}{ll}  \ f_s( Q; \alpha) \ |\alpha Q|^{2(1 - s)} & : \quad s < 1 \\ 
 \hline 
 \ f_s(Q; \alpha) \ |\alpha Q|^{2(s - 1)} & : \quad 1 < s < 2 \\
 \hline 
 (- \log(\alpha)) \ \ f_s(Q; \alpha) \ |\alpha Q|^2 & : \quad s = 2 \\
 \hline 
 \ f_s(Q; \alpha) \ |\alpha Q|^2 & : \quad 2 < s \leq \infty \end{array} \right. .
\end{align}
For $s\ne1$,  $\chi_{\rm lo}(Q)$ projects onto the set where  $|\alpha Q| \leq \alpha\lambda_s(\alpha)^{-1} = \alpha^{r}$ and \eqref{eqn:e3est} follows in this case.

Finally, consider the case where $s = 1$. In this case, $p=\min(s,1)=1$.  Here we take  $\lambda_s(\alpha)^{-1}= - \log(\alpha) >1$. To manage the logarithmic term in \eqref{eqn:minussymbolnonlocal}, note that 
\begin{align}
& | \log(|Q|) | \ |Q|^2   \leq \frac{1}{2e} \qquad {\rm when} \qquad |Q| \leq 1, \nonumber \\
{\rm and} \qquad &  | \log(|Q|) |  \leq \log(- \log(\alpha)) \qquad {\rm when} \qquad 1 < |Q| \leq \lambda(\alpha)^{-1} = - \log(\alpha).   \label{eqn:logbd}
\end{align}
We therefore address the intervals $ |Q| \leq 1$ and $|Q| \geq 1$ separately via projections. We use Lemma \ref{lemma:minus2nonlocal} and the fact that $(1 + |Q|^2)^{-2} \leq 1$ to get
{\footnotesize{
\begin{align}
& \bigg\| \ \chi_{_{\rm lo}}(Q) \  \left[ \ |Q|^{2} - M_{\alpha}^s(Q) \ \right] \ \widehat{g}(Q)  \ \bigg\|_{L^{2,a-{2} }(\mathbb{R}_Q)} \nonumber\\
 &  \leq     \bigg\| \chi ( |Q| \leq 1 ) \ \chi_{_{\rm lo}}(Q)  \  \left[ \ |Q|^{2} - M_{\alpha}^s(Q) \ \right] \ \widehat{g}(Q)  \ \bigg\|_{L^{2,a-{2} }(\mathbb{R}_Q)} 
 \quad  + \bigg\| \chi ( |Q| >  1 )  \  \chi_{_{\rm lo}}(Q)  \  \left[ \ |Q|^{2} - M_{\alpha}^s(Q) \ \right] \ \widehat{g}(Q)  \ \bigg\|_{L^{2,a-{2} }(\mathbb{R}_Q)} \nonumber \\
& \leq      \bigg\| \chi ( |Q| \leq 1 )  \  \left[ \ |Q|^{2} - M_{\alpha}^s(Q) \ \right] \ \widehat{g}(Q)  \ \bigg\|_{L^{2,a}(\mathbb{R}_Q)} 
\ + \bigg\| \chi ( |Q| > 1 ) \  \chi_{_{\rm lo}}(Q)  \ |Q|^{-2} \ \left[ \ |Q|^{2} - M_{\alpha}^s(Q) \ \right] \ \widehat{g}(Q)  \ \bigg\|_{L^{2,a }(\mathbb{R}_Q)} \label{eqn:splitlog}
\end{align}
}}
The term localized on $|Q|\le1$ is bounded, using the first estimate in \eqref{eqn:logbd} and \eqref{eqn:minussymbolnonlocal},  by
$  \chi ( |Q| \leq 1 )  \  \left| \ |Q|^{2} - M_{\alpha}^s(Q) \ \right| \lesssim \frac{1}{- \log(\alpha)}$.
The term localized to $|Q| \ge 1$ is  bounded using the second estimate in 
\eqref{eqn:logbd} and \eqref{eqn:minussymbolnonlocal}, by 
$ \chi ( |Q| \ge 1 )  \chi_{_{\rm lo}}(Q)  \ |Q|^{-2} \ \left| \ |Q|^{2} - M_{\alpha}^s(Q) \ \right| \lesssim \frac{\log(- \log(\alpha))}{(- \log(\alpha))}$. 
Lemma \ref{lemma:symbolconvlononlocal} is proved. $\Box$ \\

\begin{lemma}
\label{lemma:expconvo2nonlocal}
Suppose that $\widehat{f_1} \in L^{2,a}(\mathbb{R})$ and $e^{C |Q|^\eta} \widehat{f_1}(Q) \in L^{2,a}(\mathbb{R}_Q)$ for some $\eta, C > 0$. Further, let $\widehat{f_2}, \widehat{f_3} \in L^{2,a}(\mathbb{R})$. Then, with $  \overline{\chi}_{_{\mathcal{B}_{_{\alpha}}}} (Q) = \chi(  |Q| > \pi/\alpha )$, we have
\begin{align}
& \left\| \ \overline{\chi}_{_{\mathcal{B}_{_{\alpha}}}} \ \widehat{f_1} \ \right\|_{L^{2,a}(\mathbb{R})} \lesssim e^{- C \pi^\eta / \alpha^\eta} \ \left\| \ e^{C | Q |^\eta } \  \widehat{f_1}(Q) \right\|_{L^{2,a}(\mathbb{R}_Q)}, \label{eqn:firstchibar} \\
& \bigg\| \left( \overline{\chi}_{_{\mathcal{B}_{_{\alpha}}}} \widehat{f_1} \right) * \widehat{f_2} * \widehat{f_3} \bigg\|_{L^{2,a-2p }(\mathbb{R})} \lesssim \  e^{- C \pi^\eta / \alpha^\eta} \ \left\| \ e^{C |Q |^\eta} \ \widehat{f_1}(Q) \right\|_{L^{2,a}(\mathbb{R}_Q)} \left\| \widehat{f_2} \right\|_{L^{2,a}(\mathbb{R})} \left\| \widehat{f_3} \right\|_{L^{2,a}(\mathbb{R})}.
\end{align}
\end{lemma}
The bound \eqref{eqn:firstchibar} follows from
$
\overline{\chi}_{_{\mathcal{B}_{_{\alpha}}}} (Q) \ e^{- C |Q|^{\eta} }  e^{C |Q|^{\eta} }\leq e^{- C \pi^{\eta} / \alpha^{\eta} } e^{ C |Q|^{\eta} } . 
$
The second follows from the first and  \eqref{eqn:algebraB}. 

\medskip

\subsection{Part 2 of the proof of Proposition \ref{prop:rescalenonlocal}; Analysis of $\mathcal{R}_{2}^{\sigma}$ and derivation of estimates \eqref{eqn:Hestimatenonlocal} and \eqref{eqn:Hdiffestimatenonlocal}} \label{section:Hanalysisnonlocal}

To bound  $\mathcal{R}_{2}^{\sigma} : \mathbb{R}  \times L^{2,a}(\mathbb{R}) \mapsto L^{2,a-2p}(\mathbb{R})$, we use estimates \eqref{eqn:Ehighestlyapnonlocal} and \eqref{eqn:DEhighestnonlocal} from Proposition \ref{prop:ifthighnonlocal}, which are valid for $0<\alpha < \alpha_0 $.
%
%\begin{align}
%& \left\| \widehat{E_{_{\rm hi}}}[\alpha, \Gamma] \right\|_{L^{2,a}(\mathbb{R})} \lesssim \  \left\{ \begin{array}{ll} \alpha^{2p (1 - r)}   \ \left\| \Gamma \right\|_{L^{2,a}(\mathbb{R})} +  e^{- C / \lambda(\alpha) ^{\eta}} & : s \neq 1 \\
%\frac{1}{(- \log(\alpha)}    \ \left\| \Gamma \right\|_{L^{2,a}(\mathbb{R})} +  e^{- C / \lambda(\alpha) }& : s = 1 \end{array} \right. 
%\label{eqn:reschiboundnonlocal} \\
%&\nonumber\\
%& \left\| \ D_{\Gamma} \widehat{E_{_{\rm hi}}}[\alpha, \Gamma] \ \right\|_{L^{2,a}(\mathbb{R}) \longrightarrow L^{2,a}(\mathbb{R})} \lesssim  \left\{ \begin{array}{ll} \alpha^{2p (1 - r)}  & : s \neq 1 \\
%\frac{1}{(- \log(\alpha)} & : s = 1 \end{array} \right. .  \label{eqn:reschibound2nonlocal}
%\end{align}
%Here, $\eta = \min(2s, 1)$. 
Note from \eqref{eqn:lowrescalednonlocal} that
\begin{align}
\widehat{E_{_{\rm lo}}}(Q) = \chi_{_{\rm lo}}(Q) \widehat{E_{_{\rm lo}}}(Q) . \label{eqn:lowsupport}
\end{align}

\medskip

 For any $\widehat{f} \in L^{2,a}(\mathbb{R})$, a direct computation using \eqref{eqn:Htildedefnonlocal}, \eqref{eqn:lowsupport}, and the linearity of $R^{\sigma}_{_{\rm L}}$ and $R_{_{\rm pert}}$ in their second argument gives
\begin{align}
D_{\widehat{E_{_{\rm lo}}}} \mathcal{R}_{2}^{\sigma}[ \alpha, \widehat{E_{_{\rm lo}}} ] \ \widehat{f}(Q)  
  = & \ \chi_{_{\rm lo}}(Q) \ R_{_{\rm L}}^{\sigma} \left[ \alpha, \ \chi_{_{\rm lo}}  \widehat{f} \right] (Q) \nonumber \\
  & + \chi_{_{\rm lo}}(Q) \ R_{_{\rm L}}^{\sigma} \left[ \alpha, \left( D_{\widehat{E_{_{\rm lo}}}} \widehat{E_{_{\rm hi}}} [ \alpha, \widehat{E_{_{\rm lo}}} ] \right) \widehat{f} \right] (Q) + R_{_{\rm pert}} \left[ \alpha, \widehat{f} \right] (Q) \nonumber \\
& + D_{\widehat{E_{_{\rm lo}}}} \left( \chi_{_{\rm lo}} \ R^{\sigma}_{_{NL}} \left[ \alpha, \widehat{E_{_{\rm lo}}}  + \widehat{E_{_{\rm hi}}}  [ \alpha, \widehat{E_{_{\rm lo}}} ] \right] \right) \widehat{f}(Q)   . \label{eqn:DHtildenonlocal}
\end{align}
Here,
\begin{align}
& D_{\widehat{E_{_{\rm lo}}}} \left( \chi_{_{\rm lo}} \ R^{\sigma}_{_{NL}} \left[ \alpha, \widehat{E_{_{\rm lo}}}  + \widehat{E_{_{\rm hi}}}  [ \alpha, \widehat{E_{_{\rm lo}}} ] \right] \right) \widehat{f}(Q)   \nonumber \\
& =   \chi_{_{\rm lo}}( Q)  \  \left( \frac{1}{2 \pi} \right)^2 \ \sum_{m = -1}^1 \ e^{2 m \pi i \sigma} \ \bigg[ 6  \ S_J^{\alpha} * \widehat{E_J^{\alpha, \sigma}} * \widehat{f} (Q - 2 m \pi / \alpha)  \nonumber \\
&   + 3 \  \widehat{E_{_{\rm lo}}} *  \widehat{E_{_{\rm lo}}} * \widehat{f} (Q - 2 m \pi / \alpha) \nonumber \\
&  +  6  \ S_J^{\alpha} * \widehat{E_{_{\rm lo}}} * \left( D_{\widehat{E_{_{\rm lo}}}} \widehat{E_{_{\rm hi}}} [ \alpha, \widehat{E_{_{\rm lo}}} ]  \cdot  \widehat{f} \right) (Q - 2 m \pi / \alpha)  \nonumber \\
&  + 3 \  \widehat{E_{_{\rm lo}}} *  \widehat{E_{_{\rm lo}}} * \left( D_{\widehat{E_{_{\rm lo}}}} \widehat{E_{_{\rm hi}}} [ \alpha, \widehat{E_{_{\rm lo}}} ]  \cdot  \widehat{f} \right) (Q - 2 m \pi / \alpha) \bigg]. \label{eqn:DNLnonlocal}
\end{align}
We now proceed to bound  $\mathcal{R}_{2}^{\sigma} $, given in  \eqref{eqn:Htildedefnonlocal} and  $D_{\widehat{E_{_{\rm lo}}}} \mathcal{R}_{2}^{\sigma}$, given in \eqref{eqn:DHtildenonlocal}, as maps  from $L^{2,a}$ to $L^{2,a-2p}$. 
%We will often use $\| \widehat{f} \|_{L^{2,a-2p}(\mathbb{R})} \leq \| \widehat{f} \|_{L^{2,a}%(\mathbb{R})}$ without explicitly stating it. 

\begin{proposition} \label{prop:Rpertboundnonlocal}
Let $0 < r < 1$, and $0 < \alpha < \alpha_0$.
 $R_{_{\rm pert}} [\alpha, \widehat{f} ] $, defined in \eqref{eqn:Rpertdefnonlocal},  satisfies 
\begin{align}
& \left\| \ R_{_{\rm pert}} \left[ \alpha, \widehat{f} \right] \ \right\|_{L^{2,a-2p }(\mathbb{R})} \ \lesssim \  \mathfrak{E}_s(\alpha) \ \left\| \ \widehat{f} \ \right\|_{L^{2,a}(\mathbb{R})}. 
\end{align}
Here, $\mathfrak{E}_s(\alpha) \to 0$ as $\alpha \to 0$ is displayed in \eqref{frak-e}.
\end{proposition}

This proposition follows from Lemmata \ref{lemma:ltwoasmallnonlocal} and \ref{lemma:symbolconvlononlocal}, and Proposition \ref{eqn:algebraB}.  Note that $\mathfrak{E}_s(\alpha) = \mathfrak{d}_s(\alpha) + \lambda(\alpha)^{2p}$, where $\mathfrak{d}_s(\alpha)$ is given in \eqref{eqn:gamma3}. 

\begin{proposition} \label{prop:RLlownonlocal} Let $0 < r < 1$ and $\eta = \min(2s, 1)$. Let $R^{\sigma}_{_{\rm L}} $ be defined in \eqref{eqn:RLNLdefnonlocal} and let $\widehat{f} \in L^{2,a}(\mathbb{R})$. Then there exist $\alpha_0, C> 0$ such that for $0 < \alpha < \alpha_0$,
\begin{align}
& \left\| \ \chi_{_{\rm lo}} \ R^{\sigma}_{_{\rm L}} \left[ \alpha, \ \chi_{_{\rm lo}} \widehat{f} \right] \ \right\|_{L^{2,a-2p }(\mathbb{R})} \ \lesssim \  e^{- C/\alpha^{\eta} }  \left\| \ \widehat{f} \ \right\|_{L^{2,a}(\mathbb{R})}, \label{eqn:RLlowprop2}   \\
& \left\| \ \chi_{_{\rm lo}} \ R^{\sigma}_{_{\rm L}} \left[ \alpha, \widehat{f} \right] \right\|_{L^{2,a-2p }(\mathbb{R})} \ \lesssim   \left\| \ \widehat{f} \ \right\|_{L^{2,a}(\mathbb{R})} .  \label{eqn:RLlowprop3}
\end{align} 
where $R_L^{\sigma}$ is given in \eqref{eqn:RLNLdefnonlocal}.
 \end{proposition}

\noindent {\bf Proof of Proposition \ref{prop:RLlownonlocal}:}  We use \eqref{eqn:algebraB} extensively. 
 From \eqref{eqn:RLNLdefnonlocal}, we have for any $\widehat{f} \in L^{2,a}(\mathbb{R})$. 
\begin{align}
\chi_{\rm lo}(Q)R_{_{\rm L}}^{\sigma} \left[ \alpha, \widehat{f} \right] (Q) \equiv  & \  \chi_{\rm lo}(Q)\chi_{_{\mathcal{B}_{_{\alpha}}}}( Q)  \ \frac{3}{4 \pi^2} \ \bigg[ \ \sum_{m = -1,0,1} \ e^{2 m \pi i \sigma} \   S * S * \widehat{f} (Q - 2 m \pi / \alpha) 
 - \ \widetilde{\psi} * \widetilde{\psi}  * \widehat{f}(Q) \bigg], \label{eqn:RLdef2nonlocal}
\end{align}
which is linear in its second argument. 
First, we address the $m = 0$ term. Recall that 
 $S(Q)  = \chi_{_{\mathcal{B}_{_{\alpha}}}}(Q)\ \widetilde{\psi}(Q)$. 
Since $\chi_{_{\mathcal{B}_{_{\alpha}}}}+\overline{\chi}_{_{\mathcal{B}_{_{\alpha}}}}=1$, 
 we have $
 S * S * \widehat{f}  - \ \widetilde{\psi} * \widetilde{\psi}  * \widehat{f} = -2 \left( \overline{\chi}_{_{\mathcal{B}_{_{\alpha}}}} \widetilde{\psi} \right) * \widetilde{\psi} * \widehat{f}  + \left( \overline{\chi}_{_{\mathcal{B}_{_{\alpha}}}} \widetilde{\psi} \right) * \left( \overline{\chi}_{_{\mathcal{B}_{_{\alpha}}}} \widetilde{\psi} \right)   * \widehat{f}
$. 
Therefore, by  Lemma \ref{lemma:expconvo2nonlocal} 
\begin{align}
\left\| \ S * S * \widehat{f}   - \ \widetilde{\psi} * \widetilde{\psi}  * \widehat{f} \right\|_{L^{2,a}(\mathbb{R})} \lesssim   e^{- C/\alpha^{\eta} } \left\| \widehat{f} \right\|_{L^{2,a}(\mathbb{R})}  . \label{eqn:RLbd1nonlocal}
\end{align}
For the $m = \pm 1$ terms, first note that by \eqref{eqn:algebraB}, 
\begin{align}
\left\| S * S * \widehat{f} (Q - 2 m \pi / \alpha) \right\|_{L^{2,a}(\mathbb{R}_Q)} \leq  \left\|   \widehat{f}   \right\|_{L^{2,a}(\mathbb{R})}.\label{eqn:otherone} 
\end{align}
Estimates \eqref{eqn:RLbd1nonlocal} and \eqref{eqn:otherone} together imply \eqref{eqn:RLlowprop3}.

To prove \eqref{eqn:RLlowprop2}  we consider the $m=\pm1$ terms. The integral to be bounded 
has variables $Q,  \xi, \zeta \in \mathbb{R}$ constrained by 
 $|\xi| \leq \frac{\pi}{\alpha}$, $ |\zeta| \leq \frac{\pi}{\alpha}$,  $ |Q| \leq \frac{1}{\lambda(\alpha)}$, and $ | \xi + \zeta - Q  - 2  \pi m/ \alpha| \leq \frac{1}{\lambda(\alpha)}$. 
First assume $ m = 1$.   Then, $
 \xi + \zeta - Q - \frac{2 \pi}{\alpha} \geq - \frac{1}{\lambda(\alpha)}$ and therefore $  \xi + \zeta \geq \frac{2 \pi}{\alpha} - \frac{1}{\lambda(\alpha)}  + Q \geq \frac{2 \pi}{\alpha} \left( 1 - \frac{\alpha}{\lambda(\alpha)} \right). 
$
Similarly when $ m = -1$. Then, 
$ - \xi - \zeta + Q - \frac{2 \pi}{\alpha} \geq - \frac{1}{\lambda(\alpha)} $ and therefore  $ - \xi - \zeta \geq \frac{\pi}{\alpha} - \frac{1}{\lambda(\alpha)}  - Q \geq \frac{2 \pi}{\alpha} \left( 1 - \frac{\alpha}{\lambda(\alpha)} \right)$. 
Note also, by the expression for $\lambda_s(\alpha)$ in \eqref{lambda-alpha}, that $ \alpha/\lambda_s(\alpha)\to0$  as $\alpha \to 0$.   Therefore, in either case $m = \pm 1$, taking $\alpha$ small enough that $\alpha/ \lambda(\alpha) \leq  1/2$ gives (using Lemma \ref{lemma:tri})
$ |\xi|^{\eta} + |\zeta|^{\eta} \geq |\xi + \zeta|^{\eta} \geq \left( \frac{2\pi}{\alpha}\right)^\eta \left( 1 - \frac{\alpha}{\lambda(\alpha)} \right)^{\eta} \geq \frac{\pi^{\eta}}{ \alpha^{\eta}}$.
%& |\xi| \leq \frac{\pi}{\alpha}, \hspace{.5cm} |\zeta| \leq \frac{\pi}{\alpha} , \hspace{.5cm} |Q| \leq \frac{1}{\lambda(\alpha)}, \hspace{.5cm} | \xi + \zeta - Q +  2 m  \pi / \alpha| \leq   \frac{1}{\lambda(\alpha)}  \nonumber
Hence,  under these constraints on $Q, \xi, \zeta$, we have  $1 = e^{C |\xi|^{\eta} } \ e^{C |\zeta|^{\eta}} \ e^{- C (|\xi|^{\eta} + |\zeta|^{\eta})} \leq e^{ - C \pi^{\eta}/  \alpha^{\eta}}.$
Therefore, 
\begin{align}
&  \bigg|  \chi_{_{\rm lo}} (Q) \   S * S  * \left( \chi_{_{\rm lo}} \widehat{f} \right)  ( Q -  2 m \pi / \alpha) \ \bigg|   \nonumber \\
& \leq \ \chi_{_{\rm lo}}(Q)  \int_{\mathbb{R}} \int_{\mathbb{R}} 
  |(\chi_{_{\mathcal{B}_{_{\alpha}}}} S)(\xi)| \ |(\chi_{_{\mathcal{B}_{_{\alpha}}}}S)(\zeta) | \ |  (\chi_{_{\rm lo}} \widehat{f})(Q - \xi - \zeta - 2 m \pi/\alpha) | \ d \xi \ d \zeta \nonumber \\
& \leq e^{- C \pi^{\eta} /   \alpha^{\eta}} \  \int_{\mathbb{R}} \int_{\mathbb{R}} e^{C |\xi|^{\eta}} \ |S(\xi)| \ e^{C |\zeta|^{\eta}} \ |S(\zeta) | \ | \widehat{f}(Q - \xi - \zeta - 2 m \pi/\alpha) | \ d \xi \ d \zeta, 
\end{align}
which using \eqref{eqn:algebraB} gives, for some $C > 0$,
\begin{align}
\left\| \  \chi_{_{\rm lo}}(Q)  \   S * S *  \left( \chi_{_{\rm lo}} \widehat{f} \right) ( Q -  2 m \pi / \alpha) \ \right\|_{L^{2,a -2p }(\mathbb{R}_Q)} \lesssim e^{- C / \alpha^{\eta}} \ \left\| e^{C |Q|^{\eta} } \ S(Q) \  \right\|^2_{L^{2,a}(\mathbb{R}_Q)} \ \left\| \ \widehat{f} \ \right\|_{L^{2,a}(\mathbb{R})}
% \nonumber \\
%\lesssim e^{- C / \alpha^{\eta}} \ \left\| \ \widehat{f} \ \right\|_{L^{2,a}(\mathbb{R})} .
\label{eqn:RLbd2nonlocal}
\end{align}
Estimates \eqref{eqn:RLbd1nonlocal} and \eqref{eqn:RLbd2nonlocal} complete the proof of Proposition \ref{prop:RLlownonlocal}. $\Box$ \\

\begin{proposition} \label{prop:RNLlownonlocal} Let  $0 < r < 1$. Let $R^{\sigma}_{_{\rm NL}} $ be defined in \eqref{eqn:RLNLdefnonlocal} and recall its derivative given in \eqref{eqn:DNLnonlocal}. Then there exists $\alpha_0, C > 0$ such that for $0 < \alpha < \alpha_0$, 
\begin{align}
& \left\| \ \chi_{_{\rm lo}} \ R^{\sigma}_{_{\rm NL}} \left[ \alpha, \widehat{E_{_{\rm lo}}} + \widehat{E_{_{\rm hi}}}  [ \alpha, \widehat{E_{_{\rm lo}}} ] \right] \ \right\|_{L^{2,a-2 }(\mathbb{R})} \nonumber \\ 
& \hspace{2cm}  \lesssim e^{- C/ \lambda(\alpha)^{\eta}} + e^{- C/ \lambda(\alpha)^{\eta}} \  \left\| \  \widehat{E_{_{\rm lo}}} \ \right\|_{L^{2,a}(\mathbb{R})}  +  \ \left\| \  \widehat{E_{_{\rm lo}}} \ \right\|^2_{L^{2,a}(\mathbb{R})} +   \ \left\| \  \widehat{E_{_{\rm lo}}} \ \right\|^3_{L^{2,a}(\mathbb{R})},    \label{eqn:RNLlowprop} \\
& \left\| \ D_{\widehat{E_{_{\rm lo}}}} \left( \chi_{_{\rm lo}} \ R^{\sigma}_{_{NL}} \left[ \alpha, \widehat{E_{_{\rm lo}}}  + \widehat{E_{_{\rm hi}}}  [ \alpha, \widehat{E_{_{\rm lo}}} ] \right] \right) \ \right\|_{L^{2,a}(\mathbb{R}) \rightarrow L^{2,a -2 }(\mathbb{R})}  \nonumber \\
& \hspace{2cm}  \lesssim e^{- C/ \lambda(\alpha)^{\eta}}  +  \left\| \  \widehat{E_{_{\rm lo}}} \ \right\|_{L^{2,a}(\mathbb{R})}  +  \ \left\| \  \widehat{E_{_{\rm lo}}} \ \right\|^2_{L^{2,a}(\mathbb{R})}  .  \label{eqn:RNLlowprop2}
\end{align}
\end{proposition}
 This follows from estimates \eqref{eqn:Ehighestlyapnonlocal} and \eqref{eqn:DEhighestnonlocal}, using \eqref{eqn:algebraB}.

\bigskip 
\begin{proposition} \label{prop:RFlownonlocal} Let $p = \min(1,s)$ and $0  < r < 1$. Recall $\mathfrak{e}_s( \alpha) $, given in \eqref{frak-e} . Let $\mathcal{D}^{\sigma,\alpha}$ be defined in \eqref{eqn:Phieqnnonlocal}. There exists $\alpha_0 > 0$ such that for $0 < \alpha < \alpha_0$,
\begin{align}
& \left\| \ \chi_{_{\rm lo}} \ \mathcal{D}^{\sigma,\alpha} \left[ S \right] \ \right\|_{L^{2,a-2p }(\mathbb{R})} \ \lesssim   \left( \frac{ \alpha}{\kappa_s(\alpha)} \right)^{1/2} \ \mathfrak{e}_s(\alpha) .   \label{eqn:RFlowpropnonlocal1}
\end{align}
 
\end{proposition}

\noindent {\bf Proof of Proposition \ref{prop:RFlownonlocal}:} We follow the proof from the nearest-neighbor case \cite{Jenkinson_Weinstein_2015} and use $\| \widehat{f} \|_{L^{2,a-2p}(\mathbb{R})} \leq \| \widehat{f} \|_{L^{2,a}(\mathbb{R})}$ extensively. Recall from \eqref{eqn:Phieqnnonlocal} that
\begin{align}
\mathcal{D}^{\sigma,\alpha} \left[ S \right] (Q) & = 
 R_{_{{\rm F},1}} \left[ \alpha \right]  + R_{_{{\rm F},2}}^{\sigma} \left[ \alpha \right]
\end{align}
where we have defined
\begin{align}
& R_{_{{\rm F},1}} \left[ \alpha \right] \equiv \chi_{_{\mathcal{B}_{_{\alpha}}}}(Q) \ \left( - \left[ 1 + M_{\alpha}^s( Q) \right] \ S(Q) + \left( \frac{1}{2 \pi} \right)^2 \ (S * S * S)(Q) \right), \nonumber \\
& R_{_{{\rm F},2}}^{\sigma} \left[ \alpha \right] \equiv \chi_{_{\mathcal{B}_{_{\alpha}}}}( Q)  \  \left( \frac{1}{2 \pi} \right)^2 \ \sum_{m = \pm 1} \ e^{2 m \pi i \sigma} \ (S * S * S) (Q - 2 m \pi / \alpha) . \label{eqn:RF12defnonlocal}
\end{align}
By Lemma \ref{lemma:expconvononlocal}, we have for some $C > 0$, 
\begin{align}
\left\| \ \chi_{_{\rm lo}} \ R_{_{{\rm F}, 2}}^{\sigma} \ \right\|_{L^{2,a - 2p}(\mathbb{R})} \leq \left\|  \  R_{_{{\rm F}, 2}}^{\sigma} \ \right\|_{L^{2,a}(\mathbb{R})} \lesssim e^{- C / \alpha^{\eta}}.
\end{align}
Next, we write 
$
S (Q)  = \widetilde{\psi} (Q) -  \overline{\chi}_{_{\mathcal{B}_{_{\alpha}}}}(Q) \ \widetilde{\psi}(Q) . 
$
Substituting into \eqref{eqn:RF12defnonlocal}, we have
\begin{align}
 \chi_{_{\rm lo}}(Q) \ R_{_{{\rm F},1}} \left[ \alpha \right] = & \   \chi_{_{\rm lo}}(Q)   \ \left( - \left[ 1 + M_{\alpha}^s( Q) \right] \ \widetilde{\psi} (Q) + \left( \frac{1}{2 \pi} \right)^2 \ \widetilde{\psi}  * \widetilde{\psi} * \widetilde{\psi}  (Q) \right), \nonumber \\
& -  \chi_{_{\rm lo}}(Q)  \   \left( \frac{1}{2 \pi} \right)^2 \ \bigg[ 3\  \left(\overline{\chi}_{_{\mathcal{B}_{_{\alpha}}}}  \ \widetilde{\psi} \right) * \widetilde{\psi}  * \widetilde{\psi}  (Q) \nonumber \\
& \hspace{2cm} + 3 \ \left(\overline{\chi}_{_{\mathcal{B}_{_{\alpha}}}}  \ \widetilde{\psi} \right) * \left(\overline{\chi}_{_{\mathcal{B}_{_{\alpha}}}}  \ \widetilde{\psi} \right) * \widetilde{\psi}  (Q)  \nonumber \\
& \hspace{2cm}  + \left(\overline{\chi}_{_{\mathcal{B}_{_{\alpha}}}}  \ \widetilde{\psi} \right) * \left(\overline{\chi}_{_{\mathcal{B}_{_{\alpha}}}}  \ \widetilde{\psi} \right) * \left(\overline{\chi}_{_{\mathcal{B}_{_{\alpha}}}}  \ \widetilde{\psi} \right) (Q) \bigg] . \label{eqn:RF12}
\end{align}
To all convolutions involving $ \overline{\chi}_{_{\mathcal{B}_{_{\alpha}}}}  \ \widetilde{\psi}$, we apply Lemma \ref{lemma:expconvo2nonlocal} to get another exponentially small bound $\lesssim e^{- C \pi^{\eta} / \alpha^{\eta}}$ in $L^{2,a}(\mathbb{R})$.

We finally turn our attention to the remaining terms in \eqref{eqn:RF12}:
\begin{align}
 & \chi_{_{\rm lo}}(Q)  \ \left( - \left[ 1 + |Q|^{2p} \right] \ \widetilde{\psi}(Q) + \left( \frac{1}{2 \pi} \right)^2 \ \widetilde{\psi} * \widetilde{\psi}  * \widetilde{\psi} (Q) + \left[   |Q|^{2p} - M_{\alpha}^s( Q) \right] \widetilde{\psi}(Q) \right) \nonumber \\
& = \chi_{_{\rm lo}}(Q)  \ \left[   |Q|^{2p} - M_{\alpha}^s( Q) \right] \widetilde{\psi}(Q),
\end{align} 
We apply Proposition \ref{prop:symbolexpo} to obtain
$
\left\| \chi_{_{\rm lo}}(Q)  \ \left[   |Q|^{2p} - M_{\alpha}^s( Q) \right] \widetilde{\psi}(Q) \right\|_{L^{2,a}(\mathbb{R}_Q)} \lesssim  \left( \alpha / \kappa_s(\alpha)  \right)^{1/2} \ \mathfrak{e}_s(\alpha). 
$
 This completes the proof of Proposition \ref{prop:RFlownonlocal}. 
$\Box$ \\

 We now apply Propositions \ref{prop:Rpertboundnonlocal} through \ref{prop:RFlownonlocal}, and estimates  \eqref{eqn:Ehighestlyapnonlocal} and \eqref{eqn:DEhighestnonlocal} to \eqref{eqn:Htildedefnonlocal} and \eqref{eqn:DHtildenonlocal}.  This implies estimates \eqref{eqn:Hestimatenonlocal} and \eqref{eqn:Hdiffestimatenonlocal} and concludes the proof of Proposition \ref{prop:rescalenonlocal}.

\subsection{Solution of the low frequency equation} \label{section:solutionlownonlocal}

We may now apply the implicit function theorem, Lemma \ref{lemma:generallinearopnonlocal}, to the rescaled low frequency equation \eqref{eqn:lowrescalednonlocal}. \\

\begin{proposition}
\label{prop:solutiontolownonlocal}
Let $a > 1/2$, $p = \min(s,1)$, and $0 < r < 1$. Then there exists 
$0< \alpha_2\le\alpha_1$ such that for all $\alpha\in (0,\alpha_2)$, there exists an even (symmetric) solution $\widehat{E_{_{\rm lo}}}$ to \eqref{eqn:lowrescalednonlocal} which satisfies\begin{align}
 \left\| \widehat{E_{_{\rm lo}}} \right\|_{L^{2,a}(\mathbb{R})} \lesssim  \left( \frac{ \alpha}{\kappa_s(\alpha)} \right)^{1/2} \ \mathfrak{e}_s(\alpha) , \label{eqn:philoestnonlocal}
\end{align}
where $\mathfrak{e}_s(\alpha) \to 0 $ is given in \eqref{frak-e}. 
Furthermore, we have that $\widehat{E_{_{\rm lo}}} = \chi_{_{\rm lo}} \widehat{E_{_{\rm lo}}}$; that is, $\widehat{E_{_{\rm lo}}}(Q)$ is supported on $ Q \in \left[ - \frac{1}{\lambda(\alpha)}, \frac{1}{\lambda(\alpha)} \right]$, where $\lambda(\alpha) \to 0$ as $\alpha \to 0$ is defined by $\lambda(\alpha) = \left\{ \begin{array}{ll} \alpha^{r - 1}, 0 < r < 1 & : s \neq 1 \\ \frac{- 1}{\log(\alpha)} & : s = 1 \end{array} \right.$. 
\end{proposition}

\noindent {\bf Proof of Proposition \ref{prop:solutiontolownonlocal}:}  By Proposition \ref{prop:Lplusnonlocal}  $\widetilde{L_+}:L^{2,a}_{_{\rm even}}(\mathbb{R})\to L_{_{\rm even}}^{2,a-2p}(\mathbb{R}) $ is an isomorphism. Moreover, by Proposition \ref{prop:rescalenonlocal} the mapping $(\alpha,\widehat{E}_{\rm lo})\mapsto \mathcal{R}_2^{\sigma}[\alpha,\widehat{E}_{\rm lo}]$, maps $L^{2,a}_{_{\rm
even}}(\mathbb{R})$ to $L^{2,a-2p}_{_{\rm even}}(\mathbb{R})$, and  is continuous at $(\alpha, \widehat{E_{_{\rm lo}}} ) = (0,0)$ .  Furthermore, by choosing $\alpha < \alpha_1$ the estimates \eqref{eqn:Hestimatenonlocal} and \eqref{eqn:Hdiffestimatenonlocal} on $\mathcal{R}_2^{\sigma}[ \alpha, \widehat{E_{_{\rm lo}}} ]$ hold. Hence,  hypotheses
\eqref{eqn:HIFTestnonlocal} and \eqref{eqn:HIFTestdiffnonlocal} of Lemma \ref{lemma:generallinearopnonlocal} are satisfied. Lemma \ref{lemma:generallinearopnonlocal} implies, for $0<\alpha<\alpha_2\le\alpha_1$, the existence of 
$\widehat{E_{_{\rm lo}}}$ satisfying the bound \eqref{eqn:philoestnonlocal}. $\Box$
\bigskip

We now complete the proofs of 
Theorems \ref{th:leadingordernonlocal} and \ref{th:mainnonlocal} \label{section:onedcompletionnonlocal}. From Proposition \ref{prop:ifthighnonlocal} we obtain 
$\widehat{E_{_{\rm hi}}}[\alpha,\widehat{E_{_{\rm lo}}}] \in L_{_{\rm even}}^{2,a}(\mathbb{R})$.
Then, $\widehat{E^{\alpha,\sigma}}( Q )\ =\ \widehat{E_{_{\rm lo}}}(Q) + \widehat{E_{_{\rm hi}}} \left[ \alpha, \widehat{E_{_{\rm lo}}} \right] (Q)$ solves the corrector equation {\eqref{eqn:Eeqnnonlocal}, and 
\begin{align}
\widehat{\Phi^{\alpha,\sigma}}( Q) =  \chi_{_{\mathcal{B}_{_{\alpha}}}}(Q) \  \widetilde{\psi} (Q)  +    \widehat{E^{\alpha, \sigma}}(Q) \in L_{_{\rm even}}^{2,a}(\mathbb{R}),  
\label{Phi-sigmanonlocal}
\end{align}
is a  solution to \eqref{eqn:Phieqnnonlocal}. 
As discussed in Section \ref{section:1dproofnonlocal}, $G^{\alpha,\sigma}(q), \sigma=0,1/2$, is constructed from  $\widehat{\phi^{\sigma,\alpha}}(q)$ via
  \begin{align}
\widehat{\phi^{\sigma}}(q) & = \chi_{_{\mathcal{B}}}(q)  \widehat{\phi^{\sigma}}(q)    = \left( \frac{\kappa_s(\alpha)}{\alpha^2} \right)^{1/2} \ \left[ \chi_{_{\mathcal{B}}}(q)  \  \widetilde{\psi} \left( \frac{q}{\alpha} \right)   + \widehat{E^{\alpha, \sigma}} \left( \frac{q}{\alpha} \right) \right],
%
%  \nonumber \\
%& \left\|  \chi_{_{\mathcal{B}}}(q)  \widetilde{\psi}   \left( \frac{q}{\alpha} \right) \right\|_{L^{2,a}(\mathbb{R}_q)} \lesssim \alpha^{1/2}, \qquad  \left\| \widehat{E^{\alpha, \sigma}} \left( \frac{q}{\alpha} \right) \right\|_{L^{2,a}(\mathbb{R}_q)} \lesssim \alpha^{1/2} \ \mathfrak{e}_s(\alpha).
\label{FjEJ-boundsnonlocal}
\end{align}  
These details are similar to those in the local discrete case and we refer to
  Sections 6.4 and 6.5 of \cite{Jenkinson_Weinstein_2015} .

 \section{ Bound on the Peierles-Nabbaro barrier - Proof of Theorem \ref{th:PNnonlocal}} \label{section:PNbarriernonlocal}
 %We now prove Theorem \ref{th:PNnonlocal} in sections \ref{section:PNbarriernonlocal} through \ref{section:diffequationnonlocal}. 
 We now prove  the differences  $\mathcal{N}[G^{\alpha,{\rm on}}]-\mathcal{N}[G^{\alpha,{\rm off}}]$ and $\mathcal{H}[G^{\alpha,{\rm on}}]-\mathcal{H}[G^{\alpha,{\rm off}}]$ 
   are bounded in magnitude by a quantity of order: $(\kappa_s(\alpha)/\alpha)\cdot   \ e^{- C / \alpha^{\eta}}$, for $\alpha$ small.  
%\begin{align}
% \mathcal{N}[ G ] & = \|  G \|^2_{l^2(\mathbb{Z})} = \sum_{n \in \mathbb{Z}} |G_n|^2,   \nonumber \\
% \mathcal{H}[ G ] & = \frac{1}{2} \sum_{n \in \mathbb{Z}}  \sum_{\substack{m \in \mathbb{Z}   \\ m \neq n}} \ J_{|n - m|} \ | G_{m} - G_n |^2 -  \frac{1}{2} |G_n|^4 \nonumber \\
%& = \frac{1}{2}  \sum_{\substack{m \in \mathbb{Z}   \\ m \neq 0}} \sum_{n \in \mathbb{Z}} \ J_{|m|} \ | G_{m + n} - G_n |^2 - \frac{1}{2}  |G_n|^4.
%\end{align}
 Here, ${G}^{\alpha, {\rm on}} = \{G_n^{\alpha, {\rm on}} \}_{n \in \mathbb{Z}} $ and ${G}^{\alpha, {\rm off}} = \{G_n^{\alpha, {\rm off}} \}_{n \in \mathbb{Z}}$ are, respectively, onsite and offsite solutions of the nonlocal DNLS equation. The parameter
 $\eta=\eta(s) = \min(2s, 1)$ governs the exponential decay rate of the continuum limit ground state solitary wave $p(s)-$ FNLS; see  Proposition \ref{prop:Psinonlocal}.
%\begin{align}
%\mathbf{PN \ Barrier:} \hspace{1cm} & \bigg| \mathcal{N} [{ G^{\rm off}}] - \mathcal{N} [{G^{\rm on}} ] \bigg| \lesssim  \frac{\kappa_s(\alpha)}{\alpha} \ e^{- C/\alpha^{\eta}} \hspace{2cm} \\
%& \bigg| \mathcal{H} [{G^{\rm off}}] - \mathcal{H} [{G^{\rm on}} ] \bigg| \lesssim \frac{\kappa_s(\alpha)}{\alpha} \ e^{- C/\alpha^{\eta}} . 
%\end{align}
%The following identity allows us to equate the nonlinear term in $\mathcal{H}$ with terms involving $ \|{G} \|^2_{l^2(\mathbb{Z})} $ and the difference operator
%\begin{align}
%\sum_{\substack{m \in \mathbb{Z}   \\ m \neq 0}} \sum_{n \in \mathbb{Z}} \ J_{|m|} \ | G_{m + n} - %G_n |^2. 
%\end{align}.

\begin{proposition} \label{prop:virialnonlocal}
Suppose $G = \{G_n \}_{n \in \mathbb{Z}} $ is real-valued and solves non-local  DNLS equation
 with interaction parameter $s$:
$
0 = \kappa_s(\alpha) \ G_n - ( \mathcal{L}^s {G})_n - (G_n)^3,\ \ n\in\mathbb{Z}.  $
Then, 
\begin{align}
 \sum_{n \in \mathbb{Z}} |G_n|^4 
 & =  \kappa_s(\alpha)  \sum_n |G_n|^2 +  \frac{1}{2} \ \sum_{\substack{m \in \mathbb{Z}   \\ m \neq 0}}  \sum_{n \in \mathbb{Z}}  J_{|m|} | G_{m + n} - G_n |^2
\end{align}
\end{proposition}

\noindent {\it Proof of Proposition \ref{prop:virialnonlocal}:} Multiply the system by $G_n$, sum over all $n \in \mathbb{Z}$ and then sum by parts. $\Box$ \\

Using Proposition \ref{prop:virialnonlocal} and following the arguments of \cite{Jenkinson_Weinstein_2015} (in particular, equation (7.9) of \cite{Jenkinson_Weinstein_2015}) we obtain:

\begin{align}
& \bigg| \mathcal{N} [{G^{\rm off}}] - \mathcal{N} [{G^{\rm on}} ] \bigg| + 
\bigg| \mathcal{H} [{ G^{\rm off}}] - \mathcal{H} [{G^{\rm on}} ] \bigg| \lesssim  \left( \frac{\kappa_s(\alpha)}{\alpha} \right) \  \left\| \widehat{\Phi^{\rm off}} - \widehat{\Phi^{\rm on}}  \right\|_{L^{2,a}(\mathbb{R})}, \label{eqn:Nestnonlocal} 
\end{align}
Therefore, bounds on the PN-barrier are reduced to bounds on  $  \widehat{\Phi^{\rm off}} -  \widehat{\Phi^{\rm on}}$  in $L^{2,a}(\mathbb{R})$. 
Here, $\widehat{\Phi^{\rm on}}\equiv \widehat{\Phi^{\alpha,\sigma=0} }$
and $\widehat{\Phi^{\rm off}}\equiv\widehat{\Phi^{\alpha,\sigma=1/2}} $.
Theorem \ref{th:PNnonlocal} now follows directly from the following proposition, proved  in the next section. \\

\begin{proposition}
\label{prop:expdiffEonoffnonlocal} Let $\alpha_0  > 0$ be that prescribed in Theorem \ref{th:mainnonlocal} and fix $\eta = \min(2s, 1)$. Then for $0 < \alpha < \alpha_0$, there exists a constant $C > 0$ such that
\begin{align}
  \left\| \widehat{\Phi^{\rm off}} - \widehat{\Phi^{\rm on}} \right\|_{L^{2,a}(\mathbb{R})} \lesssim e^{- C / \alpha^{\eta}} . \label{eqn:diffexpononlocal}
\end{align}
\end{proposition}

\subsection{Estimation of the difference $ \widehat{\Phi^{\rm off}} - \widehat{\Phi^{\rm on}}$; proof of Proposition \ref{prop:expdiffEonoffnonlocal} } \label{section:diffequationnonlocal}
 The idea is to derive a non-homogeneous equation for $\widehat{\Phi^{^{\rm diff}}}\equiv \widehat{\Phi^{\rm off}} - \widehat{\Phi^{\rm on}}$. We shall prove that this equation is driven  by
 terms which are exponential small in $\alpha$ due to uniform  decay bounds on $e^{\mu |Q|^\eta}\widehat{\Phi^{\alpha,\sigma}}(Q)$,  $\eta=\min(s,1)$, for some $\mu>0$. 

\begin{proposition}
\label{prop:ediffeqn}
Let $0 < \alpha < \alpha_0$. Then $\widehat{\Phi^{^{\rm diff}}} = \widehat{\Phi^{\rm off}} - \widehat{\Phi^{\rm on}}$ solves the following \emph{linear} equation:
\begin{align}
& \left[ 1 + M^s_{\alpha}(Q) \right] \ \widehat{\Phi^{^{\rm diff}}}(Q) - \chi_{_{\mathcal{B}_{_{\alpha}}}} (Q)  \frac{1}{ 4 \pi^2} \bigg( \widehat{\Phi^{\rm off}} *  \widehat{\Phi^{\rm off}} * \widehat{\Phi^{^{\rm diff}}}   (Q)  \nonumber \\
& + \widehat{\Phi^{\rm on}} *  \widehat{\Phi^{\rm off}} *  \widehat{\Phi^{^{\rm diff}}}  (Q) + \widehat{\Phi^{\rm on}} *    \widehat{\Phi^{\rm on}} * \widehat{\Phi^{^{\rm diff}}}  (Q) =  R_{_{\rm diff}} \left[ \widehat{\Phi^{\rm off }}, \widehat{\Phi^{\rm on }} \right]( Q ), \label{eqn:Ediffnonlocal}
\end{align}
where the inhomogeneous right-hand side, involving only $m=\pm1$ side-band terms,  is given by
\begin{align}
 R_{_{\rm diff}} \left[ \widehat{\Phi^{\rm off }}, \widehat{\Phi^{\rm on }} \right]( Q ) = -
 \chi_{_{\mathcal{B}_{_{\alpha}}}}( Q ) \frac{1}{4 \pi^2} \sum_{m = \pm 1} \bigg(  & \ \widehat{\Phi^{\rm off}}  * \widehat{\Phi^{\rm off}} * \widehat{\Phi^{\rm off}}  ( Q - 2 m \pi / \alpha) \nonumber \\
& + \ \widehat{\Phi^{\rm on}} *  \widehat{\Phi^{\rm on}}  * \widehat{\Phi^{\rm on}}   ( Q - 2 m \pi / \alpha)  \bigg), \label{eqn:R3defnonlocal}
\end{align}
 %and satisfies the bound:\ $
%\left\|  R_{_{\rm diff}} \left[ \widehat{\Phi^{\rm off }}, \widehat{\Phi^{\rm on }} \right] \right\|_{L^{2,a}%(\mathbb{R})}  \lesssim e^{- C \pi^{\eta} / \alpha^{\eta}}.
% \label{eqn:residdiffestnonlocal}
%$
\end{proposition}

Recall from Theorem \ref{th:leadingordernonlocal} that $\widehat{\Phi^{\sigma, \alpha}}  \in L^{2,a}(\mathbb{R}) $ is well-defined, $ \| \widehat{\Phi^{\sigma,\alpha}} \|_{L^{2,a}(\mathbb{R}_Q)} \lesssim 1$, for $\alpha$ sufficiently small and satisfies equation \eqref{eqn:Phieqnnonlocal}. Note that $\widehat{\Phi^{\sigma,\alpha}}$ is supported on $\mathcal{B}_{_{\alpha}} = [ - \frac{\pi}{\alpha}, \frac{\pi}{\alpha}]$, an interval which grows as $\alpha \downarrow 0$. We begin by proving a uniform decay bound for $\widehat{\Phi^{\sigma, \alpha}}$. 
\medskip
\begin{proposition} \label{prop:Phinonlocal} 
For $0  < \alpha < \alpha_2$, let  $\widehat{\Phi^{\sigma, \alpha}}\in L^{2,a},\ a>1/2$  denote the onsite ($\sigma=0$) and offsite ($\sigma=1/2$) nonlocal DNLS solitary waves obtained in Theorem \ref{th:mainnonlocal}.  Let $\eta = \min(2s, 1)$. Then there exist constants 
$\mu=\mu\left(\|\widehat{\Phi^{\sigma, \alpha}}\|_{L^{2,a}}\right)$ and  $C_1 > 0$, independent of $\widehat{\Phi^{\sigma, \alpha}}$ and $\alpha$,
  such that
  \begin{align}
  \| e^{\mu |Q|^{\eta}} \widehat{\Phi^{\sigma,\alpha}} \|_{L^{2,a}(\mathbb{R}_Q)} \leq C_1 \ 
  \|\widehat{\Phi^{\sigma,\alpha}}\|_{L^{2,a}(\mathbb{R}_Q)}. 
  \end{align}
\end{proposition}

\medskip
\noindent {\bf Proof of Proposition \ref{prop:Phinonlocal}:} $\widehat{\Phi}$ solves
\begin{align}
\left( 1 + M^s_{\alpha}(Q) \right) \ \widehat{\Phi}(Q)  - \frac{1}{(2 \pi)^2} \ \sum_{ m = -1}^1 \chi_{_{\mathcal{B}_{_{\alpha}}}}(Q)  \widehat{\Phi} * \widehat{\Phi} * \widehat{\Phi} (Q - 2 m \pi / \alpha) . 
\end{align}
We shall apply Lemma \ref{lemma:expogeneral} from Appendix \ref{appendix:expogeneral} with the identifications $M(Q) \equiv M^s_{\alpha}(Q)$, $A = \mathcal{B}_{_{\alpha}}$, $\tau_m =  - 2 \pi m / \alpha$, and $ m = - 1, 0, 1$. 
We need to check that there exists $D_M > 0$ such that
\begin{align}
& (a) \quad \frac{ \chi_{_{\mathcal{B}_{_{\alpha}}}}(Q) \ |Q|^{\eta} }{ 1 + M_{\alpha}^s(Q) } \leq D_M, 
{\rm and} \quad (b) \quad \chi_{_{\mathcal{B}_{_{\alpha}}}}(Q) \ |Q| \leq   |Q - 2 m \pi/\alpha|  , \qquad m = -1, 0, 1.  \label{eqn:ablemma}
\end{align}
We need only focus on $Q \in \mathcal{B}_{_{\alpha}}$. Consider the cases $(i) \quad 1/2 \leq s \leq \infty, s \neq 1, \quad $ $(ii) \quad 1/4 < s < 1/2, s \neq 1, \quad $ $(iii) \quad s = 1$. First, $(b)$ holds for $Q \in \mathcal{B}_{_{\alpha}} = \left[ - \frac{\pi}{\alpha}, \frac{\pi}{\alpha} \right]$ easily. 

Next, we prove $(a):  $ $\| e^{\mu |Q|} \widehat{\Phi^{\sigma,\alpha}} \|_{L^{2,a}(\mathbb{R}_Q)} \leq C_1 \ 
  \|\widehat{\Phi^{\sigma,\alpha}}\|_{L^{2,a}(\mathbb{R}_Q)} $  for $s \geq 1/2$ and $s \neq 1$. Here, $\eta = \min(1 , 2s) = 1$, and we apply Lemma \ref{lemma:expogeneral} to equation \eqref{eqn:Phieqnnonlocal} for $\widehat{\Phi^{\sigma, \alpha}}$, which gives decay $\sim e^{- C |Q|}$ for $\widehat{\Phi^{\sigma, \alpha}}$. To see that the hypotheses of the lemma are satisfied, first note that for $ m \in \{-1,0,1\}$ and $Q \in \mathcal{B}_{_{\alpha}},$  $
|Q| \leq |Q - 2 m \pi / \alpha|. $  For $s \neq 1$ and $p = \min(1,s)$, recall from Lemma \ref{lemma:Mbdnonlocal} that there exists $C > 0$ such that 
$
M^s_{\alpha}( Q ) 
%= \frac{4}{\kappa_s(\alpha)} \sum_{m = 1}^{\infty} J_m \sin^2 (Q m \alpha/2)   \qquad
\geq  C |Q|^{2p},\ \  Q \in \mathcal{B}_{_{\alpha}}.
$
This implies that, by maximization over $Q$,
\begin{align}
\frac{ \chi_{_{\mathcal{B}_{_{\alpha}}}}(Q) \ |Q|}{1 + M^s_{\alpha}(Q)} \leq \frac{ \chi_{_{\mathcal{B}_{_{\alpha}}}}(Q) \ |Q|}{1 + C \ |Q|^{2p}} \leq \frac{1}{2p \ C^{1/2p}} \left( 2p-1 \right)^{(2p-1)/2p},
\end{align}
where we understand the right-hand side to mean $C^{-1}$ for $s = p = 1/2$.

Next, we prove $(a)$ for $1/4 < s < 1/2$ such that $\eta = \min(1,2s) = 2s$. We again apply Lemma \ref{lemma:expogeneral}, which gives weaker decay $\sim e^{- C |Q|^{2s}}$ for $\widehat{\Phi^{\sigma, \alpha}}$. We again recall from Lemma \ref{lemma:Mbdnonlocal} (since for $s < 1$,  $p = \min(1,s) = s$) that there exists $C > 0$ such that 
$
M^s_{\alpha}( Q )  
 \geq  C |Q|^{2s},\ \ Q \in \mathcal{B}_{_{\alpha}}.
$
This implies that
\begin{align}
\frac{\chi_{_{\mathcal{B}_{_{\alpha}}}}(Q) \ |Q|^{2s} }{1 + M^s_{\alpha}(Q)} \leq \frac{\chi_{_{\mathcal{B}_{_{\alpha}}}}(Q) \ |Q|^{2s}}{1 + C \ |Q|^{2s}} \leq \frac{1}{C}. 
\end{align}

Finally, to satisfy $(a)$ from \eqref{eqn:ablemma} when $s = 1$, we require the following proposition.

\begin{proposition} \label{prop:sequals1bd}
Let $s = 1$ and let the spectral cutoff projections $\chi_{_{\rm hi}}$ and $\chi_{_{\rm lo}}$ be as defined in \eqref{lo-hi-cutnonlocal}. Then for $\alpha$ sufficiently small, there exists some $C > 0$ such that
\begin{align}\nonumber
& (a) \quad  \chi_{_{\mathcal{B}_{_{\alpha}}}}(Q) \ \chi_{_{\rm hi}}(Q) \ M^s_{\alpha}(Q) \geq C |Q|  \qquad\textrm{and}\qquad (b) \quad \chi ( | Q| > 1 ) \ \chi_{_{\rm lo}}(Q) \ M^s_{\alpha}(Q) \geq C |Q|^2. 
\end{align}
\end{proposition}

\noindent We defer the proof until later in this section. Proposition \ref{prop:sequals1bd} implies that
\begin{align}
\frac{ \chi_{_{\mathcal{B}_{_{\alpha}}}}(Q) \ |Q|}{1 + M^s_{\alpha}(Q)} & = \chi_{_{\mathcal{B}_{_{\alpha}}}}(Q) \left( \frac{ \chi ( | Q| \leq 1 ) \  |Q|}{1 + M^s_{\alpha}(Q)} + \frac{  \chi ( | Q| > 1 )  \  |Q|}{1 + M^s_{\alpha}(Q)}  \right)  \leq 1 + \chi_{_{\mathcal{B}_{_{\alpha}}}}(Q) \left(  \frac{ \chi ( | Q| > 1 )  \  |Q|}{1 + M^s_{\alpha}(Q)} \right) \nonumber \\
& \leq  1 + \frac{ \chi ( | Q| > 1 ) \ \chi_{_{\rm lo}}(Q)\  |Q|}{ 1 + M^s_{\alpha}(Q)  } + \frac{ \chi_{_{\mathcal{B}_{_{\alpha}}}}(Q) \ \chi_{_{\rm hi}}(Q) \ |Q|}{ 1 + M^s_{\alpha}(Q)}  \nonumber \\
& \leq  1 + \frac{ \chi ( | Q| > 1 ) \ \chi_{_{\rm lo}}(Q)\  |Q|}{1 + C |Q|^2 } + \frac{ \chi_{_{\mathcal{B}_{_{\alpha}}}}(Q) \ \chi_{_{\rm hi}}(Q) \ |Q|}{1 + C |Q|}   \leq 1 +  \frac{1}{C^{1/2}} + \frac{1}{C} = D_{s = 1}, 
\end{align}
such that the hypotheses of Lemma \ref{lemma:expogeneral} are again satisfied when $s = 1$.

This completes the proof of Proposition \ref{prop:Phinonlocal}. $\Box$

\bigskip
\noindent {\bf Proof of Proposition \ref{prop:sequals1bd}:} Recall that for $s = 1$, the the spectral projections give
\begin{align}
\chi_{_{\rm lo}}(Q) \ |Q| \leq (- \log(\alpha)), \qquad \qquad \chi_{_{\rm hi}}(Q) \ |Q| \geq (- \log(\alpha)). 
\end{align} 
Furthermore, recall from Lemma \ref{lemma:Mbdnonlocal} that for $s = 1, \quad$ 
$
\chi_{_{\mathcal{B}_{_{\alpha}}}}(Q) \ M_{\alpha}^s(Q) \geq \frac{C}{(- \log(\alpha))} |Q|^2, 
$
such that
\begin{align}
\chi_{_{\mathcal{B}_{_{\alpha}}}}(Q) \ \chi_{_{\rm hi}}(Q)  \ M_{\alpha}^s(Q)   \geq \frac{C}{(- \log(\alpha))} \  \chi_{_{\rm hi}}(Q)  \ |Q|^2 \geq C |Q|. 
\end{align}
This proves $(a)$. 
Next, we use the expansion from Proposition \ref{prop:symbolconvnonlocal1} for $s = 1$ and $Q \in \mathcal{B}_{_{\alpha}} = \left[ - \frac{\pi}{\alpha}, \frac{\pi}{\alpha} \right]$:
\begin{align}
M_{\alpha}^s(Q) & = |Q|^{2} +  \frac{1}{- \log(\alpha)} \ f_s(Q ; \alpha) \ \left( \frac{3}{2} - \log(|Q|) \right) \ |Q|^2   \\
  & = |Q|^2 \ \left[ 1 +  \frac{1}{- \log(\alpha)} \ f_s(Q ; \alpha) \ \left( \frac{3}{2} - \log(|Q|) \right) \right], \qquad | f_s (Q; \alpha) | \lesssim 1. 
\end{align}
Recall that $\lambda(\alpha) =  \frac{1}{(- \log(\alpha))} $ for $s = 1$. Furthermore, recall that $\chi_{_{\rm lo}}(Q) = \chi \left( |Q| \leq \lambda(\alpha)^{-1} \right)$, where $\lambda(\alpha)^{-1} \leq \pi/\alpha$ for $\alpha$ sufficiently small, which implies that $\chi_{_{\rm lo}}(Q) \ | f_s(Q; \alpha) | \lesssim 1$. Therefore, we have
\begin{align}
& \chi ( | Q| > 1 ) \  \chi_{_{\rm lo}}(Q) \bigg|  \frac{1}{- \log(\alpha)} \ f_s(Q ; \alpha) \ \left( \frac{3}{2} - \log(|Q|) \right) \bigg| \lesssim \frac{1 + \log(- \log(\alpha))}{(- \log(\alpha))}. 
\end{align}
such that for $\alpha$ sufficiently small, there exists $C > 0$ such that
\begin{align}
& \chi ( | Q| > 1 ) \ \chi_{_{\rm lo}}(Q) \ \bigg( 1 +  \frac{1}{- \log(\alpha)} \ f_s(Q ; \alpha) \ \left( \frac{3}{2} - \log(|Q|) \right) \geq C.
\end{align}
This completes the proof of Proposition \ref{prop:sequals1bd}. $\Box$

\bigskip
Next, we derive the equation for $\widehat{\Phi^{^{\rm diff}}} = \widehat{\Phi^{\rm off}} - \widehat{\Phi^{\rm on}}$.

\begin{proposition}
\label{prop:ediffeqnnonlocal}
Let $0 < \alpha < \alpha_0$ and $\eta = \min(2s, 1)$. Then, the non-homogeneous source term 
 in equation \eqref{eqn:Ediffnonlocal} for  $\widehat{\Phi^{^{\rm diff}}}(Q)$ 
% solves the following \emph{linear} equation:
%\begin{align}
%& \left[ 1 + M^s_{\alpha}(Q) \right] \ \widehat{\Phi^{^{\rm diff}}}(q) - \chi_{_{\mathcal{B}_{_{\alpha}}}} (Q)  \frac{1}{ 4 \pi^2} \bigg( \widehat{\Phi^{\rm off}} *  \widehat{\Phi^{\rm off}} * \widehat{\Phi^{^{\rm diff}}}   (Q)  \nonumber \\
%& + \widehat{\Phi^{\rm on}} *  \widehat{\Phi^{\rm off}} *  \widehat{\Phi^{^{\rm diff}}}  (Q) + \widehat{\Phi^{\rm on}} *    \widehat{\Phi^{\rm on}} * \widehat{\Phi^{^{\rm diff}}}  (Q) =  R_{_{\rm diff}} \left[ \widehat{\Phi^{\rm off }}, \widehat{\Phi^{\rm on }} \right]( Q ), \label{eqn:Ediffnonlocal}
%\end{align}
%where the inhomogeneous right-hand side is given by
%\begin{align}
% R_{_{\rm diff}} \left[ \widehat{\Phi^{\rm off }}, \widehat{\Phi^{\rm on }} \right]( Q ) = -
% \chi_{_{\mathcal{B}_{_{\alpha}}}}( Q ) \frac{1}{4 \pi^2} \sum_{m = \pm 1} \bigg(  & \ \widehat{\Phi^{\rm off}}  * \widehat{\Phi^{\rm off}} * \widehat{\Phi^{\rm off}}  ( Q - 2 m \pi / \alpha) \nonumber \\
%& + \ \widehat{\Phi^{\rm on}} *  \widehat{\Phi^{\rm on}}  * \widehat{\Phi^{\rm on}}   ( Q - 2 m \pi / \alpha)  \bigg), \label{eqn:R3defnonlocal}
%\end{align}
 and 
 satisfies the bound:\ $
\left\|  R_{_{\rm diff}} \left[ \widehat{\Phi^{\rm off }}, \widehat{\Phi^{\rm on }} \right] \right\|_{L^{2,a}(\mathbb{R})}  \lesssim e^{- C \pi^{\eta} / \alpha^{\eta}}.
% \label{eqn:residdiffestnonlocal}
$
\end{proposition}

\medskip

\noindent {\bf Proof of Proposition \ref{prop:ediffeqnnonlocal} :} 
%We subtract equation \eqref{eqn:Phieqnnonlocal} for $\sigma = 0$ from the same equation for %$\sigma = 1/2$. 
We apply Lemma \ref{lemma:expconvononlocal}  and Proposition \ref{prop:Phinonlocal}. This gives
\[ \left\| \chi_{_{\mathcal{B}_{_{\alpha}}}}(Q) \ \widehat{\Phi^{\sigma}} * \widehat{\Phi^{\sigma}} * \widehat{\Phi^{\sigma}} ( Q - 2 m \pi /\alpha) \right\|_{L^{2,a}(\mathbb{R}_Q)} \lesssim  e^{- C \pi^{\eta} / \alpha^{\eta}}, \qquad m = \pm 1. 
\qquad\qquad\qquad\Box \]
%which gives \eqref{eqn:residdiffestnonlocal}.
 
\medskip

We now use Proposition \ref{prop:ediffeqnnonlocal} to prove the exponential bound \eqref{eqn:diffexpononlocal} on $ \widehat{\Phi^{^{\rm diff}}}$. We use a Lyapunov-Schmidt reduction argument, analogous to that used in the proof of Theorem \ref{th:leadingordernonlocal}. We summarize the argument since the details are now quite familiar. Introduce
\begin{align}
\widehat{\Phi^{\rm diff}_{_{\rm lo}}}(Q) \equiv  \chi_{_{\rm lo}}(Q) \ \widehat{\Phi^{^{\rm diff}}}(Q), \quad \quad {\rm and } \quad \quad \widehat{\Phi^{\rm diff}_{_{\rm hi}}}(Q) \equiv  \chi_{_{\rm hi}}(Q) \ \widehat{\Phi^{^{\rm diff}}}(Q).
\end{align}
Solving for  $\widehat{\Phi^{\rm diff}_{_{\rm hi}}}$ as a functional of  $\widehat{\Phi^{\rm diff}_{_{\rm lo}}}$ and  estimation of the mapping yields:
\begin{align}
\left\| \widehat{\Phi^{^{\rm diff}}_{_{\rm hi}}} \right\|_{L^{2,a}(\mathbb{R})}  \lesssim \  \left\{ \begin{array}{ll}  \alpha^{2p (1 - r)}  \left( \left\| \widehat{\Phi^{^{\rm diff}}_{_{\rm lo}}} \right\|_{L^{2,a}(\mathbb{R})} +  e^{- C \pi^{\eta}/\alpha^{\eta}}  \right) & : s \neq 1 \\
  \frac{1}{(- \log(\alpha))} \left( \left\| \widehat{\Phi^{^{\rm diff}}_{_{\rm lo}}} \right\|_{L^{2,a}(\mathbb{R})} + e^{- C \pi/ \alpha}  \right) & : s = 1 \end{array} \right. , \label{eqn:highestdiffnonlocal}
\end{align}
$\widehat{\Phi^{^{\rm diff}}_{_{\rm lo}}} $ satisfies an inhomogeneous equation forced by  $\chi_{_{\rm lo}} \ R_{_{\rm diff}} \left[ \widehat{\Phi^{\rm off }}, \widehat{\Phi^{\rm on }} \right] $ which satisfies the exponential bound of Proposition \ref{prop:ediffeqnnonlocal}. A simple bootstrap argument using \eqref{eqn:highestdiffnonlocal} and the bounds on $\widehat{\Phi^{\rm off }}$ and $\widehat{\Phi^{\rm on }}$ give
$ 
\left\| \widehat{\Phi^{^{\rm diff}}_{_{\rm lo}}} \right\|_{L^{2,a}(\mathbb{R})} \lesssim  e^{- C \pi^{\eta} / \alpha^{\eta} },
$ 
for $\alpha$ sufficiently small. Since $\widehat{\Phi^{\rm diff}}=
\widehat{\Phi^{\rm diff}_{_{\rm lo}}}+\widehat{\Phi^{\rm diff}_{_{\rm hi}}}$, the bound \eqref{eqn:highestdiffnonlocal} implies the assertion of  Proposition \ref{prop:expdiffEonoffnonlocal}. $\Box$

This completes the proof of Theorem \ref{th:PNnonlocal}.
 
\section{Higher order expansions in $\alpha$ of $G^{\alpha,{\rm on}}$ and $G^{\alpha,{\rm off}}$ } \label{section:higherorder}

In this section, we provide a general outline for how to extend the expansion of $G^{\alpha, \sigma}$ to higher order in $\alpha$, for $\alpha$ small. The structure of the expansion, and in particular the precise expansion parameter, depends on the analytic properties of $M_{\alpha}^s(Q)$, which is expressible in terms of the polylogarithms (see Appendix \ref{appendix:asymptotics}). We have

\begin{proposition} \label{prop:expansionsM}
Let $M_{\alpha}^s(Q)$ be given in \eqref{eqn:Mdefnonlocal} with $J_m^s$ given in \eqref{eqn:Jdeftrue}. Then for $|Q| \leq \pi/\alpha$, 
\begin{align}
M^s_{\alpha}(Q) = \left\{  \begin{array}{ll}  |Q|^{2s}  +  \frac{2}{C_s} \ \sum_{j = 1}^{\infty} \frac{ \zeta(1 + 2s - 2j) }{ (2j)! } \ (-1)^{j + 1} \ \alpha^{2j - 2s} \ |Q|^{2j} & : s < 1 \\ 
& \\
\hline \\
|Q|^{2} + \frac{1}{C_s \ (- \log(\alpha))} \ \left[ \frac{3}{2} - \log( |Q|)  \right]  |Q|^2 & \\
  + \frac{2}{C_s \ (- \log(\alpha))} \ \sum_{ \substack{ j = 2  } }^{\infty} \frac{ \zeta(3 - 2j) }{ (2j)! } \ (-1)^{j + 1} \ \alpha^{2j - 2} \ |Q|^{2j} & : s = 1 \\
  & \\ \hline \\
 |Q|^2 + \frac{ - 2 \Gamma(- 2s) \ \cos(\pi s)}{C_s} \ \alpha^{2s - 2} \ |Q|^{2s} & \\
+ \frac{2}{C_s} \  \sum_{j = 2}^{\infty} \frac{ \zeta(1 + 2s - 2j) }{ (2j)! } \ (-1)^{j + 1} \ \alpha^{2j - 2} \ |Q|^{2j} & : 1 < s < \infty, \  s \notin \mathbb{N}  \\
& \\ \hline \\
|Q|^2 + \frac{ 2 (-1)^s }{C_s \ (2s)!}  \  \left[ - \left( \sum_{j = 1}^{2s} \frac{1}{j} \right) + \log(\alpha) + \log( |Q|)    \right] \ \alpha^{2s - 2} \ |Q|^{2s} & \\
  + \frac{2}{C_s} \ \sum_{ \substack{ j = 2 \\ j \neq s} }^{\infty} \frac{ \zeta(1 + 2s - 2j) }{ (2j)! } \ (-1)^{j + 1} \ \alpha^{2j - 2} \ |Q|^{2j} & : 1 < s < \infty, \ s \in \mathbb{N}  \\
  & \\ \hline \\
  |Q|^2 + \frac{2}{  C_s  } \ \sum_{k = 1}^{\infty} \left( \sum_{ m = 1}^{\infty} \ m^{2k + 2} \ e^{- \gamma m} \right) \frac{ (-1)^k \ \alpha^{2k}    \ |Q|^{2k + 2} }{ (2k + 2)!} & : s = \infty
  \end{array} \right. \label{eqn:formalexpansionsnonlocal2}
\end{align}
\end{proposition}

\noindent {\bf Proof of Proposition \ref{prop:expansionsM}:} We follow the approach in Appendix \ref{appendix:asymptotics} and Proposition \ref{prop:symbolconvnonlocal1}.

Let $q = Q \alpha \in \mathcal{B} = [- \pi, \pi]$. Recall that $M^s(q) = \frac{1}{C_s} \sum_{ m = 1}^{\infty} \frac{4 \sin^2( q m / 2)}{m^{1 + 2s}}$ for $0 < s < \infty$ and $M^s(q) = \frac{1}{C_\infty} \sum_{ m = 1}^{\infty} 4 e^{- \gamma m}  \sin^2( q m / 2), \ \gamma > 0$, for $s = \infty$. Here, $C_s > 0$ is defined in \eqref{LC-def}. Also recall that $\kappa_s(\alpha)$ is defined in \eqref{eqn:kappa-def}.

{\it Case 1: Let $1 + 2s \notin \mathbb{N}, \ s \neq \infty$}. Expand $
4 \sin^2( q m / 2) = 2 - e^{i q m} - e^{- i qm} , 
$
and note that since $|q| \leq \pi < 2 \pi$, we may apply Lemma \ref{lemma:polylog1} to obtain \eqref{eqn:expansionone}:
\begin{align}
C_s \ M^s(q) & =    - 2 \Gamma(- 2s) \ \cos(\pi s) \  |q|^{2s} + 2 \sum_{j = 1}^{\infty} \frac{ \zeta(1 + 2s - 2j) }{ (2j)! } \ (-1)^{j + 1} \ |q|^{2j} .
\end{align}
Applying the rescaling $q = Q \alpha$ gives 
\begin{align}
M^s_{\alpha}(Q) = \frac{1}{  \kappa_s(\alpha)} M^s(Q \alpha) = - \frac{2 \Gamma(- 2s) \ \cos(\pi s)}{C_s \ \kappa_s(\alpha)} \ \alpha^{2s} \ |Q|^{2s} +  \ \sum_{j = 1}^{\infty} \frac{ \zeta(1 + 2s - 2j) }{ C_s \ \kappa_s(\alpha) \ (2j)! } \ (-1)^{j + 1} \alpha^{2j} \ |Q|^{2j} . \label{eqn:notintexpand}
\end{align}
The series is absolutely convergent on $Q = q/\alpha \in \left[ - \frac{\pi}{\alpha}, \frac{\pi}{\alpha} \right]$ due to Proposition \ref{prop:zeta}. To obtain \eqref{eqn:formalexpansionsnonlocal2} for $1 + 2s \notin \mathbb{N}, s \neq \infty$, recall that $C_s = - 2 \Gamma(- 2s) \ \cos(\pi s)$, $\kappa_s(\alpha) = \alpha^{2s}$ for $s < 1$ and $C_s = \zeta(2s - 1), \quad \kappa_s(\alpha) = \alpha^2$ for $s > 1$. 

{\it Case 2: Let $1 + 2s \in \mathbb{N}, \ s \neq \infty$}. Again expand $
4 \sin^2( q m / 2) = 2 - e^{i q m} - e^{- i qm} , 
$ and note that since $|q| \leq \pi < 2 \pi$, we may apply Lemma \ref{lemma:polylog2} to obtain \eqref{eqn:expansiontwo}:
\begin{align}
M^s(q) & =      \frac{ 2 \cos(\pi s) }{(2s)!} \left[ - \left( \sum_{j = 1}^{2s} \frac{1}{j} \right) + \log(|q|)    \right] \ |q|^{2s} + \frac{\pi \  \cos \left( \pi(s - 1/2) \right)}{(2s)!} \ |q|^{2s} \nonumber \\
& \hspace{2.3cm} + 2 \sum_{ \substack{ j = 1 \\ j \neq s} }^{\infty} \frac{ \zeta(1 + 2s - 2j) }{ (2j)! } \ (-1)^{j + 1} \ |q|^{2j} .
\end{align}
Applying the rescaling $q = Q \alpha$ gives 
\begin{align}
M^s_{\alpha}(Q) = \frac{1}{  \kappa_s(\alpha)} M^s(Q \alpha) = & \  \frac{ 2 \cos(\pi s) }{C_s \ \kappa_s(\alpha) \ (2s)!} \left[ - \left( \sum_{j = 1}^{2s} \frac{1}{j} \right) + \log(\alpha) + \log(|Q|)    \right] \  \alpha^{2s} \ |Q|^{2s} \nonumber \\
& + \frac{\pi \  \cos \left( \pi(s - 1/2) \right)}{C_s \ \kappa_s(\alpha) \ (2s)!} \alpha^{2s} \ |Q|^{2s}  
 + 2 \sum_{ \substack{ j = 1 \\ j \neq s} }^{\infty} \frac{ \zeta(1 + 2s - 2j) }{C_s \ \kappa_s(\alpha) \  (2j)! } \ (-1)^{j + 1} \alpha^{2j} \ |Q|^{2j} . \label{eqn:intexpand}
 \end{align}
The infinite series is absolutely convergent on $Q = q/\alpha \in \left[ - \frac{\pi}{\alpha}, \frac{\pi}{\alpha} \right]$ due to Proposition \ref{prop:zeta}. To obtain \eqref{eqn:formalexpansionsnonlocal2} for for $1 + 2s \in \mathbb{N}, s \neq \infty$, recall that $C_s = - 2 \Gamma(- 2s)  \cos(\pi s)$, $\kappa_s(\alpha) = \alpha^{2s}$ for $s = 1/2$ and $C_s = \zeta(2s - 1), \quad \kappa_s(\alpha) = \alpha^2$ for $s > 1$. For $s = 1$, recall that $C_s = 1$ and $\kappa_s(\alpha) = (- \log(\alpha)) \ \alpha^2$. We now repeat Remark \ref{remark:euler} here.

\begin{remark} Note that one of the first two terms in \eqref{eqn:intexpand} will be zero, depending on whether $s \in \mathbb{N}$ or $s = (2k + 1)/2$ for $k \in \mathbb{N}$. 

Furthermore, in the case where $s = (2k + 1)/2$ for $k \in \mathbb{N}$, the series in \eqref{eqn:intexpand} is finite. To see this, observe from Proposition \ref{prop:zeta} that the ``trivial zeros" of the zeta function on the real line lie occur at negative even integers. In the series above, this occurs when
$
1 + 2s - 2j  \leq -2 \quad \Longrightarrow \quad j \geq s + 3/2. $

We also note that for $s = (2k + 1)/2$ for $ k \in \mathbb{N}$, the expansion \eqref{eqn:intexpand} is compatible with the expansion \eqref{eqn:notintexpand} where we assumed that $1 + 2s \notin \mathbb{N}$. To see this we use Euler's reflection formula from Proposition \ref{prop:Gamma} to obtain, for $k \in \mathbb{N}$, equation \eqref{eqn:eulerremark}:
\begin{align}
\underset{s \to \frac{2k - 1}{2} }{\lim}  \ \ - 2 \ \Gamma(- 2s) \ \cos(\pi s)  = \frac{ \pi (-1)^{s - 1/2} }{ (2s)! } \bigg|_{s = \frac{2k - 1}{2}}.
\end{align}

\end{remark}

 {\it Case 3: Finally, let $s = \infty$}. We have  $M_{\alpha}^s(q) = \frac{1}{C_\infty \ \alpha^2} \sum_{ m = 1}^{\infty} 4 e^{- \gamma m}  \sin^2( Q m  \alpha/ 2), \ \gamma > 0$, $\kappa_s(\alpha) = \alpha^2$, and $Q \in [- \frac{\pi}{\alpha}, \frac{\pi}{\alpha}]$. Taylor expansion about $Q \alpha = 0$ gives
 \begin{align}
\frac{4 \sin^2 ( Q m \alpha / 2)}{\alpha^2} =  2 \sum_{k = 0}^{\infty} \frac{ (-1)^k \ \alpha^{2k} \ m^{2k + 2} \ |Q|^{2k + 2} }{ (2k + 2)!} . 
 \end{align}
 In turn, 
\begin{align}
M^s_{\alpha}(Q) = \frac{1}{C_s} \ \left( \sum_{m = 1}^{\infty} \ m^2 \ e^{- \gamma m} \right)  \ |Q|^2   + \frac{2}{  C_s } \ \sum_{k = 1}^{\infty} \left( \sum_{ m = 1}^{\infty} \ m^{2k + 2} \ e^{- \gamma m} \right) \frac{ (-1)^k \ \alpha^{2k}    \ |Q|^{2k + 2} }{ (2k + 2)!} . 
\end{align}
Observe that
\begin{align}
& \left( \sum_{m = 1}^{\infty} \ m^2 \ e^{- \gamma m} \right) =  \frac{ {\rm exp}(\gamma) (  {\rm exp}(\gamma) + 1) }{ ({\rm exp}(\gamma) - 1)^3} = C_{\infty}. 
\end{align}
This gives \eqref{eqn:formalexpansionsnonlocal2} for $s = \infty$. This completes the proof of Proposition \ref{prop:expansionsM}.   $\Box$

Consider the equation \eqref{eqn:Phieqnnonlocal}, $\mathcal{D}^{\sigma, \alpha} \left[ \widehat{\Phi} \right] = 0$,  for the rescaled $s$-nonlocal DNLS solitary wave $\widehat{\Phi}(Q)$ supported on $\mathcal{B}_{_{\alpha}}$. Consider the case where $s \notin \mathcal{N}$, $s = b_1/b_2$ where $b_1, b_2 \in \mathbb{N}$. By Lemma \ref{lemma:expconvononlocal} and Proposition \ref{prop:Phinonlocal},  we have
\begin{align}
\left\| R_1^{\sigma} [ \widehat{\Phi^{\sigma}} ]  \right\|_{L^{2,a}(\mathbb{R})} \lesssim \mathcal{O}(\alpha^{\infty}). 
\end{align}
Furthermore, we have

\begin{lemma} \label{lemma:Mtau}
Let $s \notin \mathbb{N}$ and $s = b_1/b_2$ reduced rational for $b_1,b_2 \in \mathbb{N}$. Then for $\tau = 2/b_2$ and $Q \in \mathcal{B}_{_{\alpha}}$, there exists a sequence $M_k \in \mathbb{R}, \quad k = 1, 2, \dots$, such that $\{ M_k \}_{k \in \mathbb{N}} \in l^1(\mathbb{N})$ and
\begin{align}
M_{\alpha}^s(Q) = |Q|^{2p} + \sum_{ k = 1}^{\infty} M_k \  \alpha^{j \tau} \  |Q|^{j \tau + 2p} . 
\end{align}
\end{lemma}

\noindent {\bf Proof of Lemma \ref{lemma:Mtau}:} First consider the case $1/4 < s < 1$. Then $p = \min(1,s) = s$ and by \eqref{eqn:formalexpansionsnonlocal2} with $\tau = 2/b_2$, 
\begin{align}
M_{\alpha}^s(Q)  & = |Q|^{2s} +  \frac{2}{C_s} \ \sum_{j = 1}^{\infty} \frac{ \zeta(1 + 2s - 2j) }{ (2j)! } \ (-1)^{j + 1} \ \alpha^{2j - 2s} \ |Q|^{2j-2s} \ |Q|^{2s} \nonumber \\
& = |Q|^{2s} +  \frac{2}{C_s} \ \sum_{j = 1}^{\infty} \frac{ \zeta(1 + 2s - 2j) }{ (2j)! } \ (-1)^{j + 1} \ \alpha^{\tau (b_2 j - b_1) } \ |Q|^{\tau( b_2 j - b_1)} \ |Q|^{2s} . 
\end{align} 
Taking $M_k \equiv \frac{ 2 \ \zeta(1 + 2s - 2j) }{ C_s \ (2j)! } \ (-1)^{j + 1}$ if $k = b_2 j - b_1$ and $M_k \equiv 0$ otherwise completes the proof for $1/4 < s < 1$.

Next, consider $1 < s < \infty$, $s \notin \mathbb{N}$. Then $p = \min(1,s) = 1$ and by \eqref{eqn:formalexpansionsnonlocal2} with $\tau = 2/b_2$, 
\begin{align}
M_{\alpha}^s(Q)   = & \  |Q|^2 + \frac{2}{C_s} \ \bigg(   -  \Gamma(- 2s) \ \cos(\pi s) \ \alpha^{2s - 2} \ |Q|^{2s - 2}  \nonumber \\
& \hspace{2cm} +    \sum_{j = 2}^{\infty} \frac{ \zeta(1 + 2s - 2j) }{ (2j)! } \ (-1)^{j + 1} \ \alpha^{2j - 2} \ |Q|^{2j - 2} \bigg) \ |Q|^2 \nonumber \\
= & \  |Q|^2 + \frac{2}{C_s} \ \bigg(  -  \Gamma(- 2s) \ \cos(\pi s) \ \alpha^{\tau (b_1 - b_2)} \ |Q|^{\tau (b_1 - b_2)}  \nonumber \\
& \hspace{2cm} +    \sum_{j = 2}^{\infty} \frac{ \zeta(1 + 2s - 2j) }{ (2j)! } \ (-1)^{j + 1} \ \alpha^{ \tau b_2 (j - 1)} \ |Q|^{ \tau b_2 (j - 1) } \bigg) \ |Q|^2 . 
\end{align} 
Taking $M_k = \frac{ - 2 \Gamma(- 2s) \ \cos(\pi s)}{C_s}$ if $k = b_1 - b_2$, $M_k \equiv \frac{ 2 \ \zeta(1 + 2s - 2j) }{ C_s \ (2j)! } \ (-1)^{j + 1}$ if $k = b_2 (j - 1)$ and $M_k \equiv 0$ otherwise completes the proof of Lemma \ref{lemma:Mtau} for $1 < s < \infty$, $s \notin \mathbb{N}$.
 $\Box$ \\

Using the above expansion of $M_{\alpha}^s(Q)$, we construct sequence of solutions
 to \eqref{eqn:Phieqnnonlocal}, which to arbitrary order $(\alpha^\tau)^J$, can be approximated by a sums of the form: 
\begin{align}
S_J^{\sigma}= \chi_{_{\mathcal{B}_{_{\alpha}}}}(Q) \  F_0(Q) + \sum_{ j = 1}^{J}  \chi_{_{\mathcal{B}_{_{\alpha}}}}(Q) \  \alpha^{j \tau} \ F_j (Q), \qquad J \in \mathbb{N}, \label{eqn:truncS}
\end{align}
with decreasing residuals:
 $\left\| \mathcal{D}^{\alpha,\sigma}\left[S_J^{\sigma}\right]\ \right\|_{L^{2,a}}=\mathcal{O} \left( \alpha^{\tau (J + 1)} \right)$
  and higher order corrector: 
\medskip

\begin{theorem}\label{th:SJalpha-error}
Let $s \notin \mathbb{N}$ and $s = b_1/b_2$ reduced rational for $b_1,b_2 \in \mathbb{N}$. Then there exists, for any $J \geq 1$, $S_J^{\sigma}$ given by \eqref{eqn:truncS}, such that $\left\| \mathcal{D}^{\alpha,\sigma}\left[S_J^{\sigma}\right]\ \right\|_{L^{2,a}}=\mathcal{O} \left( \alpha^{\tau (J + 1)} \right)$. Furthermore, $\widehat{\Phi^{\sigma, \alpha}}$, the solution of \eqref{eqn:Phieqnnonlocal}, is given by $\widehat{\Phi^{\sigma, \alpha}} = S_J^{\sigma} + \widehat{E^{\sigma, \alpha}_J}$, where $\| \widehat{E^{\sigma, \alpha}_J} \|_{L^{2,a}(\mathbb{R})} \lesssim \alpha^{2J + 2}$. 
\end{theorem}
\medskip

We next explain the construction of the sequence $F_j(Q),\ j\ge0$.
 The proof of Theorem \ref{th:SJalpha-error} uses a Lyapunov-Schmidt  strategy analogous to that used in the proof of Theorem \ref{th:leadingordernonlocal}. 
The steps in the proof  are analogous to those of the nearest neighbor case; see \cite{Jenkinson_Weinstein_2015}. 
\\
\medskip

To construct the sequence $F_j$, we use Lemma \ref{lemma:Mtau} and consider the related equation
\begin{align}
 \left[ 1 + M_{\alpha}^s( Q) \right]\ F( Q ) - \ \frac{1}{4 \pi^2}
\left(F *F *F\right)( Q) = 0\ . \label{eqn:continuumF}
\end{align}
 Using the power series expansion from Lemma \ref{lemma:Mtau}, we may construct  a formal power series expansion in $\alpha^{\tau}$ for  $F^\alpha(Q)$:
\begin{align}
& F^\alpha(Q)= F_0(Q) + \sum_{ j = 1}^{\infty} \alpha^{j \tau} \ F_j (Q), \label{eqn:2series}
\end{align}

Substituting \eqref{eqn:2series} into \eqref{eqn:continuumF}, we obtain a hierarchy of equations for $F_j(Q)$ at each order of $\alpha^{\tau j}, \quad j \geq 0$. Each term $F_j(Q)$ will have support on all $\mathbb{R}$ and can be shown to decay exponentially as  $|Q|\to\infty$.
The deviation of \eqref{eqn:continuumF} from the equation \eqref{eqn:Phieqnnonlocal}, $\mathcal{D}^{\sigma, \alpha} \left[ \widehat{\Phi} \right] = 0$,  for the rescaled $s$-nonlocal DNLS solitary wave $\widehat{\Phi}(Q)$ are terms of the form:
\begin{equation}
- \left(1-{\chi}_{_{\mathcal{B}_{_{\alpha}}}}(Q) \right) \left[ 1 + M_{\alpha}^s(Q) \right] F(Q) + \frac{1}{4 \pi^2} \ \left(1-{\chi}_{_{\mathcal{B}_{_{\alpha}}}}(Q) \right) (F^\alpha*F^\alpha*F^\alpha)(Q) + R^{\alpha,\sigma}_1[F^\alpha](Q) ,
\label{F-alpha}
\end{equation}
whose norms at each order in $\alpha^{\tau j}$ can be shown to be beyond all polynomial orders in $\alpha$
 as $\alpha\to0$, {\it i.e.} $\mathcal{O}(\alpha^m)$, for all $m\ge1$,  in $L^{2,a}(\mathbb{R};dQ)$ with $ a>1/2$. Therefore, we expect that if 
 $ \widehat{\Phi^{\alpha,\sigma}}(Q)$ is a solution of \eqref{eqn:Phieqnnonlocal}, then
the function  $ {\chi}_{_{\mathcal{B}_{_{\alpha}}}}(Q) \ F^\alpha(Q)$, where 
 $F^\alpha$ solves \eqref{eqn:continuumF}, formally solves  \eqref{eqn:Phieqnnonlocal} with an error which is beyond all polynomial orders in $\alpha^2$, {\it i.e.}
 \begin{equation}
\left\| \mathcal{D}^{\alpha,\sigma}\left[\ {\chi}_{_{\mathcal{B}_{_{\alpha}}}} \ F^\alpha\ \right]\ \right\|_{L^{2,a}(\mathbb{R};dQ)}
\ =\  \mathcal{O}(\alpha^\infty)
\label{sol-beyond-all}
\end{equation}

We proceed by outlining the derivation of the formal asymptotic expansion. Substituting of \eqref{eqn:2series} into \eqref{F-alpha}, we obtain a hierarchy of equations for $F_j$.
\medskip

\begin{flalign}
 & \mathcal{O}(\alpha^{0}) {\bf \ equation:} \hspace{1.5cm} \left[ 1 + | Q |^{2p} \right] \ F_0(Q) -  \left( \frac{1}{2 \pi} \right)^2 \ F_0 * F_0 * F_0  (Q) = 0. & \label{eqn:order0}
\end{flalign}
Equation \eqref{eqn:order0} is the Fourier transform of continuum FNLS \eqref{eqn:introtiNLS}. Denote by
\begin{align}
F_0(Q) = \widetilde{\psi}(Q) = \mathcal{F}_{_C} \left[ \psi \right](Q),
\end{align}
where $\psi(x)$ is the unique (up to translation)
 \emph{positive} and decaying solution of NLS. $\psi(x)$ is real-valued and radially symmetric about some point, which we take to be $x=0$. By Proposition \ref{prop:Psinonlocal}  there exists $C_0>0$ such that for $\eta = \min(1,2s)$,
 \begin{equation}
 e^{C_0 | Q|^{\eta} } \ \widetilde{\psi}(Q) \in L^{2,a}(\mathbb{R}).
 \label{C0-def}\end{equation}
 
 At each order in $\alpha^2$, we shall derive an equation for $F_j(Q)$ of the general form:
 \begin{align}
\widetilde{L_+}  \ F^{\sharp}(Q) = F^{\flat}(Q).\label{Fsharp}
\end{align}
It is important for us to understand how decay properties $F^{\flat}(Q)$ propagate to the solution 
$F^{\sharp}(Q)$.
\begin{proposition}  \label{prop:lplusinvert}
Fix $a > 1/2$ and $\eta = \min(1,2s)$. Suppose that $F^{\flat}(Q) \in L_{_{\rm even}}^{2,a}(\mathbb{R};dQ)$ and that there exists a constant $C_\flat > 0$ such that $e^{C_\flat |Q|^{\eta}} F^{\flat}(Q) \in L^{2,a}(\mathbb{R};dQ)$. Then there exists a solution of \eqref{Fsharp},  $F^{ \sharp}(Q) \in L_{_{\rm even}}^{2,a}(\mathbb{R};dQ)$.
Furthermore, we have $e^{C_\flat |Q|^{\eta}} F^{\sharp}(Q) \in L^{2,a}(\mathbb{R};dQ)$.
\end{proposition}
\medskip

 Since $F^{\flat}(Q)$ is even it is $L^2(\mathbb{R};dQ)$ orthogonal to the kernel of $\widetilde{L_+} ={\rm span}\{Q\widetilde{\psi}(Q)\}$. Therefore, $F^{\sharp}=\left( \widetilde{L_+} \right)^{-1} F^\flat\in
L_{_{\rm even}}^{2,a+2p}(\mathbb{R};dQ)$;
  see  Proposition \ref{prop:Lplusnonlocal}.  A detailed proof that the exponential decay rate is preserved is given for the case of nearest-neighbor DNLS in  \cite{Jenkinson_2015}. The idea is to break $F^{\sharp}$ into its  low ($|Q|\le\epsilon^{-1}$) and high  ($|Q|\ge\epsilon^{-1}$) frequency components, $F_{{\rm lo},\epsilon}$ and  $F_{{\rm hi},\epsilon}$. The norm $\|e^{C_\flat|Q|^{\eta}}F_{{\rm lo},\epsilon}\|_{L^{2,a}}$ is bounded in terms of $\|F^\sharp\|_{L^{2,a}}$. The norm $\|e^{C_\flat|Q| ^{\eta} }F_{{\rm hi},\epsilon}\|_{L^{2,a}}$ is controlled  by a boot-strap argument using that $\chi(|Q|\ge\epsilon^{-1})(1+|Q|^{2p})^{-1}\ F_0*F_0*F_{{\rm hi},\epsilon}$ has  $L^{2,a}-$ norm which is bounded by $\sim \epsilon^{2p} \ \|e^{C_\flat|Q|^{\eta} }F_{{\rm hi},\epsilon}\|_{L^{2,a}}$.
\medskip

We now turn to the hierarchy of equations  at order $\alpha^{\tau j}$, beginning with $j=1$.
 We find
\begin{align}
& \left[ 1 + | Q |^{2p} \right] \ F_1(Q) -   \ \frac{3}{4 \pi^2}\ \widetilde{\psi} * \widetilde{\psi} * F_1  (Q) = - M_1 \ | Q |^{2p + \tau} \ \widetilde{\psi}(Q),
\end{align}
or
\begin{flalign}
 & \mathcal{O}(\alpha^{\tau}) {\bf \ equation:} \hspace{3cm} \widetilde{L_+} \ F_1(Q) = - M_1 \ | Q |^{2p + \tau} \ \widetilde{\psi}(Q).  &  \label{eqn:order1}
\end{flalign}
Here, $L_+= (-  \Delta_x)^p+1-3\psi^2(x)$, is the linearization of the continuum NLS operator about $\psi(x)$. 
By Proposition \ref{prop:lplusinvert}, 
$
F_1(Q) = - \left( \widetilde{L_+} \right)^{-1} \ \left( M_1 \ | Q |^{2p + \tau} \ \widetilde{\psi}(Q) \right) \in L_{_{\rm even}}^{2,a}(\mathbb{R}).$ Since \eqref{eqn:order1} has real-valued forcing, $F_1$ is real-valued. 
Let $C_\flat=3C_0/4$ and note $\|e^{C_\flat|Q|^{\eta}}|Q|^{2p + \tau} \ \widetilde{\psi}\|_{L^{2,a}}\lesssim  
\|e^{C_0 |Q|^{\eta}}\widetilde{\psi}\|_{L^{2,a}}$
Therefore,  $e^{\frac{3C_0}{4}|Q|^{\eta}} F_1(Q)\in L^{2,a}(\mathbb{R};dQ)$ for $a>1/2$.

\medskip
We now proceed to  inductively construct and bound the  sequence $F_j(Q),\ j\ge1$ using Proposition \ref{prop:lplusinvert} and the following two lemmata, which are proved in detail in \cite{Jenkinson_2015} for the case $\eta = 1$. The generalization follows the same proof.

\begin{lemma} \label{lemma:expoconvogen}
Fix $a > 1/2$ and $\eta= \min(1,2s)$. Suppose that  $\widetilde{f_1}, \widetilde{f_2} \in L_{_{\rm even}}^{2,a}(\mathbb{R})$. Then $\widetilde{f_1} * \widetilde{f_2} \in L_{_{\rm even}}^{2,a}(\mathbb{R} )$. Suppose further that there exist $c_1 , c_2 > 0$ such that $e^{c_1 |Q|^{\eta} } \widetilde{f_1}(Q) \in L^{2,a}(\mathbb{R})$, $e^{c_2  |Q|^{\eta}} \widetilde{f_2}(Q) \in L^{2,a}(\mathbb{R}_Q)$. Then for $c_3 = \min(c_1, c_2)$, we have $e^{c_3 |Q|^{\eta}} \ \widetilde{f_1} * \widetilde{f_2}(Q) \in L_{_{\rm even}}^{2,a}(\mathbb{R}_Q)$ and 
\begin{align}
\left\| \ e^{c_3 |Q|^{\eta}} \ (\widetilde{f_1} * \widetilde{f_2}) (Q) \ \right\|_{L^{2,a}(\mathbb{R}_Q)} \lesssim \left\| \ e^{c_3 |Q|^{\eta}} \ \widetilde{f_1}(Q)  \ \right\|_{L^{2,a}(\mathbb{R}_Q)} \left\| \ e^{c_3 |Q|^{\eta}}  \   \widetilde{f_2} (Q) \ \right\|_{L^{2,a}(\mathbb{R}_Q)}.
\end{align}  
\end{lemma} 

Lemma \ref{lemma:expoconvogen} is a direct consequence of Lemma \ref{lemma:Ha-algebra} and appropriately distributing the exponential weights, and $c_3 |Q|^{\eta} - c_1 |Q - \xi |^{\eta} - c_2 |\xi|^{\eta} \leq 0$, which follows from Lemma \ref{lemma:tri}. 
\medskip

\begin{lemma} \label{lemma:polyexpo}
Fix $a > 1/2$, $\eta = \min(1,2s)$, and $k \in \mathbb{N}$. Suppose that  $\widetilde{f} \in L_{_{\rm even}}^{2,a}(\mathbb{R})$ and that there exists $c_1>0$ such that $e^{c_1 |Q|^{\eta}} \widetilde{f}(Q) \in L^{2,a}(\mathbb{R})$. Then
$ | Q |^{2k} \ \widetilde{f} \in L_{_{\rm even}}^{2,a}(\mathbb{R})$ and for any $0 < c_2 < c_1$, we have
\begin{align}
\left\| e^{c_2 |Q|^{\eta} } |Q|^{2k} \widetilde{f} \right\|_{L^{2,a}(\mathbb{R})} \leq e^{- \frac{2 k }{\eta}} \ \left[ \frac{ 2k }{ \eta (c_1 - c_2 ) } \right]^{2k / \eta} \  \left\| \ e^{c_1 |Q|^{\eta}}  \   \widetilde{f} (Q) \ \right\|_{L^{2,a}(\mathbb{R}_Q)}. 
\end{align}\end{lemma} \\
Lemma \ref{lemma:polyexpo} follows from $ e^{c_2 |Q|^{\eta} } |Q|^{2k} | \widetilde{f}(Q) | \leq  e^{- \frac{2 k }{\eta}} \ \left[ \frac{ 2k }{ \eta (c_1 - c_2 ) } \right]^{2k / \eta} \cdot e^{c_1 |Q|^{\eta} } | \widetilde{f}(Q) |$ and then taking the $L^{2,a}$ norm. 
 
\medskip

\begin{proposition} \label{prop:orderj}
Let $ j \geq 1$. The equation for $F_j$ at order $\mathcal{O}(\alpha^{\tau j})$, independent of $\alpha$ and $\sigma$, is given by
\begin{flalign}
 & \mathcal{O}(\alpha^{\tau j}) {\bf \ equation:} &  \widetilde{L_+} \ F_j(Q) & = - \sum_{k = 0}^{j-1} M_{j - k} \ | Q |^{ (j - k) \tau + 2p} \ F_k (Q) , & \nonumber \\
& & & + \frac{1}{( 2 \pi)^2} \sum_{ \substack{ k + l + z = j \\ 0 \leq k, l, z < j  }} \ F_k * F_l * F_z (Q) \equiv H_j \left[F_0, \dots, F_{j-1} \right](Q) , & \label{eqn:orderj}
 \end{flalign}
 and has the unique solution 
 \begin{align}
 F_j & = \left( \widetilde{L_+} \right)^{-1} \bigg( H_j \left[ F_0 , \dots, F_{j-1} \right] \bigg) \in L_{_{\rm even}}^{2,a}(\mathbb{R};dQ). 
 \end{align}
 Furthermore, $F_j$ is real-valued and $e^{C_j |Q|^{\eta} } \ F_j(Q) \in L^{2,a}(\mathbb{R};dQ)$, where
 $C_j \equiv C_0 \left( \frac{1}{2} + \frac{1}{2^{j+1}} \right) \geq \frac{C_0}{2}$ and $C_0>0$ is as in \eqref{C0-def}.   \\
\end{proposition}

\noindent {\bf Proof of Proposition \ref{prop:orderj}:} We induct to solve at each order in $\alpha^{\tau j}$. Let $F_0(Q) \equiv \widetilde{\psi}(Q)$, which solves \eqref{eqn:order0}, is real-valued, and satisfies $e^{C_0 |Q|^{\eta}} F_0(Q) \in L^{2,a}(\mathbb{R};dQ)$. Fix $m \geq2$ and assume that for $1 \leq j \leq m - 1$, $F_j(Q) \in L_{_{\rm even}}^{2,a}(\mathbb{R})$ satisfies \eqref{eqn:orderj} and is real-valued. Furthermore, assume that
\begin{align}
e^{C_j |Q|^{\eta} } F_j(Q) \in L^{2,a}(\mathbb{R};dQ), \qquad C_j \equiv C_0 \left( \frac{1}{2} + \frac{1}{2^{j+1}} \right) \geq \frac{C_0}{2}. \label{j-bounds}
\end{align}
We have already proven above that these inductive hypotheses hold for $ j = 1$. We expand
\begin{align}
M_{\alpha}(Q)  = |Q|^{2p} + \sum_{j = 1}^{\infty} \alpha^{\tau j} \  M_j \ |Q|^{\tau j + 2 p}, \hspace{2cm}  F^\alpha(Q)=   \sum_{ j = 0}^{\infty} \alpha^{\tau j} \ F_j (Q),
\end{align}
and substitute into \eqref{eqn:continuumF}. Using \eqref{eqn:order0} for $F_0(Q) \equiv \widetilde{\psi}(Q)$ 
 we obtain 

\begin{align}
& \sum_{j = 1}^{\infty} \ \alpha^{\tau j} \ \widetilde{L_+} \ F_j(Q) 
 = - \sum_{j = 1}^{\infty} \sum_{k = 0}^{\infty}   \alpha^{\tau j + \tau k} \  M_j \ |Q|^{\tau j + 2p } \  F_k(Q) + \sum_{j = 1}^{\infty} \frac{\alpha^{2j}}{( 2 \pi)^2}  \ \sum_{ \substack{ k + l + z = j \\ 0 \leq k, l, z < j  }} \ F_k * F_l * F_m (Q)  \nonumber \\
& = \sum_{j = 1}^{\infty} \alpha^{2j} \bigg( -  \sum_{k = 0}^{j-1} M_{j - k} \  | Q |^{\tau j - \tau k + 2p}  \ F_k (Q)  + \frac{1}{ 4 \pi^2 }  \ \sum_{ \substack{ k + l + z = j \\ 0 \leq k, l, z < j  }} \ F_k * F_l * F_z (Q) \bigg). \label{eqn:derive2}
\end{align}
Applying the inductive hypothesis \eqref{eqn:orderj}  for $1 \leq j \leq m - 1$ and dividing by $\alpha^{\tau m}$, \eqref{eqn:derive2} becomes
\begin{align}
&  \widetilde{L_+} \ F_m(Q)  + \alpha^{\tau} \ \bigg[  \sum_{j = m + 1}^{\infty} \ \alpha^{\tau j - \tau (m + 1)} \ \widetilde{L_+} \ F_j(Q) \bigg]
\nonumber \\
& = - \sum_{k = 0}^{m - 1} M_{m - k} \  | Q |^{\tau (m - k) + 2p} \ F_k (Q)   + \frac{1}{( 2 \pi)^2} \sum_{ \substack{ k + l + z = m \\ 0 \leq k, l, z < m  }} \ F_k * F_l * F_z (Q)    \nonumber \\
&  \ \ +  \alpha^{\tau} \ \bigg[  \sum_{j = m + 1}^{\infty} \alpha^{ \tau j - \tau (m+1)} \bigg( - \sum_{k = 0}^{j-1} \ M_{j - k} \  | Q |^{\tau (j - k) + 2p}  \ F_k (Q)  + \frac{1}{ 4 \pi^2 }  \ \sum_{ \substack{ k + l + z = j \\ 0 \leq k, l, z < j  }} \ F_k * F_l * F_z (Q) \bigg) \bigg]. \label{eqn:derive3}
\end{align}
Since $\tau j - \tau(m+1) \geq 0$ for $j \geq m + 1$, the bracketed terms with coefficient $\alpha^{\tau}$ are $\mathcal{O}(\alpha^{\tau})$. Therefore the terms of order precisely $\alpha^{\tau m}$ are given by\eqref{eqn:orderj}. This establishes the case: $ j = m$.

\medskip

We now prove that \eqref{eqn:orderj} has a solution, $F_m$ satisfying \eqref{j-bounds} with $j=m$.  
 First, applying Lemmata \ref{lemma:expoconvogen} and \ref{lemma:polyexpo} to the right hand side  of \eqref{eqn:orderj} for $j=m$, $H_m$,  we have that $H_m\in L_{_{\rm even}}^{2,a}$ with the bound:

\begin{align}
\left\| \ e^{C_m |Q|^{\eta} } \ H_m \left[  F_0 , \dots, F_{m-1} \right] (Q)  \ \right\|_{L^{2,a}(\mathbb{R}; dQ)} \lesssim  \lambda_m,\ \ 
\end{align}
where {\small{ $C_m=C_0 \left( \frac{1}{2} + \frac{1}{2^{m+1}} \right)$ and 
\begin{align}
\lambda_m  \equiv \ & 2 \ \sum_{k = 0}^{m - 1} \ e^{- \frac{2 k }{\eta}} \ \left[ \frac{ 2k }{ \eta (c_1 - c_2 ) } \right]^{2k / \eta} \ \left\| \ e^{C_k |Q|^{\eta} } \ {F_k}(Q) \ \right\|_{L^{2,a}(\mathbb{R}_Q)} \nonumber \\
& + \frac{1}{( 2 \pi)^2} \sum_{ \substack{ k + l + z = m \\ 0 \leq k, l, z < m  }} \left\| \ e^{ C_k |Q |^{\eta} } \  {F_k}(Q)  \ \right\|_{L^{2,a}(\mathbb{R}_Q)} \ \left\| \ e^{ C_l |Q |^{\eta} } \  {F_l}(Q)    \ \right\|_{L^{2,a}(\mathbb{R}_Q)}  \ \left\| \ e^{ C_z |Q |^{\eta} } \  {F_z}(Q)  \ \right\|_{L^{2,a}(\mathbb{R}_Q)}. 
\end{align}
Proposition \ref{prop:lplusinvert} implies that there exists a unique solution $F_m(Q) \in L_{_{\rm even}}^{2,a}(\mathbb{R})$ to equation \eqref{eqn:orderj} with $e^{C_m |Q|^{\eta} } F_m(Q) \in L^{2,a}(\mathbb{R})$. Finally, to see that $F_m$ is real-valued, note that equation \eqref{eqn:orderj} for $F_m$ is a linear with inhomogeneous forcing on the right-hand-side given by $H_m \left[F_0, \dots, F_{m -1} \right]$, which is necessarily real-valued for $F_j$ real-valued, $ j = 0, \dots, m-1$. This completes the proof of Proposition \ref{prop:orderj}. 
$\Box$

\appendix

\section{A subadditivity inequality} \label{appendix:subadd}
%We work extensively in $L^{2,a}(\mathbb{R}^d)$ with $a > d/2$. Here,
%\begin{align}
%L^{2,a}(A) = \left\{ f \quad : \quad \left\| ( 1 + |\cdot|^2)^{a/2} f \right\|_{L^2(A)} =  \left( \int_A (1 + |q|^2)^a |f(q)|^2 dq \right)^{1/2} < \infty \right\}. 
%\end{align}
%
%\medskip
%
%\begin{lemma} \cite{Dohnal_Uecker_2009} \label{lemma:algebra}
% Let $a > d/2$ and $f_1, f_2 \in L^{2,a}(\mathbb{R}^d)$. Then
%\begin{align}
%\left\| f_1 * f_2 \right\|_{L^{2,a}(\mathbb{R}^d)} \lesssim \left\| f_1 \right\|_{L^{2,a}(\mathbb{R}^d)} \ \left\| f_2 \right\|_{L^{2,a}(\mathbb{R}^d)}. 
%\end{align}
%\end{lemma}
%
% 
%We require the following generalized triangle inequality on several occasions. 

\begin{lemma} \label{lemma:tri}
For $x, y \in \mathbb{R}$ and $0 < p \leq 1$, we have
$
| x + y |^p \leq  |x|^p + |y|^p.
$ 
\end{lemma}

\medskip
\noindent {\it Proof of Lemma \ref{lemma:tri}:} The result is trivial for $\eta = 1$. The result also holds trivially for $x = 0$ or $y = 0$, so we assume $x \neq 0$ and $y \neq 0$. First let $x > 0$ and $y < 0$ (or $x < 0$ and $y > 0$). Then
$ 
|x + y|^\eta \leq {\rm max} \left\{ |x|^\eta, |y|^\eta \right\} \leq |x|^\eta + |y|^\eta. 
$
Now let $x > 0$ and $y > 0$. Then
\begin{align}
|x + y|^\eta  = \frac{ |x + y|}{ |x + y|^{1 - \eta}} \leq \frac{ |x| + |y|}{ |x + y|^{1 - \eta}} = \left( \frac{ |x|}{|x + y|} \right)^{1 - \eta} \ |x|^\eta +  \left( \frac{ |y| }{|x + y|} \right)^{1 - \eta} \ |y|^\eta \leq |x|^\eta + |y|^\eta,
\end{align}
where we have used that
$
|x| |x + y|^{-1} \leq 1, \quad |y| |x + y|^{-1} \leq 1, 
$ for $x > 0$, $y > 0$.   $\Box$

\section{Properties of the Discrete Fourier Transform}
\label{appendix:DFT}
In this section, we prove some general properties of the discrete Fourier transform as defined in \eqref{eqn:DFT}.
 
\medskip

\begin{lemma} \label{lemma:discderivtransform}
For $u = \{ u_n \}_{n \in \mathbb{Z}}   \in l^1(\mathbb{Z}) \cap l^2(\mathbb{Z})$ and $v_n \equiv u_{m + n} - u_n$, 
$\widehat{v}(q)  = (e^{- i q m} - 1) \ \widehat{u}(q). 
$

\end{lemma}

\begin{lemma}
\label{lemma:producttransform}
For any two functions $u, v \in l^1(\mathbb{Z}) \cap l^2(\mathbb{Z})$, and with their product given by $u \cdot v = \{ u_n v_n \}_{n \in \mathbb{Z}}, $
$
\mathcal{F}_{_D} [ u \cdot v ](q) = \widehat{u v}(q) = (2 \pi)^{-1} \ \widehat{u} *_{_1} \widehat{v}(q).
$
where the periodic convolution $*_{_1}$ is defined in \eqref{eqn:periodicconvolution}.
\end{lemma}
%
%\medskip
%
%\begin{lemma}
%\label{lemma:exponentials}
%Let  $\widehat{u}$ and $ \widehat{v}$ be \ $L^1_{_{\rm loc}}(\mathbb{R})$ functions. Then if $C$ is any constant, we have
%\begin{align}
%\left( e^{i C (\cdot)} \widehat{u} \right) *_{_1} \left( e^{i C (\cdot)} \widehat{v} \right)(q) = e^{i C q} \left( \widehat{u} *_{_1} \widehat{v}\right)(q).
%\end{align}
%\end{lemma}

\medskip

\begin{lemma}
\label{lemma:commutativity}
\begin{enumerate}
\item Assume $\widehat{u}, \widehat{v}\in L^1_{_{\rm loc}}(\mathbb{R})$. Then, for any constant $C$, 
$
\left( e^{i C (\cdot)} \widehat{u} \right) *_{_1} \left( e^{i C (\cdot)} \widehat{v} \right)(q) = e^{i C q} \left( \widehat{u} *_{_1} \widehat{v}\right)(q).
$
\item (Commutativity) For $2 \pi$-periodic functions $\widehat{u}\ ,\ \widehat{v} \in L^1_{_{\rm loc}}(\mathbb{R})$, we have
\begin{align}
\widehat{u} *_{_1} \widehat{v}(q) = \widehat{v} *_{_1} \widehat{u}(q).
\end{align}
\end{enumerate}
\end{lemma}

\begin{lemma}
\label{lemma:commutativity2}
For three functions $\widehat{u}, \widehat{v}, \widehat{w} \in L^1_{_{\rm loc}}(\mathcal{\mathbb{R}})$, we have
$
\widehat{u} *_{_1} \left[ \widehat{v} *_{_1} \widehat{w} \right] (q) = \widehat{v} *_{_1} \left[ \widehat{u} *_{_1} \widehat{w} \right](q).
$
Note that $\widehat{u}, \widehat{v}$, and $\widehat{w}$ need not be periodic.
\end{lemma}
%
%\noindent {\bf Proof of Lemma \ref{lemma:commutativity2}:} By a simple application of Fubini's theorem, we have
%\begin{align}
%\widehat{u} *_{_1} \left[ \widehat{v} *_{_1} \widehat{w} \right](q) = \int_{\mathcal{B}} \int_{\mathcal{B}} \widehat{u}(\xi) \widehat{v}(\zeta) \widehat{w}(q - \xi - \zeta) d\xi d\zeta \nonumber \\
%= \int_{\mathcal{B}} \int_{\mathcal{B}} \widehat{u}(\xi) \widehat{v}(\zeta) \widehat{w}(q - \xi - \zeta) d\zeta d\xi
%= \widehat{v} *_{_1} \left[ \widehat{u} *_{_1} \widehat{w} \right](q).\ \ \Box
%\end{align}
% \\

\begin{lemma}
\label{lemma:evenconvo}
Let $\widehat{u}$, $\widehat{v} \in L^1_{_{\rm loc}}(\mathbb{R}) $ be even such that $\widehat{u}(-q) = \widehat{u}(q)$ and $\widehat{v}(-q) = \widehat{v}(q)$ . Then $\widehat{u} *_{_{\alpha}} \widehat{v}$ is also even. That is, $\widehat{u} *_{_{\alpha}} \widehat{v}(-q) = \widehat{u} *_{_{\alpha}} \widehat{v}(q)$.
\end{lemma}
\medskip

The proof of Lemma \ref{lemma:evenconvo} also provides us with similar result for standard convolutions on the line.  
\medskip

\begin{corollary}
\label{cor:standardconvosym}
Let $\widehat{u}$, $\widehat{v} \in L^1_{_{\rm loc}}(\mathbb{R}) $ be even such that $\widehat{u}(-q) = \widehat{u}(q)$ and $\widehat{v}(-q) = \widehat{v}(q)$ . Then $\widehat{u} * \widehat{v}$ is also even. That is, $\widehat{u} *  \widehat{v}(-q) = \widehat{u} *  \widehat{v}(q)$.
\end{corollary}

\medskip

\section{Properties of the operator $\mathcal{L}$- proof of Proposition \ref{prop:Lprops}} \label{appendix:operatorL} 
In this section, we prove Proposition \ref{prop:Lprops} on  properties of the non-local linear difference operator $\mathcal{L}$ given by
\begin{align}
& (\mathcal{L} u)_n  \equiv \sum_{ m \in \mathbb{Z} } J_{|m - n|} \ (u_m - u_n), \qquad \{ J_{|m|} \}_{m \in \mathbb{Z}} \in l^1(\mathbb{Z}), \qquad J_0 = 0,\ J_m\ge0.
\end{align}
% We begin with a restatement and a proof of . 
%
%\begin{proposition} \label{prop:Lpropsapp}
%Assume $ J= \{J_m \}_{m \in \mathbb{Z}}$ is non-negative, real-valued, symmetric (such that $J_m = J_{- m}$, and in $J \in l^1(\mathbb{Z})$. 
%Then,
%\begin{enumerate}
%\item $\mathcal{L}$ maps $l^2(\mathbb{Z})$ to $l^2(\mathbb{Z})$. 
%\item $\mathcal{L}$ is self-adjoint.
%\item $-\mathcal{L}$ is non-negative.
%\item The spectrum of $-\mathcal{L}$ is continuous and equal to $[0,M_\star]$, where $M_\star = \underset{q \in [- \pi, \pi]}{ \max} \left(  4 \sum_{m = 1}^{\infty} J_m \ \sin^2 \left( \frac{qm}{2} \right) \right)$.
%\end{enumerate}
%\end{proposition}

\noindent {\it Boundness  },  $\|\mathcal{L}u\|_{l^2(\mathbb{Z})}\le\|J\|_{l^1(\mathbb{Z})}\ \|u\|_{l^2(\mathbb{Z})} $, follows by Young's inequality.
\medskip

 \noindent{\it  Self-adjointness of $\mathcal{L}$}\   on $l^2(\mathbb{Z})$ follows since
  $\left(\mathcal{L} u\right)_n= \sum_{m} \mathcal{L}_{mn}{u_m}$, where 
  $\mathcal{L}_{mn}= J_{|m-n|} - \|J\|_{l^1(\mathbb{Z})}\delta_{mn}$. 
%
%The second property holds since
%\begin{align}
%& \langle \mathcal{L} u, v \rangle  = \sum_{n \in \mathbb{Z}} \sum_{m \in \mathbb{Z}} J_{|m - n|} \ (u_m - u_n) \ \overline{v_n} =   \sum_{m \in \mathbb{Z}} \sum_{n \in \mathbb{Z}} J_{|n - m|} \ u_m \overline{ v_n } - \sum_{m \in \mathbb{Z}} \sum_{n \in \mathbb{Z}} J_{|n - m|} u_n \overline{ v_n } \nonumber \\
%& =  \sum_{m \in \mathbb{Z}} \sum_{n \in \mathbb{Z}} J_{|n - m|} \ u_m \overline{ v_n } - \sum_{m \in \mathbb{Z}} \sum_{n \in \mathbb{Z}} J_{|n - m|} u_m \overline{ v_m } =   \sum_{m \in \mathbb{Z}} \sum_{n \in \mathbb{Z}} \overline{ J_{|n - m|} \ (v_n - v_m) } \ u_m =   \langle u,  \mathcal{L}   v \rangle . 
%\end{align}

\noindent{\it  Non-negativity of $-\mathcal{L}$:}  By Young's inequality,  
$
 \langle \mathcal{L} u, u \rangle =  \sum_{n \in \mathbb{Z}} \sum_{m \in \mathbb{Z}} J_{|m - n|} \ u_m \ \overline{ u_n } - \| J \|_{l^1(\mathbb{Z})} \ \| u \|_{l^2(\mathbb{Z})}^2  \leq 0$. 
 
 \noindent{\it Dispersion relation of $-\mathcal{L}$:} This is a consequence of the following
\begin{lemma}
\label{lemma:nonlocaltransform}
Assume $ J= \{J_m \}_{m \in \mathbb{Z}}$ is non-non-negative, real-valued, symmetric (such that $J_m = J_{- m}$, and in $J \in l^1(\mathbb{Z})$. 
Then for any function $u \in l^1(\mathbb{Z}) \cap l^2(\mathbb{Z})$,
\begin{align}
\widehat{(\mathcal{L} u)}(q) = - 4 \sum_{m = 1}^{\infty} J_m \sin^2(q m/2) \ \widehat{u}(q) \ . 
\end{align}
\end{lemma}
\medskip
\noindent {\bf Proof of Lemma \ref{lemma:nonlocaltransform}:} First, observe that
{\footnotesize{
\begin{align}
\mathcal{F}_{_D} \left[ \sum_{ \substack{ m \in \mathbb{Z} \\ m \neq n }} J_{|m - n|} \ u_n \right](q) &  = \sum_{ n \in \mathbb{Z}} e^{- i q n}  \sum_{ \substack{ m \in \mathbb{Z} \\ m \neq n }} J_{|m - n|} \ u_n =  \sum_{ n \in \mathbb{Z}} e^{- i q n}  \sum_{ \substack{ m \in \mathbb{Z} \\ m \neq 0 }} J_{|m|} \ u_n \nonumber \\
& = 2 \ \left( \sum_{ m = 1}^{\infty}  J_m \right) \ \left( \sum_{ n \in \mathbb{Z}} e^{- i q n}  \  u_n \right) 
= 2 \ \left( \sum_{ m = 1}^{\infty}  J_m \right) \ \widehat{u}(q). 
\end{align}
}}
Next
{\footnotesize{
\begin{align}
\mathcal{F}_{_D} \left[ \sum_{ \substack{ m \in \mathbb{Z} \\ m \neq n }} J_{|m - n|} \ u_m \right](q) & = \sum_{ n \in \mathbb{Z}} e^{- i q n}  \sum_{ \substack{ m \in \mathbb{Z} \\ m \neq n }} J_{|m - n|} \ u_m = \sum_{ m \in \mathbb{Z} }  \  \sum_{ \substack{ n \in \mathbb{Z} \\ n \neq m } } e^{- i q n}  J_{|m- n|} \ u_m \nonumber \\
& = \sum_{ m \in \mathbb{Z} }  \  \sum_{ \substack{ n \in \mathbb{Z} \\ n \neq 0 } } e^{- i q (n + m)}  J_{|n|} \ u_m = \sum_{ m \in \mathbb{Z}} e^{- i q m} \left(  \sum_{ \substack{ n \in \mathbb{Z} \\ m \neq 0 }} J_{|n|} \ e^{- i q n} \right)  \ u_m \nonumber \\
& = \left(  \sum_{ n = 1}^{ \infty } J_{n} \ [ e^{- i q n}  + e^{i q n} ] \right)  \ \left( \sum_{ m \in \mathbb{Z}} e^{- i q m}  u_m \right)   =  \left(  \sum_{ m = 1}^{ \infty } J_{m} \ [ e^{- i q m}  + e^{i q m} ] \right) \widehat{u}(q). 
\end{align} }}
Therefore, since $4 \sin^2(q m / 2) = 2 - e^{ - i q m} -  e^{i q m} $, we have
{\footnotesize{
\begin{align}
\mathcal{F}_{_D} \left[ -\mathcal{L}u\right](q) =   \left(  \sum_{ m = 1}^{ \infty } J_{m} \ [ e^{- i q m}  + e^{i q m} - 2 ] \right) \widehat{u}(q)  = - 4 \sum_{m = 1}^{\infty} J_m \sin^2(q  m / 2) \ \widehat{u}(q). \hspace{2cm} 
\end{align}
%$\Box$
}}

\medskip

\section{Implicit Function Theorem} \label{appendix:IFT}
In this section, we state a variant of the implicit function theorem which is used in the proofs in Section \ref{section:1dproofnonlocal}.  \\

\begin{theorem}
\label{th:IFT}
(Implicit Function Theorem) Assume the following hypotheses.
\begin{enumerate}
\item $X$, $Y$, and $Z$ are Banach spaces.

 \item The mapping
$
  f:[0,1] \times X \times Y \rightarrow Z, \quad 
  (\alpha, x , y ) \mapsto f(\alpha,x,y), $ 
 satisfies $f(0, 0, 0) = 0$ and is continuous at $(0,0,0)$.

\item For all $(\alpha, x) \in [0,1] \times X$, the mapping
$
y \mapsto f(\alpha,x,y)
$
is Fr\'{e}chet differentiable which we denote $D_y f(\alpha, x,y): Y \rightarrow Z$. Furthermore, the mapping
$
(\alpha, x, y ) \mapsto  D_y f(\alpha, x,y)
$
 is continuous at $(0, 0, 0)$.

 \item $D_y f(0, 0, 0)$ is an isomorphism of $Y$ onto $Z$.
\end{enumerate}
Then there exist $\alpha_0 , \delta, \kappa > 0$ such that for $(\alpha, x) \in [0,\alpha_0) \times B_{\delta}(0) $, there exists a unique map $y_*: [0,\alpha_0) \times B_{\delta}(0) \mapsto Y$ such that $y_*[\alpha, x] \quad \text{is well-defined on} \quad [0, \alpha_0) \times B_{\delta}(0)$ and
\begin{align}
& y_*[0,0] = 0, \hspace{4.15cm}
\| y_*[\alpha, x] \|_{Y} \leq \kappa,  & \\
& \underset{(\alpha,x) \rightarrow (0,0)}{\lim} y_*[\alpha, x] = y_*[0,0] = 0, \hspace{1.2cm} 
 f(\alpha, x,y_*[\alpha,x]) = 0. &
\end{align}
Suppose also that
\begin{itemize}
\item[5.] For all $(\alpha, y) \in [0,1] \times Y$, the mapping
$
x \mapsto f(\alpha,x,y)
$
is Fr\'{e}chet differentiable, which we denote $D_x f(\alpha, x,y): Y \rightarrow Z$. Furthermore, the mapping
$
(\alpha, x, y ) \mapsto  D_x f(\alpha, x,y)
$
 is continuous at $(0, 0, 0)$.
\end{itemize}
Then $D_x y_*[\alpha,x]: X \rightarrow Y $ exists and is continuous.
\end{theorem}
\medskip
 
This variant of the standard implicit function theorem  \cite{Nirenberg_2001}
 is directly applicable to our setting.  In particular, we include the parameter $\alpha > 0$ explicitly and require only that $f(\alpha,x,z)$ and $D_x f(\alpha,x,y)$ be continuous at the origin. The complete proof may be found in \cite{Jenkinson_Weinstein_2015}.  \\

\section{Special functions ${\rm Li}_s(z)$, $\zeta(z)$, $\Gamma(z)$; the dispersion relation $M^s(q)$ of $- \mathcal{L}^s$}  \label{appendix:asymptotics}

This section contains key calculations for the proofs of Proposition \ref{prop:symbolexpo} and Lemma \ref{lemma:symbolconvlononlocal} on the asymptotics of the scaled dispersion relation $M_{\alpha}^s(Q) \ : \ M_{\alpha}^s(Q) \simeq |Q|^{2p}$ as $\alpha \to 0$ for $p = \min(1,s)$ and $0 < s \leq \infty$. The key result is given in Proposition \ref{prop:symbolconvnonlocal1}.

 Recall that $M^s(q)$ is the dispersion relation for the operator $- \mathcal{L}^s$. Recall from Lemma \ref{lemma:nonlocaltransform} that
$M^s(q) = \frac{1}{C_s} \sum_{ m = 1}^{\infty} \frac{4 \sin^2( q m / 2)}{m^{1 + 2s}}$ for $0 < s < \infty$ and $M^s(q) = \frac{1}{C_\infty} \sum_{ m = 1}^{\infty} 4 e^{- \gamma m}  \sin^2( q m / 2), \ \gamma > 0$, for $s = \infty$. As noted in Section \ref{section:nonlocalstrategy}, the factor $C_s$ and the $\alpha$-dependent rescalings are introduced so that the asymptotic coefficient of $|Q|^{2p}$ is one. 

We introduce the polylogarithm:
\begin{align}
{\rm Li}_s(z) \equiv \sum_{m = 1}^{\infty} \frac{z^m}{m^s}. 
\end{align}
for $z \in \mathbb{C}$, $s \in \mathbb{R}$. Observe in particular that for cases in which $|z| = 1$ and $s > 1$, the series ${\rm Li}_s(z)$ clearly converges. 
We also make use of the $\zeta$ and $\Gamma$ functions, defined to be the analytic continuation on $z \in \mathbb{C}$ of the functions defined by
\begin{align}
& \zeta(z) = \sum_{j = 1}^{\infty} \frac{1}{m^z}, \ \ {\rm for}\ \ {\rm Re} \ z > 1, \qquad \Gamma(z) = \int_0^{\infty} x^{z - 1} \ e^{- x} dx\ \ {\rm for}\ \ {\rm Re} \ z > 0.
\end{align}

 The following two propositions ensure that the $\zeta$ and $\Gamma$ functions are well-behaved in all of their appearances in this paper, in the sense that they are not singular or do not grow asymptotically when they appear in infinite series.

 \begin{proposition}[Properties of the $\Gamma$ function on $\mathbb{R}$,  \cite{Havil_2003}]  \label{prop:Gamma}
 { \ }  
 \begin{enumerate} 
 \item For $z \in \mathbb{R}$ fixed with $z > 0$, $\Gamma(z)$ is continuous and finite. 
 
 \item (Euler's reflection formula) For any $z \notin \mathbb{Z}$, we have
 $
 \Gamma(z) \ \Gamma(1 - z) = \frac{\pi}{\sin(\pi z)}. 
$
 \end{enumerate}
\end{proposition}

 \begin{proposition}[Properties of the $\zeta$ function on $\mathbb{R}$, \cite{Edwards_2001}] \label{prop:zeta}
 { \ }
 \begin{enumerate}
 \item For $z \in \mathbb{R}$ fixed with $z \neq 1$, $\zeta(z)$ is continuous and finite. 
 
 \item For $z \in \mathbb{R}$ fixed with $z > 1$, 
$
 | \zeta(z) | \leq 1 + \frac{1}{z - 1}. 
$
 
  \item For $z \in \mathbb{R}$, $
\underset{  z \to - \infty}{\lim} \zeta(z) = 0. $

\item (Trivial zeroes) Let $z =  - 2n$ for $n \in \mathbb{N}$. Then $\zeta(z) = 0$.

 \end{enumerate}
\end{proposition}
  
\medskip

We require the following lemmata which express the polylogarithm of an exponential function. 

\begin{lemma} \cite{Wood_1992} \label{lemma:polylog1}
Let $s \notin \mathbb{N}$, $s > 1$. Let $|\mu| <  2 \pi$. Then
\begin{align}
{\rm Li}_s \left( e^{\mu} \right) = \Gamma(1 - s) (- \mu)^{s - 1} + \sum_{j = 0}^{\infty} \frac{ \zeta(s - j) }{j!} \mu^j. 
\end{align}
\end{lemma}

\begin{lemma} \label{lemma:polylog2} \cite{Wood_1992}
Let $s \in \mathbb{N}$, $s > 1$. Let $|\mu| < 2 \pi$. Then
\begin{align}
{\rm Li}_s \left( e^{\mu} \right) = \frac{\mu^{s-1}}{(s - 1)!} \left[ \sum_{j = 1}^{s} \frac{1}{j} - \log( - \mu ) \right] + \sum_{ \substack{ j = 0 \\ j \neq s - 1}}^{\infty} \frac{ \zeta( s - j)}{j!} \mu^j. 
\end{align}
\end{lemma}

We now use Proposition \ref{prop:Gamma} through Lemma \ref{lemma:polylog2} to prove the following result on the asymptotics of the scaled dispersion relation $M_{\alpha}^s(Q)$.

\begin{proposition} \label{prop:symbolconvnonlocal1}
Let $Q \in \mathcal{B}_{_{\alpha}} = \left[ - \frac{\pi}{\alpha}, \frac{\pi}{\alpha} \right]$, let $C_s > 0$ be defined in \eqref{LC-def}, and let $\kappa_s(\alpha) > 0$ be defined in \eqref{eqn:kappa-def}.  Then there exists a function $f_s(Q;\alpha)$ such that $| f_s(Q; \alpha) | \lesssim 1$ and
 \begin{align}
 M^s_{\alpha}(Q) = \frac{1}{  \kappa_s(\alpha)} M^s(Q \alpha) = |Q|^{2p} + \left\{ \begin{array}{ll} \alpha^{2 - 2s} \ f_s( Q; \alpha) \ |Q|^2 & : \quad s < 1 \\ 
 \hline 
 \frac{1}{- \log(\alpha)} \ f_s(Q ; \alpha) \ \left( \frac{3}{2} - \log(|Q|) \right) \ |Q|^2 & : \quad s = 1 \\
 \hline 
 \alpha^{2s - 2} \ f_s(Q; \alpha) \ |Q|^{2s} & : \quad 1 < s < 2 \\
 \hline 
 (- \log(\alpha)) \ \alpha^2 \ f_s(Q; \alpha) \ |Q|^4 & : \quad s = 2 \\
 \hline 
 \alpha^2 \ f_s(Q; \alpha) \ |Q|^4 & : \quad 2 < s \leq \infty \end{array} \right. . \label{eqn:symbolexpoest3}
 \end{align}
\end{proposition}

\noindent {\bf Proof of Proposition \ref{prop:symbolconvnonlocal1}:} Let $q = Q \alpha \in \mathcal{B} = [- \pi, \pi]$. Recall that $M^s(q) = \frac{1}{C_s} \sum_{ m = 1}^{\infty} \frac{4 \sin^2( q m / 2)}{m^{1 + 2s}}$ for $0 < s < \infty$ and $M^s(q) = \frac{1}{C_\infty} \sum_{ m = 1}^{\infty} 4 e^{- \gamma m}  \sin^2( q m / 2), \ \gamma > 0$, for $s = \infty$. Here, $C_s > 0$ is to be determined.

{\it Case 1: Let $1 + 2s \notin \mathbb{N}, \ s \neq \infty$}. Expand $
4 \sin^2( q m / 2) = 2 - e^{i q m} - e^{- i qm} , 
$
and note that since $|q| \leq \pi < 2 \pi$, we may apply Lemma \ref{lemma:polylog1}:
\begin{align}
C_s \ M^s(q) & = 4 \sum_{m = 1}^{\infty} \frac{\sin^2( q m / 2) }{m^{1 + 2s}} =   \sum_{m = 1}^{\infty} \frac{2}{m ^{1 + 2s}} - \sum_{m = 1}^{\infty} \frac{e^{i q m}}{m ^{1 + 2s}} - \sum_{m = 1}^{\infty} \frac{e^{i q m}}{m ^{1 + 2s}} \nonumber \\
& =  2 \zeta(1 + 2s) - {\rm Li}_{1 + 2s}(e^{i q}) - {\rm Li}_{1 + 2s}(e^{- i q}) \nonumber \\
& = 2 \zeta(1 + 2s) - \Gamma(- 2s) (- i q)^{2s} - \sum_{j = 0}^{\infty} \frac{ \zeta(1 + 2s - j) }{j!} (- i q)^j  \nonumber \\
& \hspace{2.3cm} - \Gamma(- 2s) ( i q)^{2s} - \sum_{j = 0}^{\infty} \frac{ \zeta(1 + 2s - j) }{j!} ( i q)^j \nonumber \\
& = - \Gamma(- 2s) \ (e^{ - i \pi s} + e^{  i \pi s} ) \ |q|^{2s} -   \sum_{j = 1}^{\infty} \frac{ \zeta(1 + 2s - j) }{ j! } \ \left[ (-i)^j + i^j \right] \ q^j \nonumber \\
& = - 2 \Gamma(- 2s) \ \cos(\pi s) \  |q|^{2s} + 2 \sum_{j = 1}^{\infty} \frac{ \zeta(1 + 2s - 2j) }{ (2j)! } \ (-1)^{j + 1} \ |q|^{2j} . \label{eqn:expansionone}
\end{align}
Applying the rescaling $q = Q \alpha$ gives 
\begin{align}
M^s_{\alpha}(Q) = \frac{1}{  \kappa_s(\alpha)} M^s(Q \alpha) = - \frac{2 \Gamma(- 2s) \ \cos(\pi s)}{C_s \ \kappa_s(\alpha)} \ \alpha^{2s} \ |Q|^{2s} +  \ \sum_{j = 1}^{\infty} \frac{ \zeta(1 + 2s - 2j) }{ C_s \ \kappa_s(\alpha) \ (2j)! } \ (-1)^{j + 1} \alpha^{2j} \ |Q|^{2j} . \label{eqn:notintexpand}
\end{align}
The series is absolutely convergent on $Q = q/\alpha \in \left[ - \frac{\pi}{\alpha}, \frac{\pi}{\alpha} \right]$ due to Proposition \ref{prop:zeta}. Taking $C_s = - 2 \Gamma(- 2s) \ \cos(\pi s)$, $\kappa_s(\alpha) = \alpha^{2s}$ for $s < 1$ and $C_s = \zeta(2s - 1), \quad \kappa_s(\alpha) = \alpha^2$ for $s > 1$ gives \eqref{eqn:symbolexpoest3} for $1 + 2s \notin \mathbb{N}, s \neq \infty$. 

{\it Case 2: Let $1 + 2s \in \mathbb{N}, \ s \neq \infty$}. Again expand $
4 \sin^2( q m / 2) = 2 - e^{i q m} - e^{- i qm} , 
$ and note that since $|q| \leq \pi < 2 \pi$, we may apply Lemma \ref{lemma:polylog2}:
\begin{align}
M^s(q) & = 4 \sum_{m = 1}^{\infty} \frac{\sin^2( q m / 2) }{m^{1 + 2s}} =   \sum_{m = 1}^{\infty} \frac{2}{m ^{1 + 2s}} +- \sum_{m = 1}^{\infty} \frac{e^{i q m}}{m ^{1 + 2s}} - \sum_{m = 1}^{\infty} \frac{e^{i q m}}{m ^{1 + 2s}} \nonumber \\
& =  2 \zeta(1 + 2s) - {\rm Li}_{1 + 2s}(e^{i q}) - {\rm Li}_{1 + 2s}(e^{- i q}) \nonumber \\
& = 2 \zeta(1 + 2s) - \frac{ (i q)^{2s}}{(2s)!} \left[ \sum_{j = 1}^{2s} \frac{1}{j} - \log( - i q ) \right] - \sum_{ \substack{ j = 0 \\ j \neq 2s}}^{\infty} \frac{ \zeta(1 + 2s - j)}{j!} (i q)^j \nonumber \\
& \hspace{2.3cm} - \frac{ (- i q)^{2s}}{(2s)!} \left[ \sum_{j = 1}^{2s} \frac{1}{j} - \log( i q) \right] - \sum_{ \substack{ j = 0 \\ j \neq 2s}}^{\infty} \frac{ \zeta(1 + 2s - j)}{j!} (- i q)^j \nonumber \\
& =  - \left(   \frac{ e^{- i \pi s} + e^{i \pi s} }{(2s)!} \ \left( \sum_{j = 1}^{2s} \frac{1}{j} \right)  - \frac{e^{i \pi s} }{(2s)!} \ \left[ \log(|q|) - \frac{\pi i }{2} \right] - \frac{ e^{- i \pi s} }{(2s)!} \  \left[ \log(|q|) + \frac{\pi i }{2} \right]  \right) \  |q|^{2s} \nonumber \\
& \hspace{2.3cm} - \sum_{ \substack{ j = 1 \\ j \neq 2s}}^{\infty} \frac{ \zeta(1 + 2s - j ) }{j!} \left[ (- i)^j + i^j \right] q^j \nonumber \\
& =   \frac{ 2 \cos(\pi s) }{(2s)!} \left[ - \left( \sum_{j = 1}^{2s} \frac{1}{j} \right) + \log(|q|)    \right] \ |q|^{2s} + \frac{\pi \  \cos \left( \pi(s - 1/2) \right)}{(2s)!} \ |q|^{2s} \nonumber \\
& \hspace{2.3cm} + 2 \sum_{ \substack{ j = 1 \\ j \neq s} }^{\infty} \frac{ \zeta(1 + 2s - 2j) }{ (2j)! } \ (-1)^{j + 1} \ |q|^{2j} . \label{eqn:expansiontwo}
\end{align}
Applying the rescaling $q = Q \alpha$ gives 
\begin{align}
M^s_{\alpha}(Q) = \frac{1}{  \kappa_s(\alpha)} M^s(Q \alpha) = & \  \frac{ 2 \cos(\pi s) }{C_s \ \kappa_s(\alpha) \ (2s)!} \left[ - \left( \sum_{j = 1}^{2s} \frac{1}{j} \right) + \log(\alpha) + \log(|Q|)    \right] \  \alpha^{2s} \ |Q|^{2s} \nonumber \\
& + \frac{\pi \  \cos \left( \pi(s - 1/2) \right)}{C_s \ \kappa_s(\alpha) \ (2s)!} \alpha^{2s} \ |Q|^{2s}  
 + 2 \sum_{ \substack{ j = 1 \\ j \neq s} }^{\infty} \frac{ \zeta(1 + 2s - 2j) }{C_s \ \kappa_s(\alpha) \  (2j)! } \ (-1)^{j + 1} \alpha^{2j} \ |Q|^{2j} . \label{eqn:intexpand}
 \end{align}
The infinite series is absolutely convergent on $Q = q/\alpha \in \left[ - \frac{\pi}{\alpha}, \frac{\pi}{\alpha} \right]$ due to Proposition \ref{prop:zeta}. Let $C_s = - 2 \Gamma(- 2s)  \cos(\pi s)$, $\kappa_s(\alpha) = \alpha^{2s}$ for $s = 1/2$ and $C_s = \zeta(2s - 1), \quad \kappa_s(\alpha) = \alpha^2$ for $s > 1$. For $s = 1$, take $C_s = 1$ and $\kappa_s(\alpha) = (- \log(\alpha)) \ \alpha^2$. 
This gives \eqref{eqn:symbolexpoest3} for $1 + 2s \in \mathbb{N}, s \neq \infty$.

\begin{remark} \label{remark:euler} Note that one of the first two terms in \eqref{eqn:intexpand} will be zero, depending on whether $s \in \mathbb{N}$ or $s = (2k + 1)/2$ for $k \in \mathbb{N}$. 

Furthermore, in the case where $s = (2k + 1)/2$ for $k \in \mathbb{N}$, the series in \eqref{eqn:intexpand} is finite. To see this, observe from Proposition \ref{prop:zeta} that the ``trivial zeros" of the zeta function on the real line lie occur at negative even integers. In the series above, this occurs when
$
1 + 2s - 2j  \leq -2 \quad \Longrightarrow \quad j \geq s + 3/2. $

We also note that for $s = (2k + 1)/2$ for $ k \in \mathbb{N}$, the expansion \eqref{eqn:intexpand} is compatible with the expansion \eqref{eqn:notintexpand} where we assumed that $1 + 2s \notin \mathbb{N}$. To see this we use Euler's reflection formula from Proposition \ref{prop:Gamma} to get, for $k \in \mathbb{N}$, 
\begin{align}
\underset{s \to \frac{2k - 1}{2} }{\lim}  \ \ - 2 \ \Gamma(- 2s) \ \cos(\pi s) & = \underset{ s \to \frac{2k - 1}{2}  }{\lim} \ \ \frac{- 2 \pi \ \cos(\pi s)}{ \sin([2 s+1] \pi)  \ \Gamma(2s + 1)}  =  \underset{  s \to \frac{2k - 1}{2} }{\lim} \ \ \frac{  2 \pi \cos(\pi s)}{ \sin(2 \pi s) \ \Gamma(2s + 1)}  \nonumber \\
& = \underset{  s \to \frac{2k - 1}{2} }{\lim} \ \ \frac{    \pi  }{ \sin( \pi s) \ \Gamma(2s + 1)} = \frac{  \pi} { \sin \left( \pi (k -  1/2) \right) \ \Gamma(2k) } \nonumber \\
& = \frac{ \pi (-1)^{k + 1} }{ (2k - 1)!} = \frac{ \pi (-1)^{s - 1/2} }{ (2s)! } \bigg|_{s = \frac{2k - 1}{2}}. \label{eqn:eulerremark}
\end{align}

\end{remark}

 {\it Case 3: Finally, let $s = \infty$}. We have  $M^s(q) = \frac{1}{C_\infty} \sum_{ m = 1}^{\infty} 4 e^{- \gamma m}  \sin^2( q m / 2), \ \gamma > 0$ and $q \in [- \pi, \pi]$. Note that $\partial_q^4 \sin^2(q m /2) = - \frac{1}{2} \ m^4 \ \cos(q m)$. Expanding about $q = 0$, Taylor's Theorem implies that there exists $z \in [- \pi, \pi]$  such that
 $ 4 \sin^2( q m / 2) = m^2 \ |q|^2 -  2 \ m^4 \ \cos(z m) \ |q|^4.$ on $q \in [- \pi, \pi]$. In turn, 
 \begin{align}
 M^s(q) = \frac{1}{C_\infty} \sum_{ m = 1}^{\infty} 4 e^{- \gamma m}  \sin^2( q m / 2) = \frac{1}{C_s} \ \sum_{m = 1}^{\infty} \ m^2 \ e^{- \gamma m} \ |q|^2  - \frac{2}{  C_s} \ \sum_{ m = 1}^{\infty} m^4 \ \cos(z m) \ |q|^4. 
 \end{align}
Therefore, for $Q \in \left[ - \frac{\pi}{\alpha}, \frac{\pi}{\alpha} \right]$ and $M^s_{\alpha}(Q) = \frac{1}{  \kappa_s(\alpha)} M^s(Q \alpha)$, 
\begin{align}
M^s_{\alpha}(Q) = \frac{1}{C_s \ \kappa_s(\alpha)} \ \left( \sum_{m = 1}^{\infty} \ m^2 \ e^{- \gamma m} \right) \ \alpha^2 \ |Q|^2   - \frac{2}{  C_s \ \kappa_s(\alpha)} \ \sum_{ m = 1}^{\infty} m^4 \ \cos(z m) \ e^{- \gamma m}   \  \alpha^4 \ |Q|^4.
\end{align}
Observe that
\begin{align}
& \left( \sum_{m = 1}^{\infty} \ m^2 \ e^{- \gamma m} \right) =  \frac{ {\rm exp}(\gamma) (  {\rm exp}(\gamma) + 1) }{ ({\rm exp}(\gamma) - 1)^3} = C_{\infty}, \nonumber \\
& \bigg| \sum_{ m = 1}^{\infty} m^4 \ \cos(z m) \ e^{- \gamma m}  \bigg| \leq   \sum_{ m = 1}^{\infty} m^4  \ e^{- \gamma m}   \lesssim 1. 
\end{align}
Taking $\kappa_s(\alpha) = \alpha^2$ gives \eqref{eqn:symbolexpoest3} for $s = \infty$. This completes the proof of Proposition \ref{prop:symbolconvnonlocal1}.  $\Box$

\section{Exponential decay of solitary waves in Fourier space}  \label{appendix:expogeneral}
We are interested in bounding solutions to the following two equations:
\begin{enumerate}

\item The Fourier transform of continuum FNLS for $d = 1,2,3$ and frequency $\omega = -1$, \eqref{Psi-eqnnonlocal}
\begin{align}
 \left[ 1 + |Q|^{2p} \right] \  \widetilde{\psi}(Q) - \left( \frac{1}{2 \pi} \right)^{2d} \ \widetilde{\psi} * \widetilde{\psi} * \widetilde{\psi}(Q)  = 0 , \qquad Q \in \mathbb{R}^d. \label{eqn:NLS1}
\end{align}
Here, $K = 1$, $C_1 \equiv - (1/2 \pi)^{2d}$, $\tau_1 \equiv 0$, $A \equiv \mathbb{R}^d$, and $M(Q) \equiv |Q|^{2p}$. 

\item The rescaled Fourier transform of DNLS localized to the Brillouin zone for $d = 1$, \eqref{eqn:Phieqnnonlocal}:
\begin{align}
  \left[ 1 + M^s_{\alpha}(Q) \right] \  \widehat{\Phi}(Q) - \frac{\chi_{_{\mathcal{B}_{_{\alpha}}}}(Q)}{4 \pi^2} \sum_{m = -1}^1 e^{2 \pi i \sigma  m} \  \widehat{\Phi} * \widehat{\Phi} * \widehat{\Phi}(Q - 2 m \pi / \alpha) = 0 , \qquad Q \in \mathbb{R}. \label{eqn:DNLS1}
\end{align}
Here, $K = 3$, $C_k \equiv (-1/4 \pi^2) e^{2 \pi i \sigma m_k}$ for $m_k \in \{-1, 0, 1 \}$,  $\tau_k \equiv 2 m_k \pi / \alpha$, $A \equiv \mathcal{B}_{_{\alpha}}$, and $M(Q) \equiv M^s_{\alpha}(Q) = \frac{4}{\kappa_s(\alpha)} \sum_{m = 1}^{\infty} \frac{  \sin^2(Q m \alpha/2)}{m^{1 + 2s}}$. 
\end{enumerate}

 \bigskip

In this section, we prove the exponential decay of solutions to an equation of the general form,
\begin{align}
\left[ 1 + M(Q) \right] \ \Phi(Q) +  \chi_{_{A}}(q)  \ \sum_{k = 1}^K \     C_k \ \Phi * \Phi * \Phi (Q + \tau_k ) = 0.   \label{eqn:generaldecayeqn}
\end{align}
where $0 < \eta \leq 1$, 
$K \geq 1$ is an integer, and $C_k \in \mathbb{R}, \ \tau_k \in \mathbb{R}^d$ are constants for $1 \leq k \leq K$, $A \subset \mathbb{R}^d$ and $M(Q)$ is a continuous function which satisfies, for some constants $0 < \eta \leq 1$ and $D_M > 0$, 
\begin{align}
& \frac{ |Q|^{\eta}}{1 + M(Q)} \leq D_M,  \qquad {\rm and} \qquad  |Q| \leq |Q + \tau_k |, \qquad {\rm for} \qquad Q \in A. 
\end{align}
We obtain the exponential decay
\begin{align}
\| e^{\mu |Q|^{\eta} } \Phi(Q) \|_{L^{2,a}(\mathbb{R}^d_Q)} \lesssim \| \Phi \|_{L^{2,a}(\mathbb{R}^d)},  \qquad \mu > 0. 
\end{align}

 We follow an approach similar to that given in \cite{Frank_Lenzmann_2010} and motivated by the approach to proving exponential decay estimates in \cite{Bona_Li_1997}. In particular, we make use of the following identity.  \\

\begin{lemma} \label{lemma:abel} (Abel's identity \cite{Riordan_1968}) For any constants $c_1, c_2 \neq 0$, we have
\begin{align}
\sum_{x = 0}^{y}  \left( \begin{array}{l} y \\ x \end{array} \right) (x + c_1)^{x - 1} \ (y - x + c_2)^{y - x  - 1} = \frac{c_1 + c_2}{c_1 c_2} (y + c_1 + c_2)^{y -1}. 
\end{align}

\medskip
\begin{remark} We use $\| f g \|_{H^a(\mathbb{R}^d)} \lesssim \| f \|_{H^a(\mathbb{R}^d)} \  \| f \|_{H^a(\mathbb{R}^d)} $ since we seek to show $\| \widehat{\Phi^{\rm off}} - \widehat{\Phi^{\rm on}} \|_{L^2(\mathcal{B})} \lesssim \| \widehat{\Phi^{\rm off}} - \widehat{\Phi^{\rm on}} \|_{L^{2,a}(\mathbb{R}^d)} $. 
In \cite{Frank_Lenzmann_2010}, Frank {\it et. al.} obtain the equivalent result in $L^1(\mathbb{R})$ using Young's inequality: $ \left\| f_1 * f_2 \right\|_{L^1(\mathbb{R})} \leq \left\| f_1   \right\|_{L^1(\mathbb{R})} \ \left\| f_2   \right\|_{L^1(\mathbb{R})}$. We may also perform our analysis in $L^1(\mathbb{R})$.  
\end{remark}
\medskip

\end{lemma}

 We seek to show $\Phi(Q) \sim e^{- \mu |Q|^{\eta} }, \ Q \in \mathbb{R}^d, \ \mu> 0$. To prove this, we will estimate the moments of $\Phi$, given by $|Q|^{j \eta} \Phi(q)$, $j \in \mathbb{N}$, inductively in $L^{2,a}(\mathbb{R}^d)$, and then use these estimates to bound the Taylor expansion of $e^{\mu |Q| }$ for an appropriate choice of $C > 0$.

\begin{lemma} \label{lemma:expogeneral}
Let $a > d/2$, $0 < \eta \leq 1$, $K \geq 1$ be an integer, and $C_k \in \mathbb{R}, \ \tau_k \in \mathbb{R}^d$ be constants for $1 \leq k \leq K$. Let $A \subset \mathbb{R}^d$ and $M(Q) \geq 0$ be a continuous function which satisfy, for some constant $D_M > 0$, 
\begin{align}
& \frac{|Q|^{\eta}}{1 + M(Q)} \leq D_M, \qquad {\rm and} \qquad |Q| \leq |Q + \tau_k |, \qquad {\rm for} \qquad Q \in A. \label{eqn:hypoth}
\end{align}
Let $\Phi \in L^{2,a}(\mathbb{R}^d)$ be the solution to 
\begin{align}
\left[ 1 + M(Q) \right] \ \Phi(Q) +  \chi_{_{A}}(Q)  \ \sum_{k = 1}^K \     C_k \ \Phi * \Phi * \Phi (q + \tau_k ) = 0.   \label{eqn:generaldecayeqn2}
\end{align}
Then there exists a constant $\mu = \mu \left( \| \Phi \|_{L^{2,a}(\mathbb{R}^d)} \right) > 0$ such that
\begin{align}
\left\| e^{\mu | Q  |^{\eta} } \ \Phi(Q) \right\|_{L^{2,a}(\mathbb{R}^d_Q)} \lesssim \left\|   \Phi \right\|_{L^{2,a}(\mathbb{R}^d)}. 
\end{align}

\end{lemma}

\noindent {\it Proof of Lemma \ref{lemma:expogeneral}:} Our strategy is to estimate the moments $|Q|^{j \eta} \ \Phi(Q)$, $j \geq 0$. Observe that by Taylor expansion, for any $\mu > 0$, 
\begin{align}
& \left\| e^{\mu |Q|^{\eta} } \ \Phi(Q) \right\|_{L^{2,a}(\mathbb{R}_Q^d)}   \leq \sum_{j = 0}^{\infty}    \left\| \frac{\mu^j}{j!} \ | Q |^{j \eta}  \ \Phi(Q) \ \right\|_{L^{2,a}(\mathbb{R}_Q^d)} . \label{eqn:exposeriesratio0}
\end{align}
We then have the following proposition. \\

\begin{proposition} \label{prop:momentsgeneral}
There exists a constant $b = b \left( \left\|  \Phi \right\|_{L^{2,a}(\mathbb{R}^d)} \right) >  0$ such that $\Phi \in L^{2,a}(\mathbb{R}^d)$ satisfies the estimates
\begin{align}
\left\| \ |Q |^{j \eta} \ \Phi(Q) \ \right\|_{L^{2,a}(\mathbb{R}_Q^d)} \leq b^j \ (2j + 1)^{j - 1} \left\| \Phi \right\|_{L^{2,a}(\mathbb{R}^d)} , \quad j \in \mathbb{N}, \quad j \geq 0. \label{eqn:indhyp}
\end{align}
\end{proposition}

\noindent The proof of Proposition \ref{prop:momentsgeneral} is found below. By Proposition \ref{prop:momentsgeneral} and \eqref{eqn:exposeriesratio0}, for any $\mu > 0$, 
\begin{align}
& \left\| e^{\mu |Q|^{\eta} } \ \Phi(Q) \right\|_{L^{2,a}(\mathbb{R}_Q^d)} \leq \ \left\| \Phi \right\|_{L^{2,a}(\mathbb{R}^d)}  \  \sum_{j = 0}^{\infty} A_j, \label{eqn:exposeriesratio}
\end{align}
where
$
A_j \equiv \frac{\mu^j}{j!} \ b ^j \ (2j + 1)^{j - 1} , 
$
and $\mu$ is to be determined so that the sum converges. 
The ratio test gives
\begin{align}
\underset{ j \to \infty}{\lim} \ \frac{A_{j+1}}{A_j} = \underset{ j \to \infty}{\lim} \ \frac{b  \ \mu  \ (2j + 3)^j}{ (j + 1) \ (2j + 1)^{j-1}} = 2 b \mu e. 
\end{align}
If $\mu < \frac{1}{2 e b \left( \| \Phi \|_{L^{2,a}(\mathbb{R})} \right) } $,  \eqref{eqn:exposeriesratio} is convergent, completing the proof of Lemma \ref{lemma:expogeneral}. $\Box$ \\

\bigskip

\noindent {\bf Proof of Proposition \ref{prop:momentsgeneral}:} 
Rewrite equation \eqref{eqn:generaldecayeqn2} as 
\begin{align}
\Phi(Q) = \frac{ \chi_{_{A}}(Q) }{1 + M(Q)} \  \sum_{k = 1}^K    \ C_k \ \Phi * \Phi * \Phi (Q + \tau_k )  .   \label{eqn:generaldecayeqn3}
\end{align}
We prove \eqref{eqn:indhyp} by induction. Define
\begin{align}
W(Q) \equiv  \ \Phi * \Phi(Q). \label{eqn:Wdef}
\end{align}
Let $m \geq 1$ and multiply equation \eqref{eqn:generaldecayeqn3} by $| Q |^{m \eta} $ to get
\begin{align}
|Q|^{m \eta} \ \Phi(Q) = \frac{ \chi_{_{A}}(Q) \ |Q|^{m \eta} }{1 + M(Q)} \  \sum_{k = 1}^K    \ C_k \ W * \Phi (Q + \tau_k ) . \label{eqn:expo1}
\end{align}

Note that \eqref{eqn:indhyp} holds trivially for $j = 0$ if we choose $b > 0$. Now assume, for all $0 \leq j \leq m - 1$, that there exists a constant $b > 0$ such that \eqref{eqn:indhyp} holds. 
We will prove that \eqref{eqn:indhyp} holds for $ j = m$. 

Observe that by \eqref{eqn:hypoth},
\begin{align}
 &  \frac{ \chi_{_{A}}(Q) \ |Q|^{m \eta} }{1 + M(Q)} \leq |Q|^{(m - 1) \eta } \ \frac{ \chi_{_{A}}(Q) \  |Q|^{\eta}}{ 1 + M(Q) } = |Q|^{ (m - 1) \eta } \  \chi_{_{A}}(Q) \ D_M \leq D_M \  |Q + \tau_k |^{(m-1) \eta}.  \label{eqn:expo2} 
\end{align}
Next, by Lemma \ref{lemma:tri} and $
|\widetilde{Q}|^{m-1} \leq \sum_{l = 0}^{m - 1} \left( \begin{array}{c} m-1 \\ l \end{array} \right) | \widetilde{Q} - \xi|^l \ | \xi|^{m - l - 1}, $
\begin{align}
|\widetilde{Q}|^{(m - 1) \eta} \leq \sum_{l = 0}^{m - 1} \left( \begin{array}{c} m-1 \\ l \end{array} \right)^{\eta} | \widetilde{Q} - \zeta|^{l \eta} \ | \zeta|^{(m - l - 1) \eta} \leq  \sum_{l = 0}^{m - 1} \left( \begin{array}{c} m-1 \\ l \end{array} \right) | \widetilde{Q} - \zeta|^{l \eta} \ | \zeta|^{(m - l - 1) \eta}. \label{eqn:trianglebinomial}
\end{align}
 Combining \eqref{eqn:trianglebinomial} (for $\widetilde{Q} \equiv Q + \tau_k$) with \eqref{eqn:expo1} and \eqref{eqn:expo2}, we have
\begin{align}
|Q|^{m \eta} \ | \Phi(Q) | & \leq  D_M \   \sum_{k = 1}^K    \  |C_k|  \ |Q + \tau_k |^{(m - 1) \eta} \ \Big| \int_{\mathbb{R}^d} W(Q + \tau_k - \xi)   \Phi (\xi ) d\xi  \Big| \nonumber \\
& \leq  D_M \ \sum_{k = 1}^K    \ |C_k| \ \sum_{l = 0}^{m - 1}   \left( \begin{array}{c} m-1 \\ l \end{array} \right) \ \left( | \xi|^{l \eta}  \ |W| \right) *^{\xi} \left( |\xi|^{(m - l - 1) \eta} \ |\Phi| \right) (Q + \tau_k ), 
\end{align}
 and therefore by the inductive hypothesis \eqref{eqn:indhyp} (which holds for $0 \leq m - l - 1 \leq m - 1$), we have
 \begin{align}
 & \left\| \ |Q|^{m \eta} \   \Phi(Q)    \ \right\|_{L^{2,a}(\mathbb{R}_Q^d)} \nonumber \\
 & \leq  K \ D_a \ D_M \  \underset{ k = 1, \dots, K} {\max} \{ | C_k | \} \  \sum_{l  = 0}^{m -1} \left( \begin{array}{c} m-1 \\ l \end{array} \right) \ \left\| | \xi|^{l \eta} \ W( \xi) \ \right\|_{L^{2,a}(\mathbb{R}_{\xi}^d)} \ \left\| | \xi|^{( m - l - 1) \eta } \ \Phi (\xi) \  \right\|_{L^{2,a}(\mathbb{R}_{\xi}^d)} \nonumber \\
 &  \leq K \ D_a \ D_M \ \underset{ k = 1, \dots, K} {\max} \{ |C_k| \} \  \sum_{l  = 0}^{m -1} \left( \begin{array}{c} m-1 \\ l \end{array} \right) b_1 \ b_2^{m - l - 1} \  [2 (m - l - 1) + 1]^{m - l - 2} \nonumber \\
 & \hspace{5cm} \cdot  \left\| | \xi|^{l \eta}  \ W( \xi) \ \right\|_{L^{2,a}(\mathbb{R}_{\xi}^d)}  \ \left\| \Phi \right\|_{L^{2,a}(\mathbb{R}^d)} . \label{eqn:fullbd}
 \end{align}

Next, we must estimate $| \xi|^{l \eta} \ W( \xi)$. Equations \eqref{eqn:Wdef} and \eqref{eqn:trianglebinomial} give
\begin{align}
& |  \xi |^{l \eta} \ | W(\xi) | \leq   \ \sum_{k = 0}^l \left( \begin{array}{c} l \\ k \end{array} \right) \ \left( | \zeta |^{k \eta} | \Phi | \right) *^{\zeta} \left( |\zeta |^{( l - k ) \eta} \  | \Phi | \right) (\xi),
\end{align}
which in turn gives, by the inductive hypothesis \eqref{eqn:indhyp} (for $0 \leq k \leq l \leq m - 1$ and $0 \leq l - k \leq m - 1$) and Abel's identity \ref{lemma:abel}, 
\begin{align}
 \left\|  |  \xi |^{l \eta} \   W( \xi )  \right\|_{L^{2,a}(\mathbb{R}_{\xi}^d)}  
& \leq D_a \ \sum_{k = 0}^l \left( \begin{array}{c} l \\ k \end{array} \right) \ \left\| |\zeta |^{k \eta} \ \Phi(\zeta )  \right\|_{L^{2,a}(\mathbb{R}_{\zeta}^d)} \ \left\| |\zeta |^{(l - k) \eta} \ \Phi(\zeta)  \right\|_{L^{2,a}(\mathbb{R}_{\zeta}^d)} \nonumber \\
& \leq D_a \ \sum_{k = 0}^l \left( \begin{array}{c} l \\ k \end{array} \right)  \ b^k \ b^{l - k} \ (2k + 1)^{k - 1} \ [ 2 (l - k) + 1 ]^{l - k - 1} \ \left\| \Phi \right\|^2_{L^{2,a}(\mathbb{R}^d)} \nonumber \\
& =   D_a \ \left\| \Phi \right\|^2_{L^{2,a}(\mathbb{R}^d)}   \ b^l \ \sum_{k = 0}^l \left( \begin{array}{c} l \\ k \end{array} \right) \ (2k + 1)^{k - 1} \ [ 2 (l - k) + 1 ]^{l - k - 1} \nonumber \\
& =  D_a \ \left\| \Phi \right\|^2_{L^{2,a}(\mathbb{R}^d)}    \ b^l \ 2^{l - 2} \sum_{k = 0}^l \left( \begin{array}{c} l \\ k \end{array} \right) \ (k + 1/2)^{k - 1} \ [   l - k + 1/2 ]^{l - k - 1} \nonumber \\
& =    D_a \  \left\| \Phi \right\|^2_{L^{2,a}(\mathbb{R}^d)}     \ b^l \ 2^l \ (l + 1)^{l - 1} =  2 \ D_a   \ b^l \ \left\| \Phi \right\|^2_{L^{2,a}(\mathbb{R}^d)}   \ ( 2 l + 2 )^{l - 1}  . \label{eqn:Wbd}
\end{align}

Combining \eqref{eqn:Wbd} with \eqref{eqn:fullbd}, along with Abel's identity again, gives 
\begin{align}
 \left\| \ |Q|^{m \eta} \   \Phi(Q)    \ \right\|_{L^{2,a}(\mathbb{R}_Q^d)} & \leq   2 \ K \ D_M \ D_a^2   \ \underset{ k = 1, \dots, K} {\max} \{ | C_k | \} \ \left\| \Phi \right\|^3_{L^{2,a}(\mathbb{R}^d)}    \ \nonumber \\
 & \hspace{2cm} \cdot  \sum_{l  = 0}^{m -1} \left( \begin{array}{c} m-1 \\ l \end{array} \right)  b^{m - l - 1} \ b_2^l  \  (2l + 2)^{l -1} \  [2 (m - l - 1) + 1]^{m - l - 2}  \nonumber \\
& = 2 \ K \ D_M \ D_a^2   \ \underset{ k = 1, \dots, K} {\max} \{ | C_k | \} \ \left\| \Phi \right\|^3_{L^{2,a}(\mathbb{R}^d)}  \ b^{m - 1}  \ 2^{m - 3} \ \nonumber \\
 & \hspace{2cm} \cdot  \sum_{l  = 0}^{m -1} \left( \begin{array}{c} m-1 \\ l \end{array} \right) \ ( l + 1 )^{l -1} \  [  (m - l - 1) + 1 /2]^{m - l - 2}  \nonumber \\
 & =  2 \ K \ D_M \  D_a^2   \ \underset{ k = 1, \dots, K} {\max} \{  | C_k | \} \ \left\| \Phi \right\|^3_{L^{2,a}(\mathbb{R}^d)}  \ b^{m - 1}  \ 2^{m - 3} \ 3 \ (m - 1 + 3/2)^{m - 2} \nonumber \\
 &  =  3 \ K \ D_M \ D_a^2    \ \underset{ k = 1, \dots, K} {\max} \{ | C_k | \} \ \left\| \Phi \right\|^3_{L^{2,a}(\mathbb{R}^d)}  \ b^{m - 1}  \ (2m + 1)^{m - 2}. 
\end{align}
 To complete the proof that \eqref{eqn:indhyp} is satisfied for $j = m \geq 1$, we need to show that this quantity is bounded by
$ 
  b^m \ (2m + 1)^{m-1} \  \left\| \Phi \right\|_{L^{2,a}(\mathbb{R}^d)}. 
$
It will then follow that
\begin{align}
 \left\| |Q|^{m \eta} \ \Phi(q) \right\|_{L^{2,a}(\mathbb{R}_Q^d)} \leq  b^m (2m + 1)^{m-1} \  \left\| \Phi \right\|_{L^{2,a}(\mathbb{R}^d)} . \label{eqn:hypfinal}
\end{align}
Condition \eqref{eqn:hypfinal} is satisfied it for any $m \geq 1$, $b_2$,  dependent on $  \left\| \Phi \right\|_{L^{2,a}(\mathbb{R}^d)} $, satisfies
\begin{align}
  3 \ K \ D_M \ D_a^2 \ \underset{ k = 1, \dots, K} {\max} \{ | C_k | \} \ \left\| \Phi \right\|^2_{L^{2,a}(\mathbb{R}^d)} \leq  b \ (2m + 1), \quad \forall m \geq 1. \label{eqn:conditionind}
\end{align}
It suffices to satisfy \eqref{eqn:conditionind} for $m = 1$ by choosing $b > 0$ sufficiently large. Recall that $b >0$ satisfies  \eqref{eqn:indhyp} trivially for $j = 0$. This completes the proof of Proposition \ref{prop:momentsgeneral}. $\Box$

\bibliography{nonlocalwriteup}
\bibliographystyle{siam}

\end{document}